\documentclass[11pt, a4paper,titlepage]{article}

\usepackage{graphicx,amsmath, amsthm, amssymb}
\usepackage[integrals]{wasysym} 
\usepackage{mathtools} 
\mathtoolsset{centercolon} 

\usepackage{color,epsfig} 
\usepackage{graphics}

\usepackage[backref = page]{hyperref}								
\hypersetup{bookmarksopen=false,colorlinks,pdfstartview=FitH, citecolor=blue,runcolor=blue,anchorcolor=blue,linkcolor=blue,
					pdftitle = {PM MSc thesis}
		   		 	}

\usepackage{geometry} 
\usepackage[activate={true,nocompatibility}, final, tracking=true, kerning=true, spacing=true, factor=1100, stretch=10, shrink=10,
protrusion=true, expansion=true]{microtype}
\microtypecontext{spacing=nonfrench}

\makeatletter
\def\@normalsize{\@setsize\normalsize{10pt}\xpt\@xpt
\abovedisplayskip 10pt plus2pt minus5pt\belowdisplayskip 
\abovedisplayskip \abovedisplayshortskip \z@ 
plus3pt\belowdisplayshortskip 6pt plus3pt 
minus3pt\let\@listi\@listI}
\def\subsize{\@setsize\subsize{12pt}\xipt\@xipt}
\def\section{\@startsection {section}{1}{\z@}{1.0ex plus 1ex minus .2ex}{.2ex plus .2ex}{\large\bf}}
\def\subsection{\@startsection {subsection}{2}{\z@}{.2ex plus 1ex} {.2ex plus .2ex}{\subsize\bf}}
\makeatother

\begin{document}

\title{\Large {\bf Chiral Corrections to the Gell-Mann-Oakes-Renner Relation from\\
QCD Sum Rules}}

\author{\normalsize Preshin Moodley\\ \\\normalsize Supervisor: Emeritus Professor C.A. Dominguez\\\normalsize Co-Supervisor: Professor A. Peshier\\ \\
				{\em \normalsize Department of Physics, University of Cape Town}}

\date{\normalsize Submitted 12 October 2010}

\footnotetext{Submitted in fulfillment of the requirements for a M.Sc. degree in Theoretical Physics.}

\maketitle
\thispagestyle{empty}

\clearpage
\thispagestyle{empty}
\subsection*{\centering ABSTRACT}

We calculate the next-to-leading order corrections to the $SU(2)\otimes SU(2)$ and $SU(3)\otimes SU(3)$ Gell-Mann-Oakes-Renner relations. We use a pseudoscalar correlator calculated from Perturbative QCD up to five loops and use the QCD Finite Energy Sum Rules with integration kernels tuned to suppress the importance of the hadronic resonances. This leads to a substantial reduction in the systematic uncertainties from the experimentally unknown resonance spectral function. We use the method of Fixed Order and Fixed Renormalization Scale Perturbation Theory to compute the integrals. We calculate these corrections to be $\delta_\pi = 0.060 \pm 0.014$ and $\delta _K =0.64 \pm 0.24$. As a result of these new values we predict the value of the light quark condensate $\left\langle {0|\bar qq|0} \right\rangle  =  - \left( {266 \pm 5{\text{ MeV}}} \right)^3$ and the Chiral Perturbation Theory low energy constant $H_2^r  =  - \left( {4.9 \pm 1.8} \right) \times 10^{ - 3}$. Results from this work have been published as: J. Bordes, C.A. Dominguez, P. Moodley, J. Pe$\widetilde{\text{n}}$arrocha and K. Schilcher, J. High Ener. Phys. 05 (2010) 064.


\clearpage

\tableofcontents 

\clearpage

\listoffigures
\listoftables


\clearpage
\section{Introduction}
\label{sec:Introduction}

In recent years it has become generally accepted that the correct physical theory for the description of the strongly interacting particles is Quantum Chromodynamics (QCD). Initially there had been strong criticism against QCD since the fundamental constituents of the theory, the quarks and gluons, had never been observed in nature or in the laboratory. This is still the case today, but there have been many predictions from QCD about the bound states of these quarks (hadrons) which can be tested in the laboratory. Since there has been no analytic solution to QCD as yet, these predictions have been made possible by exploiting one of the key features of QCD, which is Asymptotic freedom and its closely related counterpart Confinement. These interesting features of QCD were discovered by Gross, Wilcek and Politzer \cite{Gross1,Gross2}. Confinement provides a theoretical explanation for why there are no hadronic colour multiplicity states observed in nature. The reason for this is, as the quarks move further away from each other, their interaction starts to increase i.e. the value of $\alpha_s$, the strong coupling (which sets the scale of the interaction), increases up to the point where there is so much energy in the two quark system that it is energetically more economical to bring two more quarks into existence from the vacuum and decrease the total energy of the system. For shorter distances between quarks the strong coupling goes to zero.

This allows us to use the tools from Perturbation theory to make an expansion of the equations of motion, which cannot be solved analytically (as yet), in terms of this small parameter $\alpha_s$. In this way we can extract information from QCD to get an approximate understanding of the behavior of the quarks and gluons.

The two main ways to compute hadronic parameters from QCD have been with the use of Lattice QCD were numerical solutions to the equations of motion are found. This is good since no approximations are made and solutions obtained are nearly exact, but since the computer is a black box we have no understanding about the subtle mechanisms at work in the theory. This is an undesirable feature of the results from Lattice QCD. The other way to extract information from the QCD equations of motion is the use of the QCD sum rules.

The idea of the QCD sum rules was developed in the late 1970's by Shifman \emph{et al} \cite{shifman1,shifman2} and has now become one of the work horses of analytic methods to extract information from QCD. QCD sum rules assume that confinement exists in QCD and tries to parameterize this effect. QCD sum rules are based on two ideas, the Operator Product Expansion (OPE) which separates the short and long distance quark-gluon interactions, and the application of Cauchy's Theorem in the complex energy (squared) plane. The short distance interactions are calculated using Perturbative QCD and the long distance interaction, which are non-perturbative effects, are parameterized in terms of the vacuum condensates.

In this project we use the method of QCD sum rules to make a direct estimate of the breaking of the $SU(2)\otimes SU(2)$ and $SU(3)\otimes SU(3)$ Gell-Mann-Oakes-Renner relations. There have been previous attempts to calculate this symmetry breaking parameter \cite{Jamin0,cad1}, but these have been indirect methods to estimate the breaking parameter $\delta_\pi$, with large uncertainties associated with them. The symmetry breaking parameter is also of great interest to the field of Chiral Perturbation Theory where $\delta_\pi$ is related to two of the theory's low energy constants $L_8^r$ and $H_2^r$.

\clearpage
\section{A Low Energy Theorem}
\label{sec:ALowEnergyTheorem}
\numberwithin{equation}{subsection}

The Gell-Mann, Oakes, Renner Relation (GMOR) is a low energy relation which comes from the global\\ $SU(N)\otimes SU(N)$ chiral symmetry which is present in the theory of strongly interacting particles. We give a proof for the relation below. For simplicity, we shall consider the case of the pion $\pi$ and hence $n=2$

\subsection{Hadronic Part}
\label{sec:HadronicPart}
We start with the hadronic pseudoscalar correlator which is the time-ordered product of the hadronic currents
\begin{eqnarray}
\left. {\psi _5 \left( {q^2 } \right)} \right|_{{\text{HAD}}}  = i\int {e^{iq \cdot x} \left\langle {0\left| {T\left[ {\partial ^\mu  A_\mu  (x)\partial ^\nu A_\nu ^\dagger (0)} \right]} \right|0} \right\rangle dx} 
\label{correl}
\end{eqnarray}

since the Fock space has the property of completeness we can insert a state between the two currents without changing the correlator
\begin{eqnarray}
\left. {\psi _5 \left( {q^2 } \right)} \right|_{{\text{HAD}}}  = i\sum\limits_n {\int {e^{iq \cdot x} \left\langle {0\left| {T\left[ {\partial ^\mu  A_\mu  (x)\left| n \right\rangle \left\langle n \right|\partial ^\nu A_\nu ^\dagger (0)} \right]} \right|0} \right\rangle } dx} 
\end{eqnarray}

we will insert only the pion state and then truncate the series. From the Partial Conserved Axial Current(PCAC) \cite{PCAC1,PCAC2,PCAC3} we have 
\begin{eqnarray}
\partial ^\mu  A_\mu  \left( x \right) = \sqrt 2 f_\pi  m _\pi ^2 \hat \varphi \left( x \right)
\label{pcac}
\end{eqnarray}

were $m _\pi$ and $f_\pi$ are the pion mass and the pion decay constant respectively. Inserting (\ref{pcac}) into the correlator (\ref{correl}) we obtain
\begin{eqnarray}
\left. {\psi _5 \left( {q^2 } \right)} \right|_{{\text{HAD}}}  = 2f_\pi ^2 m _\pi ^4 i\int {e^{iq \cdot x} \left\langle {0\left| {T\left[ {\hat \varphi \left( x \right)\hat \varphi ^\dag  \left( 0 \right)} \right]} \right|0} \right\rangle d^4 x} 
\label{correl2}
\end{eqnarray}

The right-hand-side of (\ref{correl2}) is just the Fourier transform of the vacuum-expectation-value(vev) of some operator. This operator is the Green's function/propagator for the pion and can then be written as
\begin{eqnarray}
  i\Delta \left( {x - y} \right) = \left\langle {0\left| {T\left[ {\hat \varphi \left( x \right)\hat \varphi ^\dagger \left( y \right)} \right]} \right|0} \right\rangle  
  \label{top}
\end{eqnarray}

were $x$ and $y$ are arbitrary spacetime points. We can then split the time ordered product (\ref{top}) into two terms which depend on which event occurred first.
\begin{eqnarray}
  i\Delta \left( {x - y} \right) = \Theta \left( {x_0  - y_0 } \right)\left\langle {0\left| {\hat \varphi \left( x \right)\hat \varphi \left( y \right)} \right|0} \right\rangle  + \Theta \left( {y_0  - x_0 } \right)\left\langle {0\left| {\hat \varphi \left( y \right)\hat \varphi \left( x \right)} \right|0} \right\rangle  \hfill
  \label{pionprop}
\end{eqnarray}

were $\Theta(x)$ is the Heaviside/Step function. The pion field can be written as a sum over the positive and negative energy states with appropriate creation and annihilation operators
\begin{eqnarray}
\hat\varphi \left( x \right) = \sum\limits_q {\left[ {\hat a_q \varphi _{\underset{\raise0.3em\hbox{$\smash{\scriptscriptstyle-}$}}{q} }^{( + )} \left( x \right) + \hat a^\dag _q \varphi _{ - \underset{\raise0.3em\hbox{$\smash{\scriptscriptstyle-}$}}{q} }^{( - )} \left( x \right)} \right]} 
\label{pionfield}
\end{eqnarray}

The $\varphi _{\underset{\raise0.3em\hbox{$\smash{\scriptscriptstyle-}$}}{q} }^{( + )} \left( x \right), \varphi _{ - \underset{\raise0.3em\hbox{$\smash{\scriptscriptstyle-}$}}{q} }^{( - )} \left( x \right)$ are the positive and negative energy plane wave solutions. Inserting the $\pi$ fields(\ref{pionfield}) into the propagator(\ref{pionprop}).
\begin{eqnarray}
i\Delta \left( {x - y} \right) = \Theta \left( {{x_0} - {y_0}} \right)\left\langle {0\left| {\sum\limits_q {\left[ {{{\hat a}_{\underline q }}\varphi _{\underline q }^{( + )}\left( x \right) + \hat a_{\underline q }^\dag \varphi _{-\underline {  q} }^{( - )}\left( x \right)} \right]} \sum\limits_{q'} {\left[ {{{\hat a}_{\underline {q}' }}\varphi _{\underline {q}' }^{( + )}\left( y \right) + \hat a_{\underline {q}' }^\dag \varphi _{-\underline {  q}' }^{( - )}\left( y \right)} \right]} } \right|0} \right\rangle +\nonumber \\
  + \Theta \left( {{y_0} - {x_0}} \right)\left\langle {0\left| {\sum\limits_{q'} {\left[ {{{\hat a}_{\underline {q}' }}\varphi _{\underline {q}' }^{( + )}\left( y \right) + \hat a_{\underline {q}' }^\dag \varphi _{-\underline {  q}' }^{( - )}\left( y \right)} \right]} \sum\limits_q {\left[ {{{\hat a}_{\underline q }}\varphi _{\underline q }^{( + )}\left( x \right) + \hat a_{\underline q }^\dag \varphi _{\underline { - q} }^{( - )}\left( x \right)} \right]} } \right|0} \right\rangle 
\end{eqnarray}

since there is a product of sums and the sums are over different momenta we can collect the summations together and proceed with normal multiplication.
\begin{eqnarray}
i\Delta \left( {x - y} \right) &=& \Theta \left( {{x_0} - {y_0}} \right)\left\langle {0\left| {\sum\limits_q {\sum\limits_{q'} {\left[ {\varphi _{\underline q }^{( + )}\left( x \right)\varphi _{\underline {q}' }^{( + )}\left( y \right){{\hat a}_{\underline q }}{{\hat a}_{\underline {q}' }} + \underline {\varphi _{\underline q }^{( + )}\left( x \right)\varphi _{- \underline { q}' }^{( - )}\left( y \right){{\hat a}_{\underline q }}\hat a_{\underline {q}' }^\dag }} \right]} }} \right|0} \right\rangle  + \nonumber \\ 
&& \Theta \left( {{x_0} - {y_0}} \right)\left\langle {0\left| {\sum\limits_q {\sum\limits_{q'} {\left[ {\varphi _{-\underline {  q} }^{( - )}\left( x \right)\varphi _{\underline {q}' }^{( + )}\left( y \right)\hat a_{\underline q }^\dag {{\hat a}_{\underline {q}' }} + \varphi _{-\underline {  q} }^{( - )}\left( x \right)\varphi _{-\underline {  q}' }^{( - )}\left( y \right)\hat a_{\underline q }^\dag \hat a_{\underline {q}' }^\dag } \right]} } } \right|0} \right\rangle  + \nonumber \\ 
&& \Theta \left( {{y_0} - {x_0}} \right)\left\langle {0\left| {\sum\limits_q {\sum\limits_{q'} {\left[ {\varphi _{\underline q }^{( + )}\left( x \right)\varphi _{\underline {q}' }^{( + )}\left( y \right){{\hat a}_{\underline {q}' }}{{\hat a}_{\underline q }} + \underline {\varphi _{\underline {q}' }^{( + )}\left( y \right)\varphi _{-\underline {  q} }^{( - )}\left( x \right){{\hat a}_{\underline {q}' }}\hat a_{\underline q }^\dag }} \right]} } } \right|0} \right\rangle  + \nonumber \\ 
&& \Theta \left( {{y_0} - {x_0}} \right)\left\langle {0\left| {\sum\limits_q {\sum\limits_{q'} {\left[ {\varphi _{-\underline {  q}' }^{( - )}\left( y \right)\varphi _{\underline q }^{( + )}\left( x \right)\hat a_{\underline q }'^\dag {{\hat a}_{\underline q }} + \varphi _{-\underline {  q}' }^{( - )}\left( y \right)\varphi _{-\underline {  q} }^{( - )}\left( x \right)\hat a_{\underline {q}' }^\dag \hat a_{\underline q }^\dag} \right]} } } \right|0} \right\rangle  
\end{eqnarray}

all terms in the above expansion are zero except the underlined ones. These are the only terms which contain the right combination of creation and annihilation operators which connect the vacuum to the vacuum. This can be seen if we use the commutation relation \cite{Lawrie}
\[\left[ {{{\hat a}_{\underline q }};\hat a_{\underline {q}' }^\dag } \right] = {\delta _{qq'}}\]
The propagator is then 
\begin{eqnarray}
i\Delta \left( {x - y} \right) = \Theta \left( {x_0  - y_0 } \right)\sum\limits_q {\left[ {\varphi _{\underset{\raise0.3em\hbox{$\smash{\scriptscriptstyle-}$}}{q} }^{( + )} \left( x \right)\varphi _{ - \underline q }^{( - )} \left( y \right)} \right]}  + \Theta \left( {y_0  - x_0 } \right)\sum\limits_q {\left[ {\varphi _{\underline q }^{( + )} \left( y \right)\varphi _{ - \underline q }^{( - )} \left( x \right)} \right]} 
\label{simpprop}
\end{eqnarray}

One of the sums over the momenta has been `killed' by the delta function which picked for us only one momentum state $q'=q$. We now proceed by using the plane wave solutions
\begin{eqnarray}
\varphi _{ \pm \underline q }^{( \pm )} \left( x \right) = C_{ \pm \underline q } e^{ \mp iq^\mu   \cdot x_\mu  } 
\label{planewave}
\end{eqnarray}
with the coefficient 
\begin{eqnarray}
C_{ \pm \underline q }  = \frac{1}
{{\left( {2\pi } \right)^{{\raise0.5ex\hbox{$\scriptstyle 3$}
\kern-0.1em/\kern-0.15em
\lower0.25ex\hbox{$\scriptstyle 2$}}} }}\frac{1}
{{\sqrt {2E_{ \pm \underline q } } }}
\label{coeff}
\end{eqnarray}
and the energy $E_{ \pm \underline q }  = \sqrt {\underline q ^2  + m^2 }$, since the energy is invariant of the sign of the momentum, the coefficient is in turn also invariant and thus only the exponential carries the sign which distinguishes the positive and negative plane waves. We will go from a discrete set of $q$ to the continuum and the sum over the $q$ states will change to an integral over the continuous momenta $q$. We shall also henceforth drop the sign in the energy and the coefficient. Using (\ref{planewave}) in the propagator (\ref{simpprop}) we obtain
\begin{align}
  i\Delta \left( {x - y} \right) &= \Theta \left( {x_0  - y_0 } \right)\int {C_{\underline q }^2 e^{ - iq^\mu   \cdot \left( {x_\mu   - y_\mu  } \right)} d^3 q}  + \Theta \left( {y_0  - x_0 } \right)\int {C_{\underline q }^2 e^{ - iq^\mu   \cdot \left( {y_\mu   - x_\mu  } \right)} d^3 q}  \nonumber \\
&= \int {\Theta \left( {x_0  - y_0 } \right)C_{\underline q }^2 e^{ - iE_{\underline q }  \cdot \left( {x_0  - y_0 } \right)} e^{i\underline q  \cdot \left( {\underline x  - \underline y } \right)} d^3 q}  + \int {\Theta \left( {y_0  - x_0 } \right)C_{\underline q }^2 e^{ - iE_{\underline q }  \cdot \left( {y_0  - x_0 } \right)} e^{i\underline q  \cdot \left( {\underline y  - \underline x } \right)} d^3 q}  \nonumber \\
&= \int {\Theta \left( {x_0  - y_0 } \right)C_{\underline q }^2 e^{ - iE_{\underline q }  \cdot \left( {x_0  - y_0 } \right)} e^{ - i\underline q  \cdot \left( {\underline y  - \underline x } \right)} d^3 q}  + \underline {\int {\Theta \left( {y_0  - x_0 } \right)C_{\underline q }^2 e^{ - iE_{\underline q }  \cdot \left( {y_0  - x_0 } \right)} e^{i\underline q  \cdot \left( {\underline y  - \underline x } \right)} d^3 q} }   
\label{prop1}	
\end{align}
making the change of variables $q \to  - q$ for the second term which is underlined in (\ref{prop1}) and noting the earlier comment of the invariance of the energy and coefficients under a change in sign in the momenta we arrive at
\begin{eqnarray}  
i\Delta \left( {x - y} \right) &=& \int {\Theta \left( {x_0  - y_0 } \right)C_{\underline q }^2 e^{ - iE_{\underline q }  \cdot \left( {x_0  - y_0 } \right)} e^{ - i\underline q  \cdot \left( {\underline y  - \underline x } \right)} d^3 q}  +^* \int {\Theta \left( {y_0  - x_0 } \right)C_{\underline q }^2 e^{ - iE_{\underline q }  \cdot \left( {y_0  - x_0 } \right)} e^{ - i\underline q  \cdot \left( {\underline y  - \underline x } \right)} d^3 q}  \nonumber \\
   &=& \int {C_{\underline q }^2 e^{ - i\underline q  \cdot \left( {\underline y  - \underline x } \right)} \left[ {\Theta \left( {x_0  - y_0 } \right)e^{ - iE_{\underline q }  \cdot \left( {x_0  - y_0 } \right)}  + \Theta \left( {y_0  - x_0 } \right)e^{ - iE_{\underline q }  \cdot \left( {y_0  - x_0 } \right)} } \right]d^3 q}   
\label{prop2}
\end{eqnarray}

* A subtle point to note about the change of variables here: even though there will be an overall negative sign coming from the differential  $-d^3 q$, we suppress the integrals signs denoting only one instead of a multitude, in this case there are three integrals that need to be done but with the change of variables we also swapped the three integration bounds which in turn gives $(-1)^3$ and with the negative sign from the differential leads to an overall positive sign.
\begin{eqnarray} 
X \equiv \Theta \left( {x_0  - y_0 } \right)e^{ - iE_{\underset{\raise0.3em\hbox{$\smash{\scriptscriptstyle-}$}}{q} } \cdot\left( {x_0  - y_0 } \right)}  + \Theta \left( {y_0  - x_0 } \right)e^{ - iE_{\underset{\raise0.3em\hbox{$\smash{\scriptscriptstyle-}$}}{q} } \cdot\left( {y_0  - x_0 } \right)} 
\label{X}
\end{eqnarray}

leading to  
\begin{eqnarray}  
i\Delta \left( {x - y} \right) = \int {C_{\underline q }^2 e^{ - i\underline q  \cdot \left( {\underline y  - \underline x } \right)} X d^3 q}   
\label{prop3}
\end{eqnarray}

We need (\ref{prop3}) to be a Lorentz covariant object\footnote{Lorentz Covariance is an extremely powerful idea which has come to be accepted as an imperative component of any modern Quantum Field Theory. This elegant idea was developed by Einstein for use in the Special Theory of Relativity. There was earlier mention of the idea by Mach but no mathematical formulation was developed.} so we will need to go to the complex plane ($\mathbb{C}$) to accomplish this task. We can write $X$ as
\begin{eqnarray}   
X = \frac{{2E}}{{2\pi i}}\int\limits_{ - \infty }^\infty  {\frac{{e^{ - i\left( {x_0  - y_0 } \right)q_0 } }}{{m_\pi ^2  - q^2  - i\varepsilon '}}dq_0 } 
\label{finX1}
\end{eqnarray}

See Appendix \ref{AStepTowardsCovariance} for details. Substituting (\ref{finX1}) into (\ref{prop3}) we obtain
\begin{eqnarray} 
  i\Delta \left( {x - y} \right) &=& \int {\frac{{2E}}{{2\pi i}}C_{\underline q }^2 \frac{{e^{ - i\left( {x_0  - y_0 } \right)q_0 } e^{i\underline q  \cdot \left( {\underline x  - \underline y } \right)} }}{{m_\pi ^2  - q^2  - i\varepsilon '}}dq_0 d^3 q}  \nonumber \\
   &=& \int {\frac{{2E}}{{2\pi i}}C_{\underline q }^2 \frac{{e^{ - i\left[ {q_0 \left( {x_0  - y_0 } \right) - \underline q  \cdot \left( {\underline x  - \underline y } \right)} \right]} }}
			{{m_\pi ^2  - q^2  - i\varepsilon '}}dq_0 d^3 q}  \nonumber \\
   &=& \int {\frac{{2E}}{{2\pi i}}C_{\underline q }^2 \frac{{e^{ - iq_\mu  \left( {x^\mu   - y^\mu  } \right)} }}{{m_\pi ^2  - q^2  - i\varepsilon '}}d^4 q}  \nonumber \\
   &=& \int {\frac{{2E}}{{2\pi i}}C_{\underline q }^2 \frac{{e^{ - iq \cdot \left( {x - y} \right)} }}{{m_\pi ^2  - q^2  - i\varepsilon '}}d^4 q} 
\label{prop4}
\end{eqnarray}
inserting the coefficient (\ref{coeff}) into (\ref{prop4}) leads to
\begin{eqnarray}   
	i\Delta \left( {x - y} \right) &=& \int {\frac{{2E}}{{2\pi i}}\frac{1}{{\left( {2\pi } \right)^3 }}\frac{1}{{2E}}\frac{{e^{ - iq \cdot \left( {x - y} \right)} }}{{m_\pi ^2  - q^2  -i\varepsilon '}}d^4 q}  \nonumber \\
   &=& \frac{1}{{\left( {2\pi } \right)^4 }}\int {e^{ - iq \cdot \left( {x - y} \right)} \frac{{ - i}}{{m_\pi ^2  - q^2  - i\varepsilon '}}d^4 q}  \nonumber\\
 &=& \frac{1}{{\left( {2\pi } \right)^4 }}\int {e^{ - iq \cdot \left( {x - y} \right)} \frac{i}{{q^2  - m_\pi ^2  + i\varepsilon '}}d^4 q} 
 \label{prop5}
\end{eqnarray}
(\ref{prop5}) is now a Lorentz covariant object and retains its form in all inertial reference frames. We note now that the right-hand-side of (\ref{prop5}) is just the Fourier transform of the propagator from momentum space to coordinate space. The propagator in momentum is then simply
\begin{eqnarray}   
\Delta \left( q \right) = \frac{1}{{q^2  - m_\pi ^2  + i\varepsilon '}}
 \label{prop6}
\end{eqnarray}

Now looking back at the correlator (\ref{correl2}) and the definition of the propagator (\ref{top}) which is shown below
\begin{eqnarray}
\left. {\psi _5 \left( {q^2 } \right)} \right|_{{\text{HAD}}}  = 2f_\pi ^2 m_\pi ^4 i\int {e^{iq \cdot x} \left\langle {0\left| {T\left[ {\hat \varphi \left( x \right)\hat \varphi ^\dag  \left( 0 \right)} \right]} \right|0} \right\rangle d^4 x} 
\label{correl3}
\end{eqnarray}
\begin{eqnarray}
  i\Delta \left( {x - y} \right) = \left\langle {0\left| {T\left[ {\hat \varphi \left( x \right)\hat \varphi ^\dagger \left( y \right)} \right]} \right|0} \right\rangle  
  \label{top2}
\end{eqnarray}
substituting (\ref{top2}) into (\ref{correl3}) we obtain
\begin{eqnarray}
\left. {\psi _5 \left( {q^2 } \right)} \right|_{{\text{HAD}}}  = -2f_\pi ^2 m_\pi ^4 \int {e^{iq \cdot x} \Delta \left( x - y \right) d^4 x} 
\label{correl4}
\end{eqnarray}
we note now that the right-hand-side of (\ref{correl4}) is just the Fourier transform of the propagator from coordinate space to momentum space but we know what this transform is in momentum space, it is (\ref{prop6}) for the pion $\pi^0$, which leads to 
\begin{eqnarray}
\left. {\psi _5 \left( {q^2 } \right)} \right|_{{\text{HAD}}}  = -2f_\pi ^2 m_\pi ^4   \frac{1}{{q^2  - m_\pi ^2  + i\varepsilon '}} 
\label{correl5}
\end{eqnarray}
and finally in the low energy limit and $\varepsilon ' \rightarrow 0$
\begin{eqnarray}
\left. {\psi _5 \left( {0 } \right)} \right|_{{\text{HAD}}}  = 2f_\pi ^2 m_\pi ^2 
\label{correl6}
\end{eqnarray}


\subsection{QCD Part}
\label{sec:QCDPart}

We now turn to the QCD correlator. Starting from the axial two point correlating function of quark currents
\begin{eqnarray}
\Pi _{\mu \nu }^5  = i\int {e^{iq \cdot x} \left\langle {0\left| {T\left[ {A_\mu  \left( x \right)A_\nu ^\dag  \left( 0 \right)} \right]} \right|0} \right\rangle d^4 x} 
\label{axial2}
\end{eqnarray}

the time ordered product can be expressed differently as
\begin{eqnarray}
T\left[ {A_\mu  \left( x \right)A_\nu ^\dag  \left( 0 \right)} \right] = \left[ {A_\mu  \left( x \right);A_\nu ^\dag  \left( 0 \right)} \right]\Theta \left( {x_0 } \right) + A_\nu ^\dag  \left( 0 \right)A_\mu  \left( x \right)
\label{top3}
\end{eqnarray}

(See Appendix \ref{TimeOrderedProducts} for proof.) Inserting the time ordered product (\ref{top3}) into (\ref{axial2}) we obtain
\begin{eqnarray}
\Pi _{\mu \nu }^5  = i\int {e^{iq \cdot x} \left\langle {0\left| \left[ {A_\mu  \left( x \right);A_\nu ^\dag  \left( 0 \right)} \right]\Theta \left( {x_0 } \right) + A_\nu ^\dag  \left( 0 \right)A_\mu  \left( x \right) \right|0} \right\rangle d^4 x} 
\label{axial3}
\end{eqnarray}
we now apply the momentum operator
\begin{eqnarray}
	q^\mu   =  -i\frac{\partial}{{\partial x_\mu}} =  i\partial^\mu 
\label{momop}
\end{eqnarray}
to (\ref{axial3})
\begin{eqnarray}
q^\mu \Pi _{\mu \nu }^5  = -i^2 \partial ^\mu \int {e^{iq \cdot x} \left\langle {0\left| \left[ {A_\mu  \left( x \right);A_\nu ^\dag  \left( 0 \right)} \right]\Theta \left( {x_0 } \right) + A_\nu ^\dag  \left( 0 \right)A_\mu  \left( x \right) \right|0} \right\rangle d^4 x} 
\label{axial4}
\end{eqnarray}

since the fields are local the integral in (\ref{axial4}) yields a function of $q$, acting on it with the momentum operator gives zero which leads to 
\begin{eqnarray}
  0 &=&   -{i^2}\int {{\partial ^\mu }\left\{ {{e^{iq\cdot x}}\left\langle {0\left| {\left[ {{A_\mu }\left( x \right);A_\nu ^\dag \left( 0 \right)} \right]\Theta \left( {{x_0}} \right) + A_\nu ^\dag \left( 0 \right){A_\mu }\left( x \right)} \right|0} \right\rangle } \right\}{d^4}x}  \nonumber \\
   &=& {q^\mu }\underline {i\int {{e^{iq\cdot x}}\left\langle {0\left| {\left[ {{A_\mu }\left( x \right);A_\nu ^\dag \left( 0 \right)} \right]\Theta \left( {{x_0}} \right) + A_\nu ^\dag \left( 0 \right){A_\mu }\left( x \right)} \right|0} \right\rangle {d^4}x} }  \nonumber \\
   &+& \int {{e^{iq\cdot x}}\left\langle {0\left| {\underline {\left[ {{\partial ^\mu }{A_\mu }\left( x \right);A_\nu ^\dag \left( 0 \right)} \right]\Theta \left( {{x_0}} \right)}  + \left[ {{A_\mu }\left( x \right);A_\nu ^\dag \left( 0 \right)} \right]{\partial ^\mu }\Theta \left( {{x_0}} \right) + \underline {A_\nu ^\dag \left( 0 \right){\partial ^\mu }{A_\mu }\left( x \right)} } \right|0} \right\rangle {d^4}x}  
\label{axial6}
\end{eqnarray}

we note that the first underlined term in (\ref{axial6} ) is $\Pi _{\mu \nu }^5$ in (\ref{axial3}) and interestingly the second and third underlined terms together form a time-ordered product
\begin{eqnarray}
\left[ {{\partial ^\mu }{A_\mu }\left( x \right);A_\nu ^\dag \left( 0 \right)} \right]\Theta \left( {{x_0}} \right) + A_\nu ^\dag \left( 0 \right){\partial ^\mu }{A_\mu }\left( x \right) = T\left[ {{\partial ^\mu }{A_\mu }\left( x \right)A_\nu ^\dag \left( 0 \right)} \right]
\label{axial7}
\end{eqnarray}
which leads to 
\begin{align}
  0 &= {q^\mu }\Pi _{_{\mu \nu }}^5 + \int {{e^{iq\cdot x}}\left\langle {0\left| {T\left[ {{\partial ^\mu }{A_\mu }\left( x \right)A_\nu ^\dag \left( 0 \right)} \right]} \right|0} \right\rangle {d^4}x}  + \int {{e^{iq\cdot x}}\left\langle {0\left| {\left[ {{A_\mu }\left( x \right);A_\nu ^\dag \left( 0 \right)} \right]{\partial ^\mu }\Theta \left( {{x_0}} \right)} \right|0} \right\rangle {d^4}x}  \nonumber \\
&= {q^\mu }\Pi _{_{\mu \nu }}^5 + \int {{e^{iq\cdot x}}\left\langle {0\left| {T\left[ {{\partial ^\mu }{A_\mu }\left( x \right)A_\nu ^\dag \left( 0 \right)} \right]} \right|0} \right\rangle {d^4}x}  + \int {{e^{iq\cdot x}}\left\langle {0\left| {\left[ {{A_0}\left( x \right);A_\nu ^\dag \left( 0 \right)} \right]{\partial ^0}\Theta \left( {{x_0}} \right)} \right|0} \right\rangle {d^4}x} 
\label{axial8}	
\end{align}
but the derivative of the Heaviside function leads to the delta function i.e. ${\partial ^0}\Theta \left( {{x_0}} \right) = \delta \left( {{x_0}} \right)$ and defining the following functions:
\begin{eqnarray}
  \Pi _{_\nu }^5 &\equiv&  - \int {{e^{iq\cdot x}}\left\langle {0\left| {T\left[ {{\partial ^\mu }{A_\mu }\left( x \right)A_\nu ^\dag \left( 0 \right)} \right]} \right|0} \right\rangle {d^4}x} \label{pinu}  \\
  \Delta _{_\nu }^5 &\equiv& \int {{e^{iq\cdot x}}\left\langle {0\left| {\left[ {{A_0}\left( x \right);A_\nu ^\dag \left( 0 \right)} \right]\delta \left( {{x_0}} \right)} \right|0} \right\rangle {d^4}x} \label{deltanu}
\end{eqnarray}
and substituting into (\ref{axial8}) gives us the Ward(I) identity\footnote{the `I' we have attached to this Ward identity serves the purpose of distinguishing it from the Ward(II) identity which will appear later. It is not a convention and will not be found in other articles on the topic}
\begin{eqnarray}
0 = {q^\mu }\Pi _{_{\mu \nu }}^5 - \Pi _{_\nu }^5 + \Delta _{_\nu }^5
\label{ward1}
\end{eqnarray}

We can express (\ref{pinu}) as
\begin{eqnarray}
\Pi _{_\nu }^5 = \int {{e^{ - iq\cdot x}}\left\langle {0\left| {T\left[ {A_\nu ^\dag \left( x \right){\partial ^\mu }{A_\mu }\left( 0 \right)} \right]} \right|0} \right\rangle {d^4}x}
\label{Pinu3}
\end{eqnarray}

See Appendix \ref{PreparatoryWorkI} for details. We now apply the momentum operator (\ref{momop}) to (\ref{Pinu3}). Since the fields are local the derivative yields zero
\begin{eqnarray}  
0  &=&  - i\partial ^\nu  \int {e^{ - iq\cdot x} \left\langle {0\left| {T\left[ {A_\nu ^\dag  \left( x \right)\partial ^\mu  A_\mu  \left( 0 \right)} \right]} \right|0} \right\rangle d^4 x}  \nonumber \\
   &=&  - i\int {\partial ^\nu  \left\{ {e^{ - iq\cdot x} \left\langle {0\left| {T\left[ {A_\nu ^\dag  \left( x \right)\partial ^\mu  A_\mu  \left( 0 \right)} \right]} \right|0} \right\rangle } \right\}d^4 x}  \nonumber \\
   &=&  - i\int {\left\{ { - iq^\nu  e^{ - iq\cdot x} \left\langle {0\left| {T\left[ {A_\nu ^\dag  \left( x \right)\partial ^\mu  A_\mu  \left( 0 \right)} \right]} \right|0} \right\rangle  + e^{ - iq\cdot x} \left\langle {0\left| {\partial ^\nu  T\left[ {A_\nu ^\dag  \left( x \right)\partial ^\mu  A_\mu  \left( 0 \right)} \right]} \right|0} \right\rangle } \right\}d^4 x}  \nonumber \\
   &=&  - q^\nu  \underline {\int {e^{ - iq\cdot x} \left\langle {0\left| {T\left[ {A_\nu ^\dag  \left( x \right)\partial ^\mu  A_\mu  \left( 0 \right)} \right]} \right|0} \right\rangle d^4 x} }  - i\int {e^{ - iq\cdot x} \left\langle {0\left| {\partial ^\nu  T\left[ {A_\nu ^\dag  \left( x \right)\partial ^\mu  A_\mu  \left( 0 \right)} \right]} \right|0} \right\rangle d^4 x}  
\label{partpinu}
\end{eqnarray}

we first note that the underlined term in (\ref{partpinu}) is the $\Pi _{_\nu }^5$ in (\ref{Pinu3}) and the time-ordered product can be expressed as
\begin{eqnarray}  
T\left[ {A_\nu ^\dag  \left( x \right)\partial ^\mu  A_\mu  \left( 0 \right)} \right] = \left[ {A_\nu ^\dag  \left( x \right);\partial ^\mu  A_\mu  \left( 0 \right)} \right]\Theta \left( {x_0 } \right) + \partial ^\mu  A_\mu  \left( 0 \right)A_\nu ^\dag  \left( x \right)
\label{top5}
\end{eqnarray}

now acting on (\ref{top5}) by the partial derivative operator we obtain
\begin{eqnarray} 
 \partial ^\nu  T\left[ {A_\nu ^\dag  \left( x \right)\partial ^\mu  A_\mu  \left( 0 \right)} \right] &=& \left[ {\partial ^\nu  A_\nu ^\dag  \left( x \right);\partial ^\mu  A_\mu  \left( 0 \right)} \right]\Theta \left( {x_0 } \right) + \left[ {A_\nu ^\dag  \left( x \right);\partial ^\mu  A_\mu  \left( 0 \right)} \right]\partial ^\nu  \Theta \left( {x_0 } \right) + \partial ^\mu  A_\mu  \left( 0 \right)\partial ^\nu  A_\nu ^\dag  \left( x \right) \nonumber \\
   &=& \left[ {\partial ^\nu  A_\nu ^\dag  \left( x \right);\partial ^\mu  A_\mu  \left( 0 \right)} \right]\Theta \left( {x_0 } \right) + \left[ {A_0^\dag  \left( x \right);\partial ^\mu  A_\mu  \left( 0 \right)} \right]\partial ^0 \Theta \left( {x_0 } \right) + \partial ^\mu  A_\mu  \left( 0 \right)\partial ^\nu  A_\nu ^\dag  \left( x \right) \nonumber \\
   &=& \underline {\left[ {\partial ^\nu  A_\nu ^\dag  \left( x \right);\partial ^\mu  A_\mu  \left( 0 \right)} \right]\Theta \left( {x_0 } \right)}  + \left[ {A_0^\dag  \left( x \right);\partial ^\mu  A_\mu  \left( 0 \right)} \right]\delta \left( {x_0 } \right) + \underline {\partial ^\mu  A_\mu  \left( 0 \right)\partial ^\nu  A_\nu ^\dag  \left( x \right)} \nonumber \\
\label{ptop}
\end{eqnarray}

the underlined terms of (\ref{ptop}) forms a new time-ordered product
\begin{eqnarray} 
\left[ {\partial ^\nu  A_\nu ^\dag  \left( x \right);\partial ^\mu  A_\mu  \left( 0 \right)} \right]\Theta \left( {x_0 } \right) + \partial ^\mu  A_\mu  \left( 0 \right)\partial ^\nu  A_\nu ^\dag  \left( x \right) = T\left[ {\partial ^\nu  A_\nu ^\dag  \left( x \right)\partial ^\mu  A_\mu  \left( 0 \right)} \right]
\label{top6}
\end{eqnarray}

so the time-ordered product (\ref{ptop}) is 
\begin{eqnarray}
\partial ^\nu  T\left[ {A_\nu ^\dag  \left( x \right)\partial ^\mu  A_\mu  \left( 0 \right)} \right] = T\left[ {\partial ^\nu  A_\nu ^\dag  \left( x \right)\partial ^\mu  A_\mu  \left( 0 \right)} \right] + \left[ {A_0^\dag  \left( x \right);\partial ^\mu  A_\mu  \left( 0 \right)} \right]\delta \left( {x_0 } \right)
\label{ptop2}
\end{eqnarray}

substituting (\ref{ptop2}) into (\ref{partpinu}) we obtain
\begin{align}
0 =  - q^\nu  \Pi _\nu ^5  - i\int {e^{ - iq\cdot x} \left\langle {0\left| {T\left[ {\partial ^\nu  A_\nu ^\dag  \left( x \right)\partial ^\mu  A_\mu  \left( 0 \right)} \right]} \right|0} \right\rangle d^4 x}  - i\int {e^{ - iq\cdot x} \left\langle {0\left| {\left[ {A_0^\dag  \left( x \right);\partial ^\mu  A_\mu  \left( 0 \right)} \right]\delta \left( {x_0 } \right)} \right|0} \right\rangle d^4 x} 
\label{qnupinu2}	
\end{align}
we now define a new function
\begin{eqnarray}
\Delta _1^5  \equiv  - i\int {e^{ - iq\cdot x} \left\langle {0\left| {\left[ {A_0^\dag  \left( x \right);\partial ^\mu  A_\mu  \left( 0 \right)} \right]\delta \left( {x_0 } \right)} \right|0} \right\rangle d^4 x} 
\label{delta51}
\end{eqnarray}

and then we arrive at 
\begin{eqnarray}
0 =  - q^\nu  \Pi _\nu ^5  - i\int {e^{ - iq\cdot x} \left\langle {0\left| {T\left[ {\partial ^\nu  A_\nu ^\dag  \left( x \right)\partial ^\mu  A_\mu  \left( 0 \right)} \right]} \right|0} \right\rangle d^4 x}  + \Delta _1^5  
\label{qnupinu3}
\end{eqnarray}

to proceed further with the analysis we need to cast second term of (\ref{qnupinu3}) into a recognizable form. We can express (\ref{qnupinu3}) as

\begin{eqnarray}
 - i\int {e^{ - iq\cdot x} \left\langle {0\left| {T\left[ {\partial ^\nu  A_\nu ^\dag  \left( x \right)\partial ^\mu  A_\mu  \left( 0 \right)} \right]} \right|0} \right\rangle d^4 x}  = i\int {e^{iq\cdot x} \left\langle {0\left| {T\left[ {\partial ^\mu  A_\mu  \left( x \right)\partial ^\nu  A_\nu ^\dag  \left( 0 \right)} \right]} \right|0} \right\rangle d^4 x} 
\label{qcd2}
\end{eqnarray}

See Appendix \ref{PreparatoryWorkII} for details, the right-hand-side of (\ref{qcd2}) is a vaguely familiar function and if we refer back to (\ref{correl}) which is shown below
\begin{eqnarray}
\left. {\psi _5 \left( {q^2 } \right)} \right|_{{\text{HAD}}}  = i\int {e^{iq \cdot x} \left\langle {0\left| {T\left[ {\partial ^\mu  A_\mu  (x)\partial ^\nu A_\nu ^\dagger (0)} \right]} \right|0} \right\rangle dx} 
\label{hadcorrel}
\end{eqnarray}

so
\begin{eqnarray}
 - i\int {e^{ - iq\cdot x} \left\langle {0\left| {T\left[ {\partial ^\nu  A_\nu ^\dag  \left( x \right)\partial ^\mu  A_\mu  \left( 0 \right)} \right]} \right|0} \right\rangle d^4 x}  = \psi _5 \left( q^2\right)
\label{pinupsi5}
\end{eqnarray}

now going back and substituting (\ref{pinupsi5}) into (\ref{qnupinu3}) we arrive at the Ward(II) identity 
\begin{eqnarray}
0 =  - q^\nu  \Pi _\nu ^5  + \psi _5  + \Delta _1^5 
\label{ward2}
\end{eqnarray}


\subsection{The Gell-Mann-Oakes-Renner Relation}
\label{sec:TheGellMannOakesRennerRelation}

We have now set the stage and all the pieces are ready to be put together to demonstrate the Gell-Mann Oakes Renner relation. We start with the Ward(I) identity (\ref{ward1})
\begin{eqnarray}
0 = {q^\mu }\Pi _{_{\mu \nu }}^5 - \Pi _{_\nu }^5 + \Delta _{_\nu }^5
\label{1ward1}
\end{eqnarray}

Multiplying both sides of the (\ref{1ward1}) by $q^\nu$
\begin{eqnarray}
0 = q^\nu  q^\mu  \Pi _{\mu \nu }^5  - q^\nu  \Pi _\nu ^5  + q^\nu  \Delta _\nu ^5 
\label{qward1}
\end{eqnarray}

now substituting the Ward(II) identity (\ref{ward2}) into (\ref{qward1})
\begin{eqnarray}
0 = q^\nu  q^\mu  \Pi _{\mu \nu }^5  - \psi _5  - \Delta _1^5  + q^\nu  \Delta _\nu ^5 
\label{qward2}
\end{eqnarray}

and rearranging we obtain
\begin{eqnarray}
q^\nu  \left( {q^\mu  \Pi _{\mu \nu }^5  + \Delta _\nu ^5 } \right) - \psi _5  = \Delta _1^5 
\label{reward2}
\end{eqnarray}

If we take the limit on both sides of (\ref{reward2}) as $q^\nu \to 0$
\begin{eqnarray}
\mathop {\lim }\limits_{q^\nu   \to 0} \left[ {q^\nu  \left( {q^\mu  \Pi _{\mu \nu }^5  + \Delta _\nu ^5 } \right) - \psi _5 } \right] = \mathop {\lim }\limits_{q^\nu   \to 0} \Delta _1^5 
\label{reward3}
\end{eqnarray}

Since the terms in the round brackets are analytic functions this just gives 
\begin{eqnarray}
\mathop { - \lim }\limits_{q^\nu   \to 0} \psi _5  = \mathop {\lim }\limits_{q^\nu   \to 0} \Delta _1^5 
\label{reward4}
\end{eqnarray}

but the limit of the right-hand-side
\begin{eqnarray}  
\mathop {\lim }\limits_{q^\nu   \to 0} \Delta _1^5  &=&  - i\mathop {\lim }\limits_{q^\nu   \to 0} \int {e^{ - iq\cdot x} \left\langle {0\left| {\left[ {A_0^\dag  \left( x \right);\partial ^\mu  A_\mu  \left( 0 \right)} \right]\delta \left( {x_0 } \right)} \right|0} \right\rangle d^4 x}  \nonumber \\
   &=&  - i\int {\left\langle {0\left| {\left[ {A_0^\dag  \left( {0,\underline x } \right);\partial ^\mu  A_\mu  \left( 0,\underline 0 \right)} \right]} \right|0} \right\rangle d^3 x}  
\label{limdel}
\end{eqnarray}

So we can now express the currents and their derivatives as a product of quark fields
\begin{eqnarray}
    \partial ^\mu  A_\mu  \left( {0;\underline 0 } \right) &=& i\left( {m_i  + m_j } \right)\overline \psi  _j \left( 0 \right)\gamma _5 \psi _i \left( 0 \right) \label{partA} \\
  A_\mu  \left( {0,\underline x } \right) &=& \overline \psi  _j \left( 0 \right)\gamma _\mu  \gamma _5 \psi _i \left( 0 \right) 
\label{Amu}
\end{eqnarray}

The $\mu=0$ component of (\ref{Amu}) is :
\begin{eqnarray}
A_0 \left( {0,\underline x } \right) = \overline \psi  _j \left( 0 \right)\gamma _0 \gamma _5 \psi _i \left( 0 \right)
\label{Amu0}
\end{eqnarray}

and the conjugate field can be expressed as :
\begin{eqnarray}
\overline \psi  _j  = \psi _j ^\dag \gamma _0 
\label{conj}
\end{eqnarray}

Some useful properties of the gamma matrices are listed below
\begin{eqnarray}  
\gamma _0 \gamma _0  &=& 1 \label{gamma2} \\
	\gamma _5 \gamma _5 &=& 1 \label{gamma52}\\
	\gamma _5 ^ \dag &=& \gamma _5 \label{gamma5dag} \\
  \left[ {\gamma _0 ,\gamma _5 } \right] &=& 0  
\label{comgamma}
\end{eqnarray}

Using properties (\ref{conj}-\ref{gamma5dag}) we obtain
\begin{eqnarray}  A_0  &=& \psi _j^\dag  \gamma _0 \gamma _0 \gamma _5 \psi _i  = \psi _j^\dag  \gamma _5 \psi _i  \nonumber \\
  A_0^\dag   &=& \left( {\psi _j^\dag  \gamma _5 \psi _i } \right)^\dag   = \psi _i^\dag  \gamma _5^\dag  \left( {\psi _j^\dag  } \right)^\dag   = \psi _i^\dag  \gamma _5 \psi _j \label{Amudag} \\
	\partial ^\mu  A_\mu   &=& i\left( {m_i  + m_j } \right)\psi _j^\dag  \gamma _0 \gamma _5 \psi _i 
	\label{partAmu}
\end{eqnarray}

and substituting (\ref{Amudag}) and (\ref{partAmu}) into the commutator of (\ref{limdel}), we get
\begin{eqnarray}
\left[ {A_0^\dag  \left( {0,\underline x } \right);\partial ^\mu  A_\mu  \left( {0;\underline 0 } \right)} \right] &=& \left[ {\psi _i^\dag  \gamma _5 \psi _j ;i\left( {m_i  + m_j } \right)\psi _j^\dag  \gamma _0 \gamma _5 \psi _i } \right] \nonumber \\
&=& i\left( {m_i  + m_j } \right)\left[ {\psi _i^\dag  \gamma _5 \psi _j ;\psi _j^\dag  \gamma _0 \gamma _5 \psi _i } \right]
	\label{commutator}
\end{eqnarray}

We can move the gamma matrices around, but being careful not to move them past each other since they do not commute with each other, we arrive at
\begin{eqnarray}
  \left[ {\psi _i^\dag  \gamma _5 \psi _j ;\psi _j^\dag  \gamma _0 \gamma _5 \psi _i } \right] &=& \psi _i^\dag  \gamma _5 \psi _j \psi _j^\dag  \gamma _0 \psi _i \gamma _5  - \psi _j^\dag  \gamma _0 \psi _i \psi _i^\dag  \gamma _5 \psi _j \gamma _5  \nonumber \\
   &=& \left( {\psi _i^\dag  \gamma _5 \psi _j \psi _j^\dag  \gamma _0 \psi _i  - \psi _j^\dag  \gamma _0 \psi _i \psi _i^\dag  \gamma _5 \psi _j } \right)\gamma _5  \nonumber \\
   &=& \left[ {\psi _i^\dag   \psi _j \gamma _5;\psi _j^\dag   \psi _i \gamma _0 } \right]\gamma _5  
	\label{commutator1}
\end{eqnarray}

We now use an operator identity 
\begin{eqnarray}  
\left[ {A^\dag  B;C^\dag  D} \right] &=& A^\dag  \left\{ {B;C^\dag  } \right\}D - C^\dag  \left\{ {D;A^\dag  } \right\}B - \left\{ {A^\dag  ;C^\dag  } \right\}BD + C^\dag  A^\dag  \left\{ {B;D} \right\} \label{comidentity} 
\end{eqnarray}

were $A,B,C,D$ are Grassman \cite{grassman} or Fermion fields. We can express the right-hand-side of (\ref{comidentity}) as
\begin{eqnarray} 
\left[ {A^\dag  B;C^\dag  D} \right] = A^\dag  D\delta _{BC}  - C^\dag  B\delta _{AD}  - BD\delta _{AC}  + C^\dag  A^\dag  \delta _{BD} 
\end{eqnarray}

and considering $B=C$ and $A=D$ $\Rightarrow$ $\delta _{AC}= \delta _{BD} =0$ so 
\begin{eqnarray} 
\left[ {A^\dag  B;C^\dag  D} \right] = A^\dag  D\delta _{BC}  - C^\dag  B\delta _{AD} 
\label{comiden}
\end{eqnarray}

Now comparing the terms in the commutator of (\ref{commutator1}) and (\ref{comiden}) we see that
\begin{eqnarray}   
	A^\dag   &=& \psi _i^\dag   \hfill \\
  B &=& \psi _j \gamma _5  \hfill \\
  C^\dag   &=& \psi _j^\dag   \hfill \\
  D &=& \psi _i \gamma _0  
\end{eqnarray}

together with 
\begin{eqnarray}
  \delta _{BC}  &=& \delta \left( {\underline x } \right)\gamma _5  \hfill \\
  \delta _{AD}  &=& \delta \left( {\underline x } \right)\gamma _0  
\end{eqnarray}

We obtain
\begin{eqnarray}
  \left[ {\psi _i^\dag  \psi _j \gamma _5 ;\psi _j^\dag  \psi _i \gamma _0 } \right] &=& \psi _i^\dag  \psi _i \gamma _0 \delta \left( {\underline x } \right)\gamma _5  - \psi _j^\dag  \psi _j \gamma _5 \delta \left( {\underline x } \right)\gamma _0  \nonumber \\
   &=& \left[ {\psi _i^\dag  \psi _i \gamma _0 \gamma _5  - \psi _j^\dag  \psi _j \gamma _5 \gamma _0 } \right]\delta \left( {\underline x } \right) \nonumber \\
   &=& \left[ {\psi _i^\dag  \gamma _0 \psi _i \gamma _5  + \psi _j^\dag  \gamma _0 \psi _j \gamma _5 } \right]\delta \left( {\underline x } \right) \nonumber \\
   &=& \left[ {\bar \psi _i \psi _i  + \bar \psi _j \psi _j } \right]\delta \left( {\underline x } \right)\gamma _5  
  \label{workedcom}
\end{eqnarray}

Substituting (\ref{workedcom}) into (\ref{commutator1}) and noting property (\ref{gamma52}), which is in turn substituted into (\ref{commutator}) and finally substituting back into (\ref{limdel}), we obtain
\begin{eqnarray}  
  \mathop {\lim }\limits_{q^\nu   \to 0} \Delta _1^5  &=&  - i^2 \left( {m_i  + m_j } \right)\int {\left\langle {0\left| {\bar \psi _i \psi _i  + \bar \psi _j \psi _j } \right|0} \right\rangle \delta \left( {\underline x } \right) d^3 x}  \nonumber \\
   &=& \left( {m_i  + m_j } \right)\left\langle {0\left| {\bar \psi _i \psi _i  + \bar \psi _j \psi _j } \right|0} \right\rangle 
  \label{delcomplete}
\end{eqnarray}

and in the case of the $SU(2)\otimes SU(2)$ symmetry with just the up(u) and down(d) quarks we have
\begin{eqnarray}  
\mathop {\lim }\limits_{q^\nu   \to 0} \Delta _1^5  = \left( {m_u  + m_d } \right)\left\langle {0\left| {\bar uu + \bar dd} \right|0} \right\rangle 
  \label{su2}
\end{eqnarray}

so referring back to (\ref{reward4}) we have
\begin{eqnarray}
\mathop { - \lim }\limits_{q^\nu   \to 0} \psi _5  = \left( {m_u  + m_d } \right)\left\langle {0\left| {\bar uu + \bar dd} \right|0} \right\rangle 
\label{gmor1}
\end{eqnarray}

but the left-hand-side of (\ref{gmor1}) also has a low energy limit which was worked out in (\ref{correl6}) so we finally arrive at the Gell-Mann Oakes Renner low energy relation
\begin{eqnarray}
 - 2f_\pi ^2 m_\pi ^2  = \left( {m_u  + m_d } \right)\left\langle {0\left| {\bar u u + \bar d d} \right|0} \right\rangle 
\label{gmor}
\end{eqnarray}

In the chiral limit the up and down quarks have zero mass and we expect that there would be three massless Goldstone boson's when this symmetry is broken. Indeed we do find these three boson's the $\pi^-,\pi^0,\pi^+$ but their masses are not zero. This is an indication to us that the Chiral symmetry is only an approximate symmetry and that the quarks are not massless. Looking at (\ref{gmor}) we see that in the limit of $m_u = m_d = 0$ the symmetry is exact and the pion's have zero mass. This provides an explanation for the existence of the light mass mesons we observe from experiment.  
\clearpage
\numberwithin{equation}{section}
\section{The QCD Finite Energy Sum Rule}
\label{sec:TheQCDFESR}

In this project we will estimate how much the Gell-Mann-Oakes-Renner relation in (\ref{gmor}) is broken with the inclusion of chiral corrections to (\ref{gmor}). The amount the symmetry is broken by is measured by the parameter $\delta_\pi$ which is given by the relation
\begin{eqnarray}
\psi _5 \left( 0 \right) \equiv  - \left( {m_u  + m_d } \right)\left\langle {0|\bar uu + \bar dd|0} \right\rangle  = 2f_\pi ^2 M_\pi ^2 \left( {1 - \delta _\pi  } \right)
\label{breakgmor}
\end{eqnarray}

We calculate $\psi _5 \left( 0 \right)$ using the QCD Sum rules and perturbation theory and thus make a direct estimate of $\delta_\pi$

\begin{figure}[h]
	\centerline{
	\mbox{\includegraphics[width=5.5in]{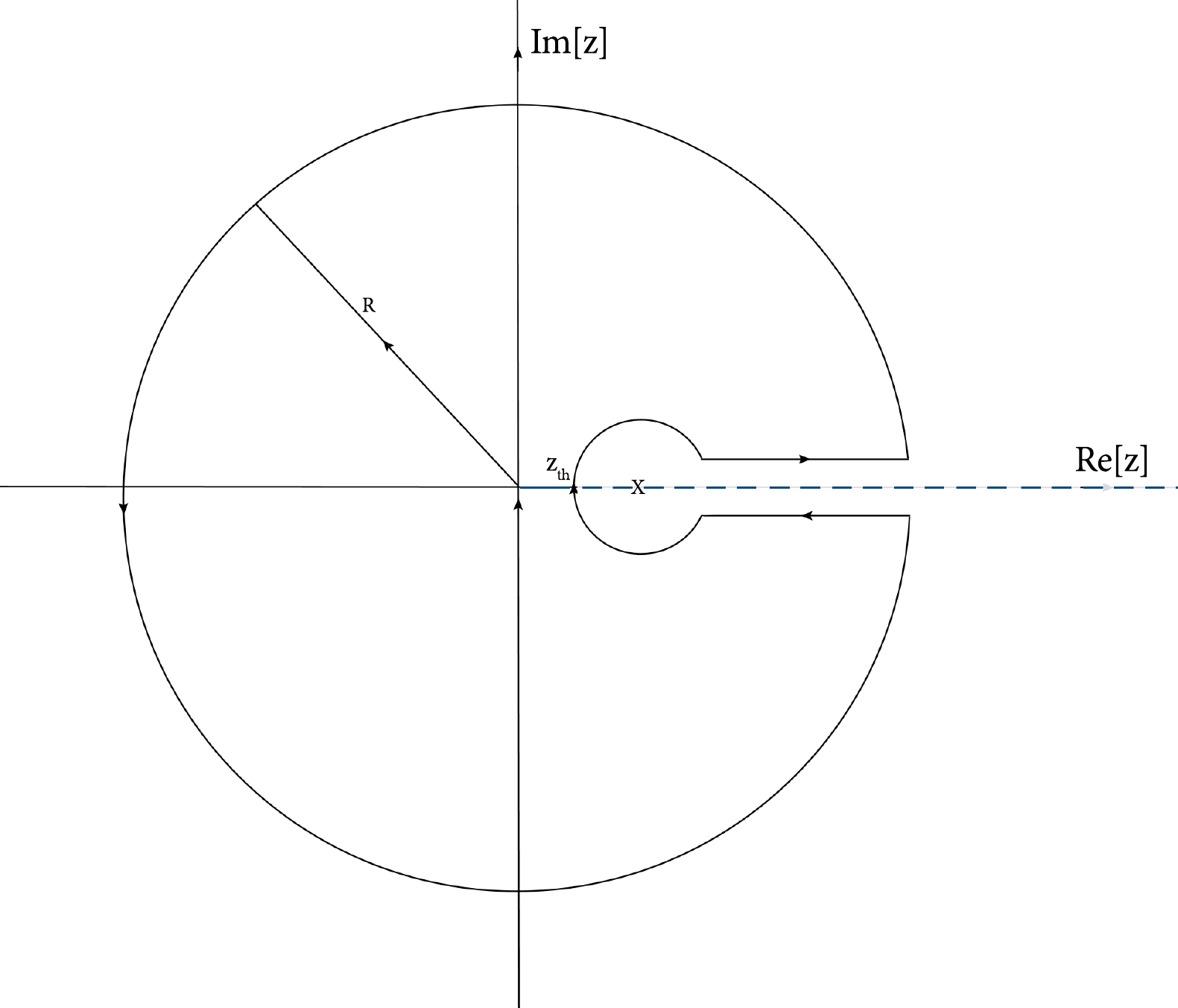}}
	}
	\caption{Integration contour used for correlator. X indicates a pole and the broken line indicates a branch cut in the complex plane along the positive real axis corresponding to the resonances}
	\label{contour}
\end{figure}

We will derive the Finite Energy Sum Rule(FESR) \cite{Colangelo} in this section. We start by using the Cauchy-Goursat Theorem on the pseudoscalar correlator $\psi_5 (z)$ which has a branch cut along the positive real axis in the complex plane We will choose the integration contour suitably so that $\psi_5 (z)$ is an \emph{analytic} function in the region of integration which will be as shown in figure \ref{contour}. The integration region is to the left of the bounding contour if we follow the contour in an anti-clockwise manner. In this way we avoid the poles which are on the positive real axis and we can consider $\psi_5 (z)$ to be \emph{analytic}

We will also introduce a kernel $\Delta(z)$ which is an arbitrary complete(the zeroth and first order terms are included) quadratic polynomial of $z$. We introduce $\Delta(z)$ to reduce the systematic uncertainties which are associated to the resonances, by constraining $\Delta(z)=0$ at the two resonance peaks. This fixes for us the coefficients of the kernel uniquely. 
\begin{align}
&\psi _5 \left( {z_0 } \right)\Delta \left( {z_0 } \right) \nonumber\\
&= \frac{1}{{2\pi i}}\mathop{\oint \mkern-0.8mu}\limits_C  {\frac{{\psi _5 \left( z \right)\Delta \left( z \right)}}
	{{z - z_0 }}dz}  \nonumber \\
&= \mathop {\lim }\limits_{\varepsilon  \to 0} \frac{1}{{2\pi i}}\int\limits_{z_{th} }^R {\frac{{\psi _5 \left( {z + i\varepsilon } \right)\Delta \left( {z + i\varepsilon } \right)}}{{z - z_0 }}dz}  			+ \frac{1}{{2\pi i}}\mathop{\oint \mkern-0.8mu}\limits_{C\left( R \right)}  {\frac{{\psi _5 \left( z \right)\Delta \left( z \right)}}{{z - z_0 }}dz}  + \mathop {\lim 												}\limits_{\varepsilon  \to 0} \frac{1}{{2\pi i}}\int\limits_R^{z_{th} } {\frac{{\psi _5 \left( {z - i\varepsilon } \right)\Delta \left( {z - i\varepsilon } \right)}}{{z - z_0 }}dz}  \nonumber \\
&= \frac{1}{{2\pi i}}\mathop{\oint \mkern-0.8mu}\limits_{C\left( R \right)}  {\frac{{\psi _5 \left( z \right)\Delta \left( z \right)}}{{z - z_0 }}dz}  + \mathop {\lim 													}\limits_{\varepsilon  \to 0} \frac{1}{{2\pi i}}\int\limits_{z_{th} }^R {\frac{{\psi _5 \left( {z + i\varepsilon } \right)\Delta \left( {z + i\varepsilon } \right)}}{{z - z_0 }}dz}  - \mathop 				{\lim }\limits_{\varepsilon  \to 0} \frac{1}{{2\pi i}}\int\limits_{z_{th} }^R {\frac{{\psi _5 \left( {z - i\varepsilon } \right)\Delta \left( {z - i\varepsilon } \right)}}{{z - z_0 }}dz}  						\nonumber \\
&= \frac{1}{{2\pi i}}\mathop{\oint \mkern-0.8mu}\limits_{C\left( R \right)}  {\frac{{\psi _5 \left( z \right)\Delta \left( z \right)}}{{z - z_0 }}dz}  + \mathop {\lim 													}\limits_{\varepsilon  \to 0} \frac{1}{{2\pi i}}\int\limits_{z_{th} }^R {\frac{{\psi _5 \left( {z + i\varepsilon } \right)\Delta \left( {z + i\varepsilon } \right) - \psi _5 \left( {z - 							i\varepsilon } \right)\Delta \left( {z - i\varepsilon } \right)}}{{z - z_0 }}dz}  
\label{FESR}	
\end{align}

The residue from the pole is included in the hadronic spectral function. Since $\Delta(z)$ is a single valued analytic function, in the limit
\[
\mathop {\lim }\limits_{\varepsilon  \to 0} \Delta \left( {z + i\varepsilon } \right) = \mathop {\lim }\limits_{\varepsilon  \to 0} \Delta \left( {z - i\varepsilon } \right) = \Delta \left( z \right)
\]

${\psi _5 \left( {z + i\varepsilon } \right)}$ and ${\psi _5 \left( {z - i\varepsilon } \right)}$ are the functions evaluated in the two different branches or Riemann sheets of the function and we have that
\[
\mathop {\lim }\limits_{\varepsilon  \to 0} \psi _5 \left( {z + i\varepsilon } \right) \ne \mathop {\lim }\limits_{\varepsilon  \to 0} \psi _5 \left( {z - i\varepsilon } \right)
\]

we can now write the (\ref{FESR}) as
\begin{eqnarray}  
\psi _5 \left( {z_0 } \right)\Delta \left( {z_0 } \right) = \frac{1}{{2\pi i}}\mathop{\oint \mkern-0.8mu}\limits_{C\left( R \right)}  {\frac{{\psi _5 \left( z \right)\Delta \left( z 			\right)}}{{z - z_0 }}dz}  + \frac{1}{{2\pi i}}\int\limits_{z_{th} }^R {\frac{{\Delta \left( z \right)}}{{z - z_0 }}\mathop {\lim }\limits_{\varepsilon  \to 0} \left[ {\psi _5 \left( {z + 							i\varepsilon } \right) - \psi _5 \left( {z - i\varepsilon } \right)} \right]dz} 
\label{FESR2}
\end{eqnarray}

let us now consider an analytic function $f(z+i\varepsilon)$, we can Laurent expand this function around $z$ as
\begin{eqnarray}
f\left( {z + i\varepsilon } \right) = \sum\limits_{n = 0}^\infty  {\frac{{f^{\left( n \right)} \left( {i\varepsilon } \right)}} {{n!}}z^n } 
\label{laurent1}
\end{eqnarray}

were $f^{\left( n \right)}\left( {i\varepsilon } \right)$ is the n$^{th}$ derivative of the function $f(z)$ evaluated at the point $z=i\varepsilon$. We can do the same for the function $f(z-i\varepsilon)$ which is
\begin{eqnarray}
f\left( {z - i\varepsilon } \right) = \sum\limits_{n = 0}^\infty  {\frac{{f^{\left( n \right)} \left( {-i\varepsilon } \right)}} {{n!}}z^n } 
\label{laurent2}
\end{eqnarray}

We can now evaluate the difference of these functions 
\begin{eqnarray}  
f\left( {z + i\varepsilon } \right) - f\left( {z - i\varepsilon } \right) &=& \sum\limits_{n = 0}^\infty  {\frac{{f^{\left( n \right)} \left( {i\varepsilon } \right)}}{{n!}}z^n }  - \sum\limits_{n = 							0}^\infty  {\frac{{f^{\left( n \right)} \left( { - i\varepsilon } \right)}}{{n!}}z^n }  \nonumber \\
   &=& \sum\limits_{n = 0}^\infty  {\frac{{z^n }}{{n!}}} \left[ {f^{\left( n \right)} \left( {i\varepsilon } \right) - f^{\left( n \right)} \left( { - i\varepsilon } \right)} \right] \nonumber \\
   &=& \sum\limits_{n = 0}^\infty  {\frac{{z^n }}{{n!}}} \frac{{d^n }}{{dz^n }}\left[ {\left. {f\left( z \right)} \right|_{z = i\varepsilon }  - \left. {f\left( z \right)} \right|_{z =  - 										i\varepsilon } } \right] \nonumber \\
   &=& \sum\limits_{n = 0}^\infty  {\frac{{z^n }}{{n!}}\frac{{d^n }}{{dz^n }}\left. {\left[ {f\left( z \right) - f\left( {\bar z} \right)} \right]} \right|_{z = i\varepsilon } }  \nonumber \\
   &=& \sum\limits_{n = 0}^\infty  {\frac{{z^n }}{{n!}}\frac{{d^n }}{{dz^n }}\left. {\left[ {f\left( z \right) - \overline {f\left( z \right)} } \right]} \right|_{z = i\varepsilon } }  \label{schwartz} \\
   &=& \sum\limits_{n = 0}^\infty  {\frac{{z^n }}{{n!}}\frac{{d^n }}{{dz^n }}\left. {2i\operatorname{Im} \left[ {f\left( z \right)} \right]} \right|_{z = i\varepsilon } }  \nonumber \\
   &=& 2i\operatorname{Im} \left[ {\sum\limits_{n = 0}^\infty  {\frac{{z^n }}{{n!}}\left. {\frac{{d^n }}{{dz^n }}f\left( z \right)} \right|_{z = i\varepsilon } } } \right] \nonumber \\
   &=& 2i\operatorname{Im} \left[ {\sum\limits_{n = 0}^\infty  {\frac{{z^n }}{{n!}}f^{\left( n \right)} \left( {i\varepsilon } \right)} } \right] \nonumber \\   
   &=& 2i\operatorname{Im} \left[ {f\left( {z + i\varepsilon } \right)} \right]
\label{laurent3}
\end{eqnarray}

We have used the Schwartz Reflection Principle \cite{JordansLemma1,JordansLemma2,JordansLemma3} in (\ref{schwartz}) to analytically continue the correlator onto the other side of the branch cut. We now take the limit $\varepsilon \Rightarrow 0$ to obtain 
\begin{eqnarray}  
\mathop {\lim }\limits_{\varepsilon  \to 0} \left[ {f\left( {z + i\varepsilon } \right) - f\left( {z - i\varepsilon } \right)} \right] = \mathop {\lim }\limits_{\varepsilon  \to 0} 2i\operatorname{Im} \left[ {f\left( {z + i\varepsilon } \right)} \right] = 2i\operatorname{Im} \left[ {f\left( z \right)} \right]
\label{laurent4}
\end{eqnarray}

Now analogously we can do the same for the $\psi_5 (z)$ to get
\begin{eqnarray}
\mathop {\lim }\limits_{\varepsilon  \to 0} \left[ {\psi _5 \left( {z + i\varepsilon } \right) - \psi _5 \left( {z - i\varepsilon } \right)} \right] = 2i\operatorname{Im} \left[ {\psi _5 \left( z \right)} \right]
\label{laurent5}
\end{eqnarray}

and substituting (\ref{laurent5}) into (\ref{FESR2}) we obtain 
\[
\psi _5 \left( {z_0 } \right)\Delta \left( {z_0 } \right) = \frac{1}{{2\pi i}}\mathop{\oint \mkern-0.8mu}\limits_{C\left( R \right)}  {\frac{{\psi _5 \left( z \right)\Delta \left( z \right)}}{{z - z_0 }}dz}  + \frac{1}{\pi }\int\limits_{z_{th} }^R {\frac{{\Delta \left( z \right)}}{{z - z_0 }}\operatorname{Im} \left[ {\psi _5 \left( z \right)} \right]dz} 
\]

and in the low energy limit we get
\begin{eqnarray}
\psi _5 \left( 0 \right)\Delta \left( 0 \right) = \frac{1}{{2\pi i}}\mathop{\oint \mkern-0.8mu}\limits_{C\left( R \right)}  {\frac{{\Delta \left( z \right)}}
	{z}\psi _5 \left( z \right)  dz}   + \frac{1}{\pi }\int\limits_{z_{th} }^R {\frac{{\Delta \left( z \right)}}{z}\operatorname{Im} \left[ {\psi _5 \left( z \right)} \right]dz} 
\label{FESR3}
\end{eqnarray}

Equation (\ref{FESR3}) is referred to as the Finite Energy Sum Rule. The size of the radius of the circle used in the contour integration in (\ref{FESR3}) hints to us which correlating function will be used, if we choose the radius to be large enough then 
we can use the Perturbative QCD (pQCD) correlator for high energies for $\psi_5 (z)$ in the contour integral on the circle \cite{tech4,tech5}. For the second term in (\ref{FESR3}) we shall use the Hadronic spectral function. 
\begin{eqnarray}
\psi _5 \left( 0 \right)\Delta \left( 0 \right) = \frac{1}{{2\pi i}}\mathop{\oint \mkern-0.8mu}\limits_{C\left( R \right)}  {\frac{{\Delta \left( z \right)}}{z}\left. {\psi _5 \left( 			z \right)} \right|_{PQCD} dz}  + \frac{1}{\pi }\int\limits_{z_{th} }^R {\frac{{\Delta \left( z \right)}}{z}\left. {\operatorname{Im} \left[ {\psi _5 \left( z \right)} \right]} \right|_{HAD} dz} 
\label{FESR4}
\end{eqnarray}

For the case of the pion, the hadronic spectral function is the sum of the pion pole term and the resonance terms i.e.
\begin{eqnarray}
\frac{1}{\pi }\left. {\operatorname{Im} \left[ {\psi _5 \left( z \right)} \right]} \right|_{HAD}  = 2f_\pi ^2 M_\pi ^4 \delta \left( {z - M_\pi ^2 } \right) + \frac{1}{\pi }\left. {\operatorname{Im} \left[ {\psi _5 \left( z \right)} \right]} \right|_{RES} 
\label{hadres}
\end{eqnarray}

substituting (\ref{hadres}) into the sum rule (\ref{FESR4}) to get
\begin{eqnarray}
\psi _5 \left( 0 \right)\Delta \left( 0 \right) = 2f_\pi ^2 M_\pi ^2 \Delta \left( {M_\pi ^2 } \right) + \frac{1}{\pi }\int\limits_{z_{th} }^R {\frac{1}{z}\Delta \left( z \right)\left. {\operatorname{Im} \left[ {\psi _5 \left( z \right)} \right]} \right|_{RES} dz}  + \frac{1}{{2\pi i}}\mathop{\oint \mkern-0.8mu}\limits_{C\left( R \right)}  {\frac{1}
{z}\Delta \left( z \right)\left. {\psi _5 \left( z \right)} \right|_{PQCD} dz} 
\label{FESR5}
\end{eqnarray}

For convenience we define two new functions to simplify the work
\begin{eqnarray}  
\left. {\delta _5 \left(R\right)} \right|_{RES} &\equiv & \frac{1}{\pi }\int\limits_{z_{th} }^R {\frac{1}{z}\Delta \left( z \right)\left. {\operatorname{Im} \left[ {\psi _5 \left( z \right)} \right]} 								\right|_{RES} dz}  \label{delta5had} \\
\left. {\delta _5 \left( R \right)} \right|_{PQCD}  &\equiv & \frac{1}{{2\pi i}}\mathop{\oint \mkern-0.8mu}\limits_{C\left( R \right)}  {\frac{1}{z}\Delta \left( z \right)\left. {\psi _5 \left( z \right)} \right|_{PQCD} dz} \label{delta5QCD} 
\end{eqnarray}

the FESR which we will use in the subsequent analysis is then 
\begin{eqnarray} 
\psi _5 \left( 0 \right)\Delta \left( 0 \right) = 2f_\pi ^2 M_\pi ^2 \Delta \left( {M_\pi ^2 } \right) + \left. {\delta _5 \left( R \right)} \right|_{RES}  + \left. {\delta _5 \left( R \right)} \right|_{PQCD} 
\label{FESR6}
\end{eqnarray}

\clearpage

\section{Integration in the Complex Plane}
\label{sec:FixedOrderAndContourImprovedPerturbationTheory}
\numberwithin{equation}{subsection}

We will use three methods to evaluate the $\left. {\delta _5 \left( R \right)} \right|_{PQCD}$ of (\ref{delta5QCD}). We will use the method of Fixed Order Perturbation Theory (FOPT) and the method of Fixed Renormalization Scale/Frozen Order Perturbation Theory (FSPT).

The corresponding Perturbative QCD expression for (\ref{delta5QCD}) can be improved with the help of the Renormalization Group (RG) improvement. In the course of calculating (\ref{delta5QCD}) we will also need to perform the contour integration over the circle. We now have a choice of which order the integral and the renormalization group improvement will be performed. 

\subsection{Fixed Order Perturbation Theory}
\label{sec:FixedOrderPerturbationTheory}

Fixed Order Perturbation Theory \cite{Jamin,paper,tech1,tech2} refers to first evaluating the contour integral and then performing the renormalization group improvement. The integrals are performed in the complex energy squared plane (z-plane) on a contour which is a circle with fixed energy (constant radius). Since the strong coupling $\alpha_s \left(z\right)$ and the masses of the quarks $m_q \left(z\right)$ depend only on the magnitude of the complex variable $z$ i.e. depend only on $R$ which is the squared energy in the complex plane, these quantities can be taken out of the contour integral. Only after the integrals are done, we then implement the renormalization group improvement by setting the renormalization scale equal to the radius of the circle i.e. $\mu ^2 = R$, so that all the logarithmic terms in the form $\ln(R/\mu^2)$ vanish. 

The Perturbative QCD correlator given in \cite{paper,paper2,paper3,paper4,paper5,paper6} is 
\begin{eqnarray}
\left. {\psi _5 \left( {q^2 } \right)} \right|_{PQCD}  = m_q^2 \left( {\left| q \right|^2 } \right)\left[ { - q^2 \Pi _0 \left( {q^2 } \right) + m_q \left( {\left| q \right|^2 } \right)\Pi _2 \left( {q^2 } \right) + \frac{{C_u }}
{{ - q^2 }}\left. {\left\langle {m_q \bar uu} \right\rangle } \right|_{\mu _0 }  + \sum\limits_{j = 1}^3 {\frac{{C_j }}
{{ - q^2 }}\left\langle {\hat O_j } \right\rangle }  +  \ldots } \right]
\label{pqcdcorrel}
\end{eqnarray}

were $\mu_0$ is the scale at which the condensate is measured.
\begin{eqnarray}  
\Pi _0 \left( {q^2 } \right) &=& \frac{1} {{16\pi ^2 }}\left[ { - 12 + 6\ln \left( {\frac{{ -q^2 }}{{\mu ^2 }}} \right) + \left( {\frac{{\alpha _s \left( {\left| q \right|^2 }\right)}} {\pi 									}}\right)\left( {24\zeta \left(3\right) -\frac{{131}}{2}+34\ln \left( {\frac{{-q^2 }}{{\mu ^2 }}} \right) - 6\ln ^2 \left( {\frac{{ - q^2 }}{{\mu ^2 }}} \right)} \right)} \right] \nonumber \\
  &+& \frac{1}{{16\pi ^2 }}\left( {\frac{{\alpha _s \left( {\left| q \right|^2 } \right)}}{\pi }} \right)^2 \left[ {{\text{A}_3}\ln \left( {\frac{{ - q^2 }}{{\mu ^2 }}} \right) + {\text{B}_3}\ln ^2 						\left( {\frac{{ - q^2 }}{{\mu ^2 }}} \right) + {\text{C}_3}\ln ^3 \left( {\frac{{ - q^2 }}{{\mu ^2 }}} \right)} \right] \nonumber \\
  &+& \frac{1}{{16\pi ^2 }}\left( {\frac{{\alpha _s \left( {\left| q \right|^2 } \right)}}{\pi }} \right)^3 \left[ {A\ln \left( {\frac{{ - q^2 }}{{\mu ^2 }}} \right) + B\ln ^2 \left( {\frac{{ - q^2 }}
			{{\mu ^2 }}} \right) + C\ln ^3 \left( {\frac{{ - q^2 }}{{\mu ^2 }}} \right) + D\ln ^4 \left( {\frac{{ - q^2 }}{{\mu ^2 }}} \right)} \right] \nonumber \\
  &+& \frac{1}{{16\pi ^2 }}\left( {\frac{{\alpha _s \left( {\left| q \right|^2 } \right)}}{\pi }} \right)^4 \sum\limits_{i = 1}^5 {H_i L^i }  
  \label{pqcdpi0}
\end{eqnarray}

with $H_i$ are a set of real numbers and $L^i = \rm{ln^i}(-q^2/\mu^2)$ and 
\begin{align}
\Pi _2 \left( {q^2 } \right) = \frac{1}{{16\pi ^2 }}\left\{ { - 12 + 12\ln \left( {\frac{{ - q^2 }}{{\mu ^2 }}} \right) + \frac{{\alpha _s \left( {\left| {q^2 } \right|} \right)}}{\pi }\left[ { - 					1000 + 64\ln \left( {\frac{{ - q^2 }}{{\mu ^2 }}} \right) - 24\left[ {\ln \left( {\frac{{ - q^2 }}{{\mu ^2 }}} \right)} \right]^2  + 48\zeta \left( 3 \right)} \right]} \right\}
\label{pqcdpi2}	
\end{align}

substituting (\ref{pqcdpi0}) and (\ref{pqcdpi2}) into (\ref{pqcdcorrel}) and using $\Delta \left(z \right)=1-a_0 z - a_1 z^2$ as the integration kernel in (\ref{delta5QCD}) we obtain
\begin{eqnarray}  
\left. {\delta _5 \left( R \right)} \right|_{PQCD}  = \left. {\delta _5 \left( R \right)} \right|_{\Pi _0 }  + \left. {\delta _5 \left( R \right)} \right|_{\Pi _2 }  + \left. {\delta _5 \left( R \right)} \right|_{C_u }  + \left. {\delta _5 \left( R \right)} \right|_{C_j } 
  \label{deltapqcd}
\end{eqnarray}

with
\begin{eqnarray}  
\left. {\delta _5 \left( R \right)} \right|_{\Pi _0 }  = \left. {\delta _5 } \right|_{1{\text{loop}}}  + \left. {\delta _5 } \right|_{{\text{2loop}}}  + \left. {\delta _5 } \right|_{{\text{3loop}}}  + \left. {\delta _5 } \right|_{{\text{4loop}}}  + \left. {\delta _5 } \right|_{{\text{5loop}}} 
\end{eqnarray}

See explicit forms of these functions which have been calculated in the Appendix \ref{PerturbativeQCDCalculations}.

With the definitions given below we can calculate numerically the behavior of $\left. {\delta _5 \left( R \right)} \right|_{PQCD}$ as a function of the energy $R$.

The coefficients of the QCD Beta function \cite{paper5} series for $n$ quark types:
\begin{eqnarray}
  \beta _0  &=& 0.25\left( {11 - \frac{2}{3}n} \right) \nonumber \\
  \beta _1  &=& \frac{1}{{16}}\left( {102.0 - \frac{{38}}{3}n} \right) \nonumber \\
  \beta _2  &=& \frac{1}{{64}}\left( {\frac{{2857.0}}{2} - \frac{{5033}}{{18}}n + \frac{{325}}{{54}}n^2 } \right) \nonumber \\
  \beta _3  &=& \frac{1}{{256.0}}\left( {\frac{{149753}}{{6.0}} + 3564\zeta (3) - \left( {\frac{{1078361}}{{162.0}} + \frac{{6508}}{{27}}\zeta (3)} \right)n + \left( {\frac{{50065}}{{162.0}} 															+\frac{{6472}}{{81}}\zeta (3)} \right)n^2  + \frac{{1093}}{{729.0}}n^3 } \right) \nonumber
\end{eqnarray}

$\Lambda \equiv \Lambda_{\rm{QCD}}$ for the definitions found in \cite{paper,paper6}:
\begin{eqnarray}
  L[z,\Lambda ] &=& \ln\left[ {\frac{z}{{\Lambda ^2 }}} \right] \nonumber \\
  a1[z,\Lambda ] &=&  - \frac{{\beta _1 }}{{\beta _0 ^2 }}\ln [L[z,\Lambda ]] \nonumber \\
  a2[z,\Lambda ] &=& \frac{1}{{\beta _0 ^2 }}\left( {\frac{{\beta _1 ^2 }}{{\beta _0 ^2 }}\left( {\ln [L[z,\Lambda ]]^2  - \ln [L[z,\Lambda ]] - 1} \right) + \frac{{\beta _2 }}{{\beta _0 }}} \right) 											\nonumber \\
  a3[z,\Lambda ] &=& \frac{1}{{\beta _0 ^3 }}\left( {\frac{{\beta _1 ^3 }}{{\beta _0 ^3 }}\left( { - \ln [L[z,\Lambda ]]^3  + 2.5\ln [L[z,\Lambda ]]^2  + 2\ln [L[z,\Lambda ]] - 0.5} \right) - 														3\frac{{\beta _1 \beta _2 }}{{\beta _0 ^2 }}\ln [L[z,\Lambda ]] + 0.5\frac{{\beta _3 }}{{\beta _0 }}} \right) \nonumber \\
  \alpha [z,\Lambda ] &=& \frac{\pi }{{\beta _0 L[z,\Lambda ]}}\left( {1 + \frac{{a1[z,\Lambda ]}}{{L[z,\Lambda ]}} + \frac{{a2[z,\Lambda ]}}{{L[z,\Lambda ]^2 }} + \frac{{a3[z,\Lambda ]}}{{L[z,\Lambda ]^3 }}} \right) \label{formulaalpha}
\end{eqnarray}

The coefficients of the QCD Gamma function series for quark masses:
\begin{eqnarray}
  \gamma _0  &=& 1 \nonumber \\
  \gamma _1  &=& \frac{1}{{16}}\left( {\frac{{202}}{3} - \frac{{20}}{9}n} \right)  \nonumber \\
  \gamma _2  &=& \frac{1}{{64}}\left[ {1249.0 - \left( {\frac{{2216}}{{27.0}} + \frac{{160}}{3}\zeta \left( 3 \right)} \right)n - \frac{{140}}{{81.0}}n^2 } \right]  \nonumber \\
	\gamma _3  &=& \frac{1}{{256}}\left[ {\left( {\frac{{4603055}}{{162.0}} + \frac{{135680}}{{27}}\zeta \left( 3 \right) - 8800\zeta \left( 5 \right)} \right) + \left( { - \frac{{91723}} {{27}} - 											\frac{{34192}}{9}\zeta \left( 3 \right) + 880\zeta \left( 4 \right) + \frac{{18400}}{9}\zeta \left( 5 \right)} \right)n} \right] \nonumber \\
						&+& \frac{1}{{256}}\left[ {\left( {\frac{{5242}}{{243}} + \frac{{800}}{9}\zeta \left( 3 \right) - \frac{{160}}{3}\zeta \left( 3 \right)} \right)n^2  + \left( { - \frac{{332}}{{243}} + 										\frac{{64}}{{27}}\zeta \left( 3 \right)} \right)n^3 } \right] \nonumber
\end{eqnarray}

\begin{eqnarray}  
d1[z,\Lambda ] &=&  - \frac{{\gamma _0 \beta _1 }}{{\beta _0 ^3 }}{\text{ln}}[L[z,\Lambda ]] \nonumber \\
d2[z,\Lambda ] &=& \frac{{\gamma _0 }}{{\beta _0 ^3 }}\left( {\frac{\beta_1 ^2 }{{\beta _0 ^2 }}\left( {{\text{ln}}[L[z,\Lambda ]]^2  - {\text{ln}}[L[z,\Lambda ]] - 1} \right) 																		+\frac{\beta_2}{{\beta _0 }}} \right) + 0.5\frac{{\gamma _0 }}{{\beta _0 }}\left( {\frac{{\gamma _0 }}{{\beta _0 }} - 1} \right)\frac{{{\beta_1}^2 }}{{\beta _0 ^4 																}}{\text{ln}}[L[z,\Lambda ]]^2 \nonumber \\
d3[z,\Lambda ] &=& \frac{{\gamma _0 }}{{\beta _0 ^4 }}\left( {\frac{{{\beta_1}^3 }}{{\beta _0 ^3 }}\left( { - {\text{ln}}[L[z,\Lambda ]]^3  + 2.5{\text{ln}}[L[z,\Lambda]]^2 + 2{\text{ln}} 														[L[z,\Lambda]] - 0.5} \right) - 3\frac{{\beta _1 \beta _2 }}{{\beta _0 ^2 }}{\text{ln}}[L[z,\Lambda ]] + 0.5\frac{{\beta _3 }}{{\beta _0 }}} \right) \nonumber\\
							 &+&0.5\frac{{\gamma _0 }}{{\beta _0 }}\left({\frac{{\gamma _0 }}{{\beta _0 }} - 1} \right)\left( { - 2\frac{{\beta _1 }}{{\beta _0 ^4 }}} \right)\left( {\frac{{\beta _1 ^2 }}{{\beta _0 								^2 }}\left( {{\text{ln}}[L[z,\Lambda ]]^2  -{\text{ln}}[L[z,\Lambda ]] - 1} \right) + \frac{{\beta _2 }}{{\beta _0 }}} \right){\text{ln}}[L[z,\Lambda ]] \nonumber \\
							 &-& \frac{1}{6}\frac{{\gamma _0}}{{\beta _0 }}\left( {\frac{{\gamma _0 }}{{\beta _0 }} - 1} \right)\left( {\frac{{\gamma _0 }}{{\beta _0 }} - 2} \right)\frac{{\beta _1 ^3 }}{{\beta _0 ^6 }}{\text{ln}}[L[z,\Lambda ]]^3 \nonumber
\end{eqnarray}

\begin{eqnarray}   
c1 &=& \frac{{\gamma _1 }}{{\beta _0 }} - \frac{{\gamma _0 \beta _1 }}{{\beta _0 ^2 }} \nonumber \\
c2 &=& 0.5\left( {\frac{{\gamma _2 }}{{\beta _0 }} - \frac{{\gamma _1 \beta _1 }}{{\beta _0 ^2 }} + \frac{{\gamma _0 }}{{\beta _0 ^2 }}\left( {\frac{{\beta _1 ^2 }}{{\beta _0 }} - \beta _2 } \right)} 										\right) \nonumber \\
c3 &=& \frac{1}{3}\left( {\frac{{\gamma _0 }}{{\beta _0 ^2 }}\left( {\frac{{\beta _1 \beta _2 }}{{\beta _0 }} - \frac{{\beta _1 }}{{\beta _0 }}\left( {\frac{{\beta _1 ^2 }}{{\beta _0 }} - \beta _2 } 			\right)-\beta _3 } \right) + \frac{{\gamma _1 }}{{\beta _0 ^2 }}\left( {\frac{{\beta _1 ^2 }}{{\beta _0 }} - \beta _2 } \right) - \frac{{\beta _1 \gamma _2 }}{{\beta _0 ^2 }} + \frac{{\gamma _3 }}
	{{\beta _0 }}} \right) \nonumber 
\end{eqnarray}

For convenience we have split the right-hand-side of the quark mass function into a product of the invariant quark mass $\widehat m_\Lambda$ and the functional dependence on the energy $z$ and the $\Lambda_{\rm{QCD}}$ scale in the $U\left( {z,\Lambda} \right) $
\begin{align}
U\left( {z,\Lambda} \right)=& \frac{1}{{\left( {\frac{1}{2}L\left[ {z,\Lambda } \right]} \right)^{\frac{{\gamma _0 }}{{\beta _0 }}} }} \Biggl\{ 1 + \left( {\frac{{c1}}{{\beta _0 }} + d1\left[ {z,\Lambda } 				\right]} \right)\frac{1}{{L[z,\Lambda ]}} +\nonumber \\
&+ \left[ {\left( {c1\frac{{a1\left[ {z,\Lambda } \right]}}{{\beta _0 }} + \frac{{0.5c1^2  + c2}}{{\beta _0 ^2 }}} \right) + d2\left[ {z,\Lambda } 						\right] + c1\frac{{d1\left[ {z,\Lambda } \right]}}{{\beta _0 }}} \right]\frac{1}{{L\left[ {z,\Lambda } \right]^2 }}  \nonumber \\
+& \left[ 								{c1\frac{{a2\left[ {z,\Lambda } \right]}}{{\beta _0 }} + \frac{{\left( {c2 + 0.5c1^2 } \right)}}{{\beta _0 ^2 }}\left( {d1\left[ {z,\Lambda } \right] - 2\frac{{\beta _1 }}{{\beta _0 ^2 }}\ln 					\left[ {L\left[ {z,\Lambda } \right]} \right]} \right) + \frac{{c3 + c1c2 + \frac{1}{6}c1^3 }}{{\beta _0 ^3 }}} \right]\frac{1}{{L\left[ {z,\Lambda } \right]^3 }} \nonumber \\
+& \left[ 								{c1\frac{{d1\left[ {z,\Lambda } \right]a1\left[ {z,\Lambda } \right]}}{{\beta _0 }} + c1\frac{{d2\left[ {z,\Lambda } \right]}}{{\beta _0 }} + d3\left[ {z,\Lambda } \right]} \right]\frac{1}
{{L\left[ {z,\Lambda } \right]^3 }} \Biggr\}	
\end{align}
%

were

\begin{eqnarray} 
m\left( {z,\Lambda } \right) = \widehat m_\Lambda  U\left( {z,\Lambda } \right)
\end{eqnarray}

\subsection{Fixed Renormalization Scale Perturbation Theory}
\label{sec:FixedRenomalizationSchemePerturbationTheory}

The method of Fixed Renormalization Scale Perturbation Theory\footnote{work done in collaboration with J.M. Bordes and J. Pe$\widetilde{\text{n}}$arrocha at the Departamento de Fisica Teorica, Universitat de Valencia, and Instituto de Fisica Corpuscular, Centro Mixto Universitat de Valencia-CSIC between November 2009 - March 2010} \cite{tech3,tech6,menke} is very similar to the method of Fixed Order Perturbation Theory. In this case we generalize the integration kernel from a simply second order polynomial to a $n^{th}$ order polynomial and instead of choosing the renormalization scale to be radius of the circle in the complex energy plane we will instead choose a fixed scale which will be varied over a large interval of $\mu^2=2-50$ GeV$^2$. The method is implemented by using the Legendre polynomials \cite{dominguez2} and making use of their orthogonality property. We first generalize the kernel to 
\begin{eqnarray}
\tilde \Delta _n \left( z \right) \equiv \frac{{P_n \left[ {x\left( z \right)} \right]}}
{{P_n \left[ {x\left( {M_\pi ^2 } \right)} \right]}} = \sum\limits_{j = 0}^n {a_j z^j } 
\label{gendelta}
\end{eqnarray}

were $P_n$ is the Legendre polynomial which can be generated by the Rodriguez Formula
\begin{eqnarray}P_n \left( x \right) = \frac{1}
{{2^n n!}}\left( {\frac{d}
{{dx}}} \right)^n \left( {x^2  - 1} \right)^n 
\label{rod}
\end{eqnarray}

We use the orthogonality of the polynomials i.e.
\[
\int\limits_{ - 1}^1 {P_n \left( x \right)P_m \left( x \right)dx}  = 0
\]
for $n \neq m$ by rescaling the domain of integration as

\[
x\left( z \right) \equiv \frac{{2z - \left( {z_0  - z_{th} } \right)}}
{{z_{th}  - z_0 }}
\]

with $z \in \left[ {z_{th} ;z_0 } \right]$ yields the following constraints on (\ref{gendelta})
\begin{eqnarray}  
\tilde \Delta _n \left( {M_\pi ^2 } \right) &=& 1 \label{constraintsa} \\
\int\limits_{z_{th} }^{z_0 } {z^k \tilde \Delta _n \left( z \right)dz}  &=& 0{\text{       }}k \in \mathbb{N},k < n 
\label{constraintsb}
\end{eqnarray}

Imposing the orthogonality condition (\ref{constraintsb}) forces the generalized polynomial $\tilde \Delta _n \left( z \right)$ to minimize the contributions of the continuum in the range $z \in \left[ {z_{th} ;z_0 } \right]$.

We then substitute (\ref{gendelta}) into (\ref{delta5QCD}) and (\ref{delta5had}) and choose allow $\mu^2$ to vary in the range $2-50$ GeV$^2$. In our case we found the best(largest stable domain) results for $\mu^2=4$ GeV$^2$. We varied the value of $\mu^2$ in the range $2-50$ GeV$^2$ to obtain an uncertainty in the value of the $\delta_\pi$

\clearpage

\section{The Resonance Contribution $\left. {\delta _5 \left(R\right)} \right|_{RES}$}
\label{sec:TheResonanceContributionLeftDelta5LeftRRightRightRES}
\numberwithin{equation}{section}
To evaluate the resonance term (\ref{delta5had}) we require information about the hadronic spectral function beyond the pion pole. At the current time this information is not available from experiment and we only have information about the mass and width of the first two excited states of the pion i.e. the $\Pi(1300)$ \cite{gluonc} and the $\Pi(1800)$ excitations. To proceed further we shall have to model the hadronic spectral function by using threshold constraints from Chiral Perturbation Theory ($\chi$-PT) which are imposed on a linear combination of Breit-Wigner profiles to describe the two excitations of the pion. Alas by introducing this model we will also introduce systematic uncertainties and solutions obtained thereafter are dependent on the model used. However, we can reduce these systematic uncertainties by introducing analytic integration kernels that are forced to be zero at the resonance points. This will reduce the effect of the systematic uncertainties in the neighborhood of the resonances. We have introduced the quadratic polynomial $\Delta(z)$ earlier precisely because of this!

We shall start by modeling the hadronic spectral function as
\begin{eqnarray} 
\frac{1}
{\pi }\left. {\operatorname{Im} \left[ {\psi _5 \left( z \right)} \right]} \right|_{RES}  = \frac{1}
{\pi }\left. {\operatorname{Im} \left[ {\psi _5 \left( z \right)} \right]} \right|_{3\pi } BW\left( z \right)
  \label{hadspec1}
\end{eqnarray}

were
\begin{eqnarray} 
\frac{1}
{\pi }\left. {\operatorname{Im} \left[ {\psi _5 \left( z \right)} \right]} \right|_{3\pi }  = \frac{1}
{{3 \cdot 2^8 \pi ^4 }}\frac{{M_\pi ^4 }}
{{f_\pi ^2 }}z\Theta \left( z \right)
  \label{pagelschpt}
\end{eqnarray}

is the constraint from $\chi$-PT \cite{pagels}. The Breit-Wigner profile BW(z) will be in the form 
\begin{eqnarray} 
BW\left( z \right) = \frac{{BW_1 \left( z \right) + \kappa BW_2 \left( z \right)}}
{{1 + \kappa }}
  \label{bw1}
\end{eqnarray}

the $\kappa$ is just a weighting parameter which controls the relative weight of the resonances with $\kappa \in \mathbb{R}^+$. $\kappa = 0.5$ was chosen for the computational work since the width of the $\Pi(1300)$ is twice as broad as the width of the $\Pi(1812)$ resonance and thus the $\Pi(1812)$ resonance would only contribute half as much as the $\Pi(1300)$ resonance. The $BW_1 \left( z \right)$ and $BW_2 \left( z \right)$ are isolated Breit-Wigner resonances. We normalize the Breit-Wigner profile at zero to be $BW(0)=1$, this implies that $BW_1 (0)=1$ and $BW_2 (0)=1$. This sets the form of the Breit-Wigner resonances we will have to use 
\begin{eqnarray} 
BW_k \left( z \right) = \frac{{M_k^2 \left( {M_k^2  + \Gamma _k^2 } \right)}}
{{\left( {z - M_k^2 } \right)^2  + M_k^2 \Gamma _k^2 }}
  \label{bwk}
\end{eqnarray}

we can now proceed with the calculation, substituting (\ref{pagelschpt}) and (\ref{bw1}) into (\ref{hadspec1}) we obtain
\begin{eqnarray}
\frac{1}{\pi }\left. {\operatorname{Im} \left[ {\psi _5 \left( z \right)} \right]} \right|_{RES}  = \frac{1}
{{3 \cdot 2^8 \pi ^4 }}\frac{{M_\pi ^4 }}
{{f_\pi ^2 }}\frac{1}
{{1 + \kappa }}z\Theta \left( z \right)\left[ {BW_1 \left( z \right) + \kappa BW_2 \left( z \right)} \right]
  \label{hadspec2}
\end{eqnarray}

for convenience we define a new parameter $\lambda$ which is just a collection of all the constants 
\[
\lambda  \equiv \frac{1}
{{3 \cdot 2^8 \pi ^4 }}\frac{{M_\pi ^4 }}
{{f_\pi ^2 }}\frac{1}
{{1 + \kappa }}
\]

and 
\begin{eqnarray}
\frac{1}
{\pi }\left. {\operatorname{Im} \left[ {\psi _5 \left( z \right)} \right]} \right|_{RES}  = \lambda z\Theta \left( z \right)\left[ {BW_1 \left( z \right) + \kappa BW_2 \left( z \right)} \right]
  \label{hadspec3}
\end{eqnarray}

we can now calculate the resonance contribution in this model by substituting (\ref{hadspec3}) into (\ref{delta5had}) to get
\begin{eqnarray}  
\left. {\delta _5 \left( R \right)} \right|_{RES}  &=& \int\limits_{z_{th} }^R {\frac{1}
					{z}\Delta \left( z \right)\lambda z\Theta \left( z \right)\left[ {BW_1 \left( z \right) + \kappa BW_2 \left( z \right)} \right]dz}  \nonumber \\
   		&=& \lambda \int\limits_{9M_\pi ^2 }^R {\frac{1}
					{z}\Delta \left( z \right)z\Theta \left( z \right)BW_1 \left( z \right)dz}  + \lambda \kappa \int\limits_{9M_\pi ^2 }^R {\frac{1}
					{z}\Delta \left( z \right)z\Theta \left( z \right)BW_2 \left( z \right)dz}  \nonumber \\
   		&=& \lambda \left[ {G_1 \left( R \right) - G_1 \left( {9M_\pi ^2 } \right)} \right] + \lambda \kappa \left[ {G_2 \left( R \right) - G_2 \left( {9M_\pi ^2 } \right)} \right] 
\end{eqnarray}

were the function $G_k \left( x \right)$ is 
\begin{align}
G_k \left( z \right) = M_k^2 \left( {M_k^2  + \Gamma _k^2 } \right) \Biggl\{\frac{1}{{M_k \Gamma _k }}&\left[ {1 - a_0 M_k^2  - a_1 \left( {M_k^4  - M_k^2 \Gamma _k^2 } \right)} \right]\arctan \left( {\frac{{z - M_k^2 }}{{M_k \Gamma _k }}} \right) +\nonumber\\
&- \left({\frac{1}
		{2}a_0  + a_1 M_k^2 } \right)\ln \left[ {\left( {z - M_k^2 } \right)^2  + M_k^2 \Gamma _k^2 } \right] - a_1 \left( {z - M_k^2 } \right) \Biggr\} \nonumber \\	
\end{align}

which is derived in Appendix \ref{TheResonanceIntegrals}. The inclusion of the analytic integration kernel $\Delta(z)$ attempts to suppress the  uncertainties from the resonance regions which is poorly understood. Even with this suppression we cannot eliminate entirely these uncertainties. To get a handle on how good the suppression is we make a Taylor expansion of the $\delta_5$ or equivalently the $G_k (z)$ around the resonance peak. For convenience we shall choose $k=1$ and keeping only the first order terms in the expansion we obtain to leading order
\begin{eqnarray}
\delta _5 \left( R \right) \approx \lambda M_1^2 \left( {M_1^2  + \Gamma _1^2 } \right)\frac{{M_1^2  - M_2^2 }}
{{2M_1^2 M_2^2 }}\left\{ {\left( {\frac{{R - M_1^2 }}
{{M_1 \Gamma _1 }}} \right) + \ln \left( {M_1 \Gamma _1 } \right)^2 } \right\}
\end{eqnarray}

The coefficient of the braces is the gradient of this line and sets the size on the contribution from the resonances. The magnitude of this coefficient is 
\[
\lambda M_1^2 \left( {M_1^2  + \Gamma _1^2 } \right)\frac{{M_1^2  - M_2^2 }}
{{2M_1^2 M_2^2 }} = 1.2 \times 10^{ - 7} 
\]

We see that not only in the immediate neighborhood ($R \approx M_1 ^2$) is the suppression effective but due to the tiny coefficient, the contribution from the resonances is effectively eliminated!

\clearpage
\section{Results}
\label{sec:Results}
We consider two cases,
\begin{itemize}
\item  $SU(2)\otimes SU(2)$ for the Pions
\item  $SU(3)\otimes SU(3)$ case for the Kaons
\end{itemize}

of which the FOPT results will be of main interest.We use the particle data from the Particle Data Group \cite{pdgdata,CADMass} for the $SU(2)\otimes SU(2)$ Pion case
\begin{itemize}
\item  $m_{\pi} =   139.642400 \pm 0.00035$ MeV
\item  $f_{\pi} =   92.21     \pm 0.14$ MeV
\item  $m_{\pi}^* = 1300 			\pm 100$ MeV
\item  $\Gamma_{\pi}^* = 400 			\pm 200$ MeV
\item  $m_{\pi}^{**} = 1812 			\pm 14$ MeV
\item  $\Gamma_{\pi}^{**} = 207 			\pm 13$ MeV
\item  $\overline{m}_u(2\rm{GeV}) = 2.9 			\pm 0.2$ MeV
\item  $\overline{m}_d(2\rm{GeV}) = 5.3 			\pm 0.4$ MeV
\item  $\hat{m}_u(2\rm{GeV}) = 3.8-3.9$ MeV
\item  $\hat{m}_d(2\rm{GeV}) = 6.4-7.2$ MeV
\end{itemize}

$SU(3)\otimes SU(3)$ Kaon\cite{dominguez2,CADMass} case
\begin{itemize}
\item  $m_{K} =   493.677 \pm 0.005$ MeV
\item  $m_{K}^* = 1460 			$ MeV
\item  $\Gamma_{K}^* = 250 	$ MeV
\item  $m_{K}^{**} = 1830 	$ MeV
\item  $\Gamma_{K}^{**} = 250 	$ MeV
\item  $\overline{m}_u(2\rm{GeV}) = 2.9 			\pm 0.2$ MeV
\item  $\hat{m}_u(2\rm{GeV}) = 3.8-3.9$ MeV
\item  $\hat{m}_s(2\rm{GeV}) = 130-150$ MeV
\end{itemize}

We note here that for the plots presented below we will be searching for the so-called stable region. This is the domain on the plot where the function varies the slowest. This domain then defines the stability region. Once we have identified the stability region we read of the average value of the function over this domain. We shall use either $\Lambda_{\rm{QCD}} =0.365$ GeV or $\Lambda_{\rm{QCD}} =0.397$ GeV. It has become standard practice to make predictions about hadronic parameters at these scales.
\clearpage

\subsection{Frozen Order Results}
\label{sec:FOResults}

\subsubsection{$SU(2)\otimes SU(2)$}
\label{SU2OtimesSU2}

\begin{figure}[h]
	\centerline{
	\mbox{\includegraphics[width=6in]{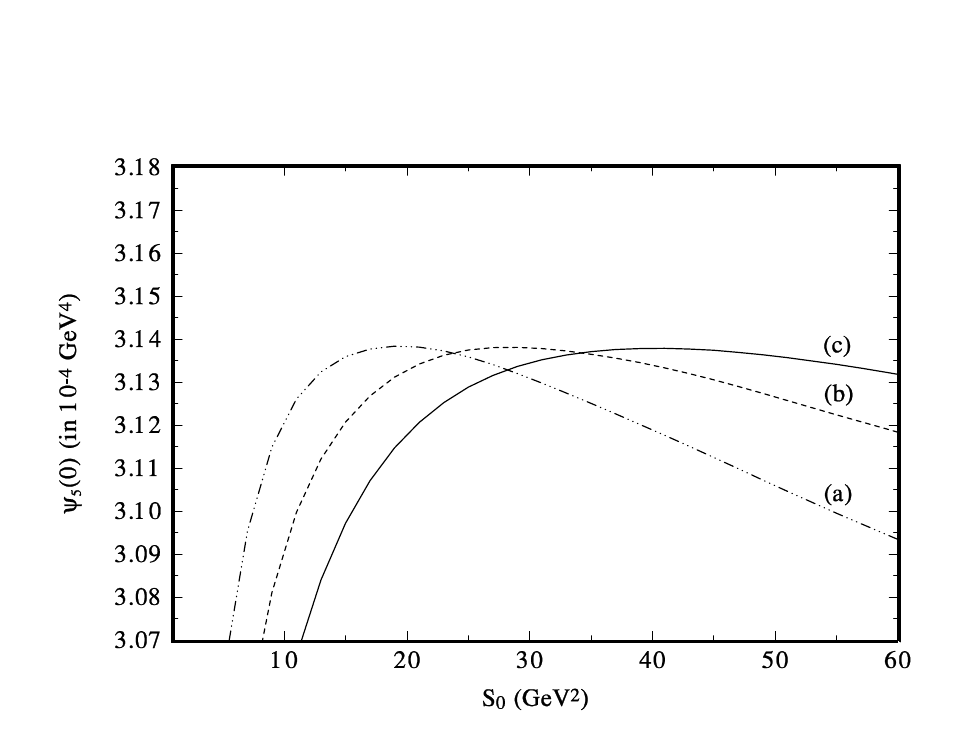}}
	}
	\caption{Results for $\psi_5 (0)$(GeV$^4$) \cite{dominguez2} as a function of the energy $s$ in FSPT with $\mu^2=4$GeV$^2$. a, b and c correspond to polynomial orders 4, 5 and 6 respectively.}
	\label{CIPTpsi5nores}
\end{figure}

From the Fixed Scale Perturbation Theory we see that the stability is over a small interval instead of the whole integration range. From the stable intervals we extract a value of $\delta_\pi =0.056 \pm 0.011$

\clearpage
\subsection{FOPT Results}
\label{sec:FOPTResults}

\subsubsection{$SU(2)\otimes SU(2)$}
\label{FOPTSU2OtimesSU2}

\begin{figure}[h]
	\centerline{
	\mbox{\includegraphics[width=9in]{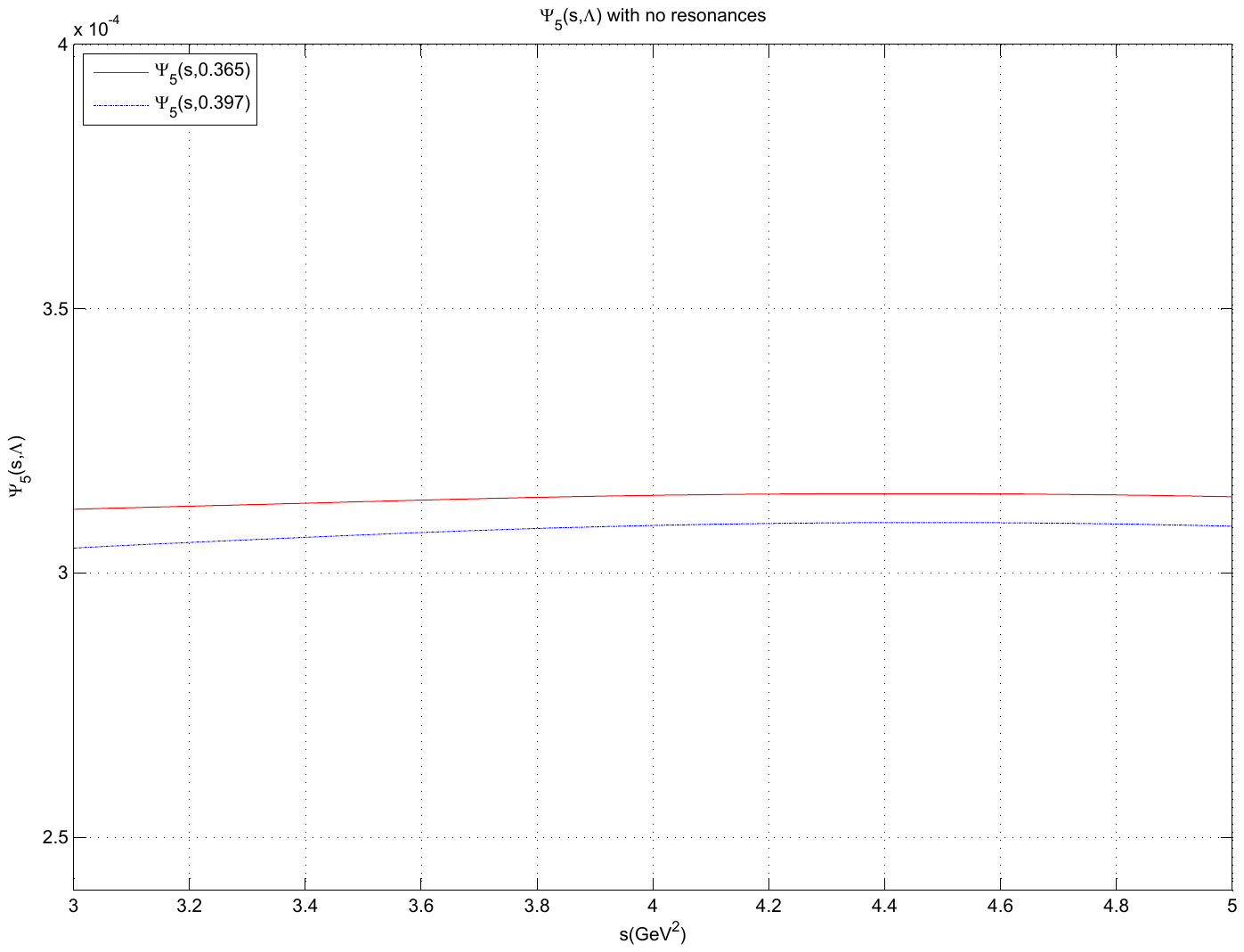}}
	}
	\caption{Results for $\psi_5 (0)$(GeV$^4$) as a function of the energy $s$ in FOPT. We have neglected the two resonances. Starting from the top we have $\psi_5 (s,0.365)$ and $\psi_5 (s,0.397)$ respectively}
	\label{FOPTpsi1}
\end{figure}

\clearpage

\begin{figure}[h]
	\centerline{
	\mbox{\includegraphics[width=9in]{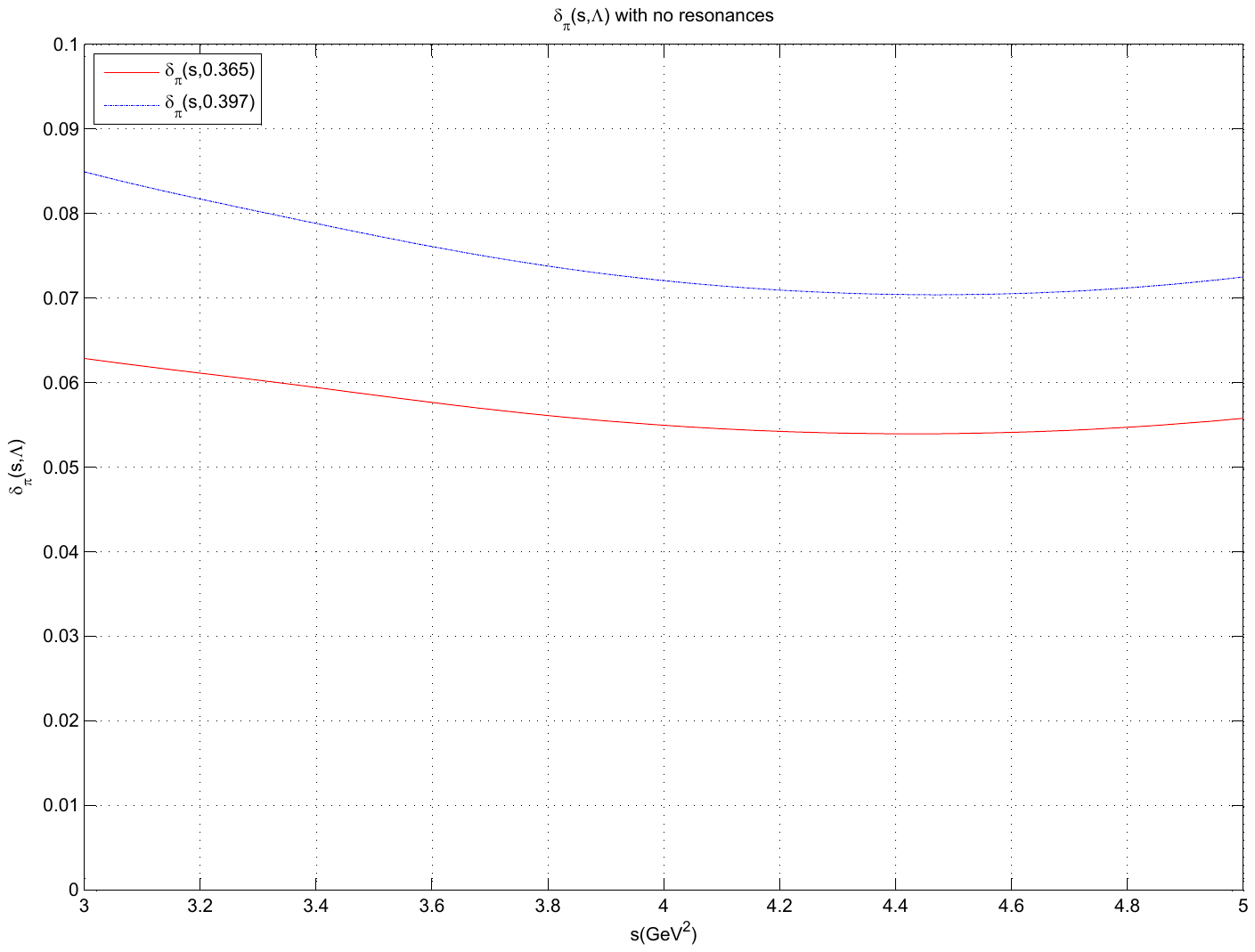}}
	}
	\caption{Results for $\delta_\pi $  as a function of the energy $s$ in FOPT. We have neglected the two resonances. Starting from the top we have $\delta_\pi (s,0.397)$ and $\delta_\pi (s,0.365)$ respectively}
	\label{FOPTdeltapi1}
\end{figure}

\clearpage

\begin{figure}[h]
	\centerline{
	\mbox{\includegraphics[width=9in]{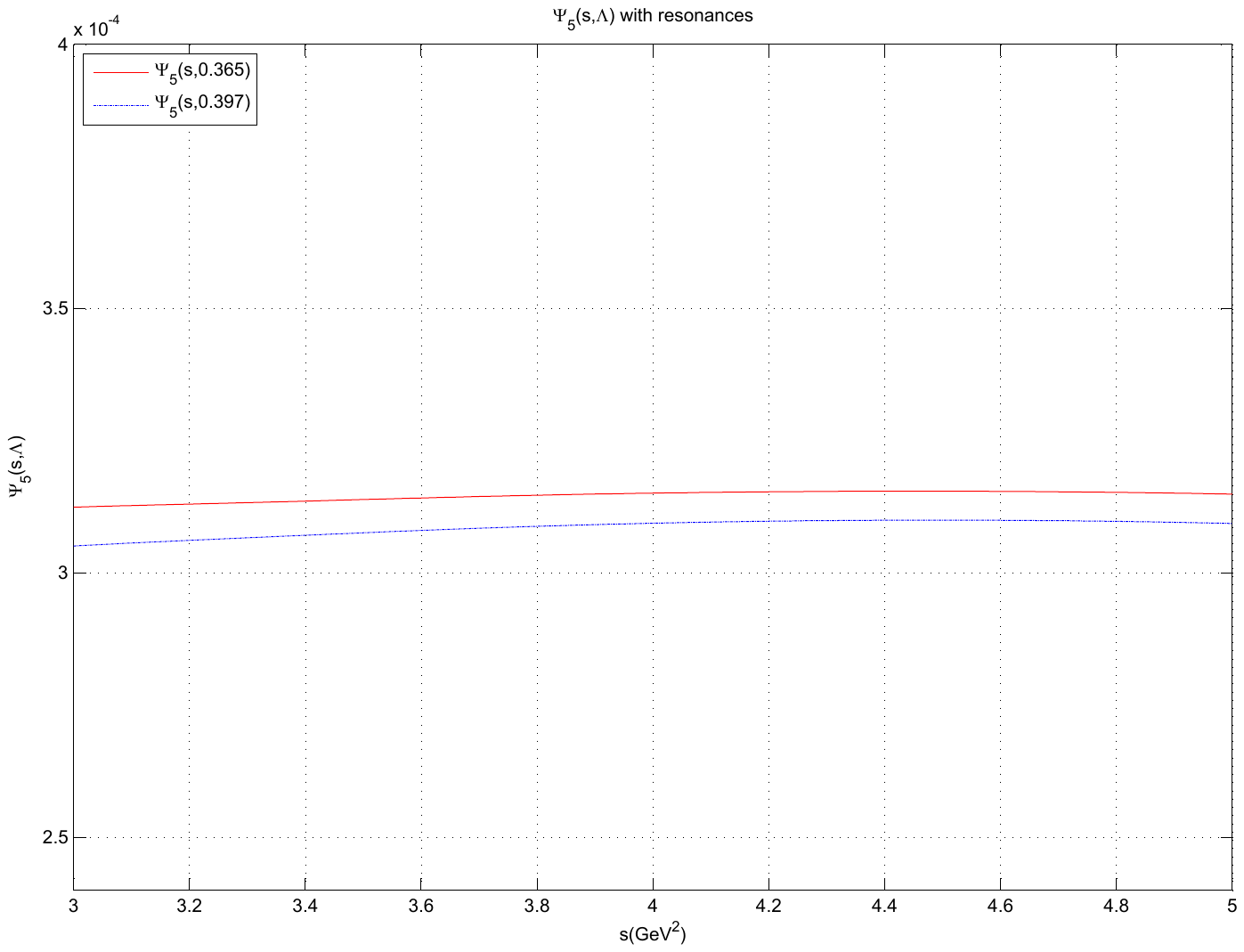}}
	}
	\caption{Results for $\psi_5 (0)$(GeV$^4$) as a function of the energy $s$ in FOPT. We have included the two resonances with the weighting factor $\kappa=0.5$. Starting from the top we have $\psi_5 (s,0.365)$ and $\psi_5 (s,0.397)$ respectively}
	\label{FOPTpsi2}
\end{figure}

\clearpage

\begin{figure}[h]
	\centerline{
	\mbox{\includegraphics[width=9in]{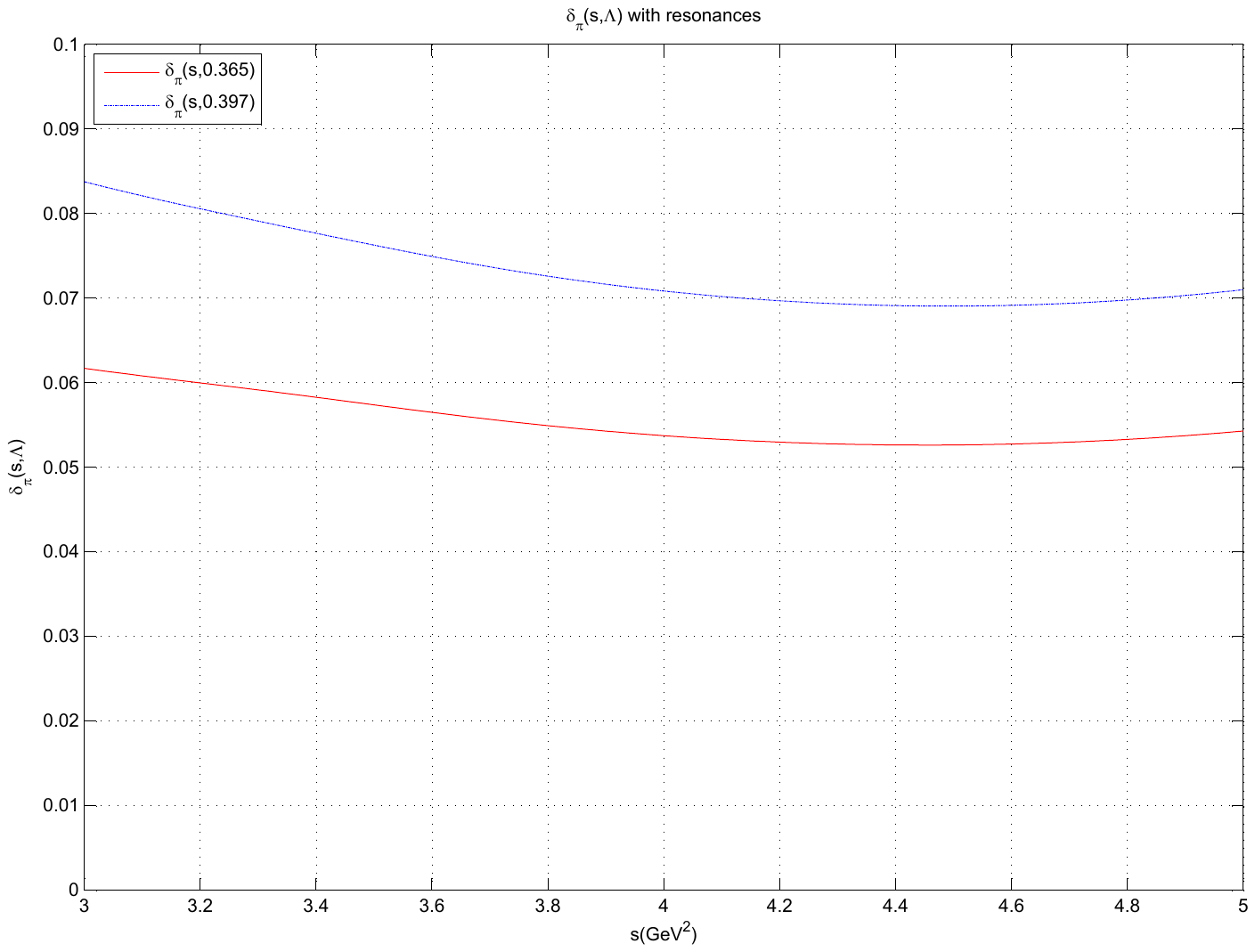}}
	}
	\caption{Results for $\delta_\pi $  as a function of the energy $s$ in FOPT. We have included the two resonances with the weighting factor $\kappa=0.5$. Starting from the top we have $\delta_\pi (s,0.397)$ and $\delta_\pi (s,0.365)$ respectively}
	\label{FOPTdeltapi2}
\end{figure}

\clearpage

\begin{figure}[h]
	\centerline{
	\mbox{\includegraphics[width=9in]{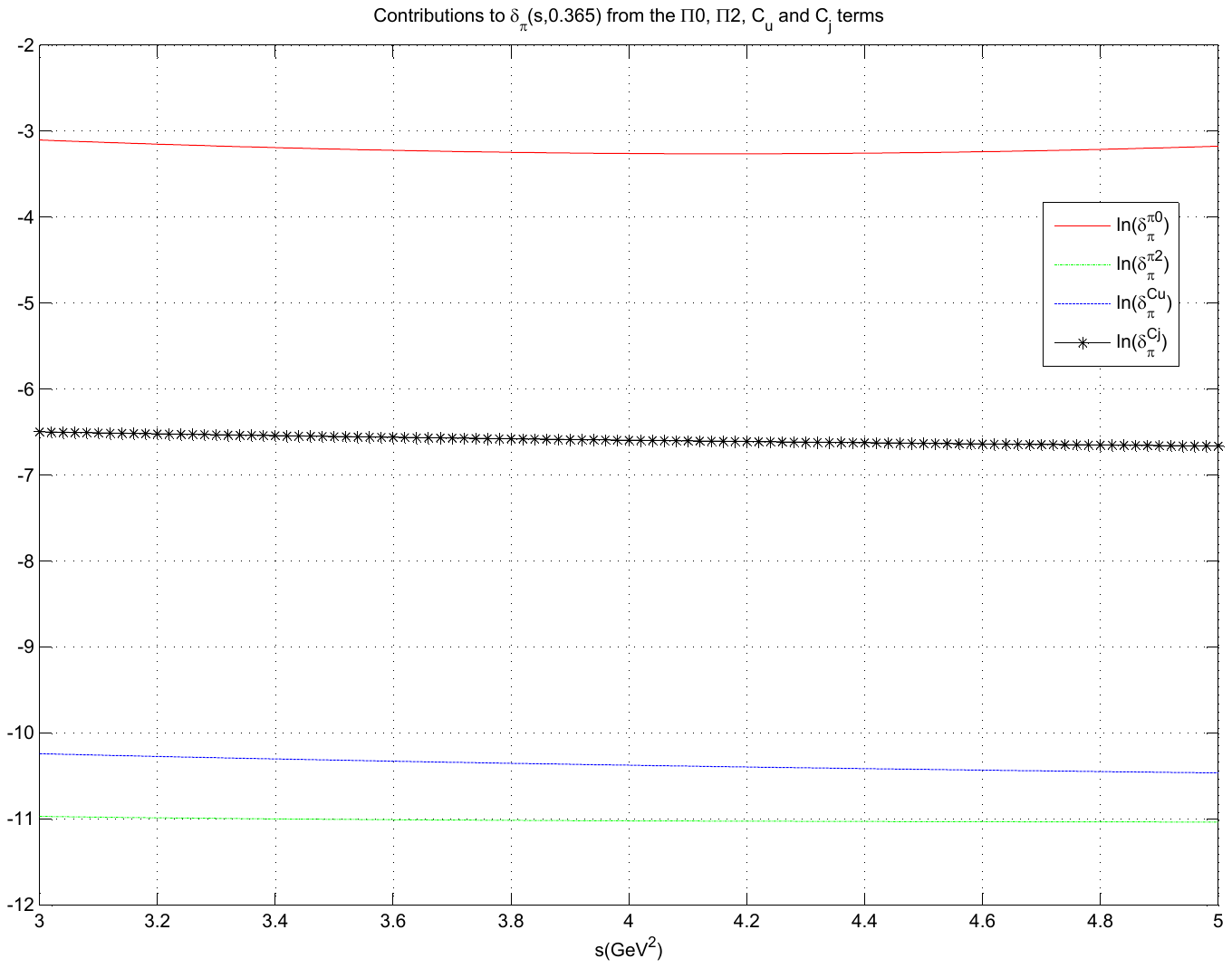}}
	}
	\caption{Plot of the various contributors to $\delta_\pi $ for $\Lambda_{QCD} = 0.365$ GeV. Starting from the top we have $\ln(\delta_{\pi} ^{\pi 0})$, $\ln(\delta_{\pi} ^{C_j})$, $\ln(\delta_{\pi} ^{C_u})$ and $\ln(\delta_{\pi} ^{\pi 2})$ respectively}
	\label{FOPTpioncontL}
\end{figure}

The various contributions are defined as
\begin{eqnarray}
  \psi _5  &=& 2f_\pi ^2 M_\pi ^2 \Delta \left( {M_\pi ^2 } \right) + \delta _{5PQCD}  + \delta _{5RES}  \nonumber \\
  \delta _\pi   &=& 1 - \frac{{\psi _5 }} {{2f_\pi ^2 M_\pi ^2 }} = 1 - \frac{{2f_\pi ^2 M_\pi ^2 \Delta \left( {M_\pi ^2 } \right) + \delta _{5PQCD}  + \delta _{5RES} }}
{{2f_\pi ^2 M_\pi ^2 }} \nonumber \\
   &=& \underbrace {1 - \frac{{2f_\pi ^2 M_\pi ^2 \Delta \left( {M_\pi ^2 } \right) + \delta _{5RES} }}
{{2f_\pi ^2 M_\pi ^2 }}}_{ = \delta _\pi ^0 } - \underbrace {\frac{{\left. {\delta _5 } \right|_{\Pi 0} }}
{{2f_\pi ^2 M_\pi ^2 }}}_{ = \delta _\pi ^{\Pi 0} } - \underbrace {\frac{{\left. {\delta _5 } \right|_{\Pi 2} }}
{{2f_\pi ^2 M_\pi ^2 }}}_{ = \delta _\pi ^{\Pi 2} } - \underbrace {\frac{{\left. {\delta _5 } \right|_{Cu} }}
{{2f_\pi ^2 M_\pi ^2 }}}_{ = \delta _\pi ^{Cu} } - \underbrace {\frac{{\left. {\delta _5 } \right|_{Cj} }}
{{2f_\pi ^2 M_\pi ^2 }}}_{ = \delta _\pi ^{Cj} } \nonumber \\
   &=& \delta _\pi ^0  - \delta _\pi ^{\Pi 0}  - \delta _\pi ^{\Pi 2}  - \delta _\pi ^{Cu}  - \delta _\pi ^{Cj} 
\end{eqnarray}

\clearpage
\begin{figure}[h]
	\centerline{
	\mbox{\includegraphics[width=7.0in]{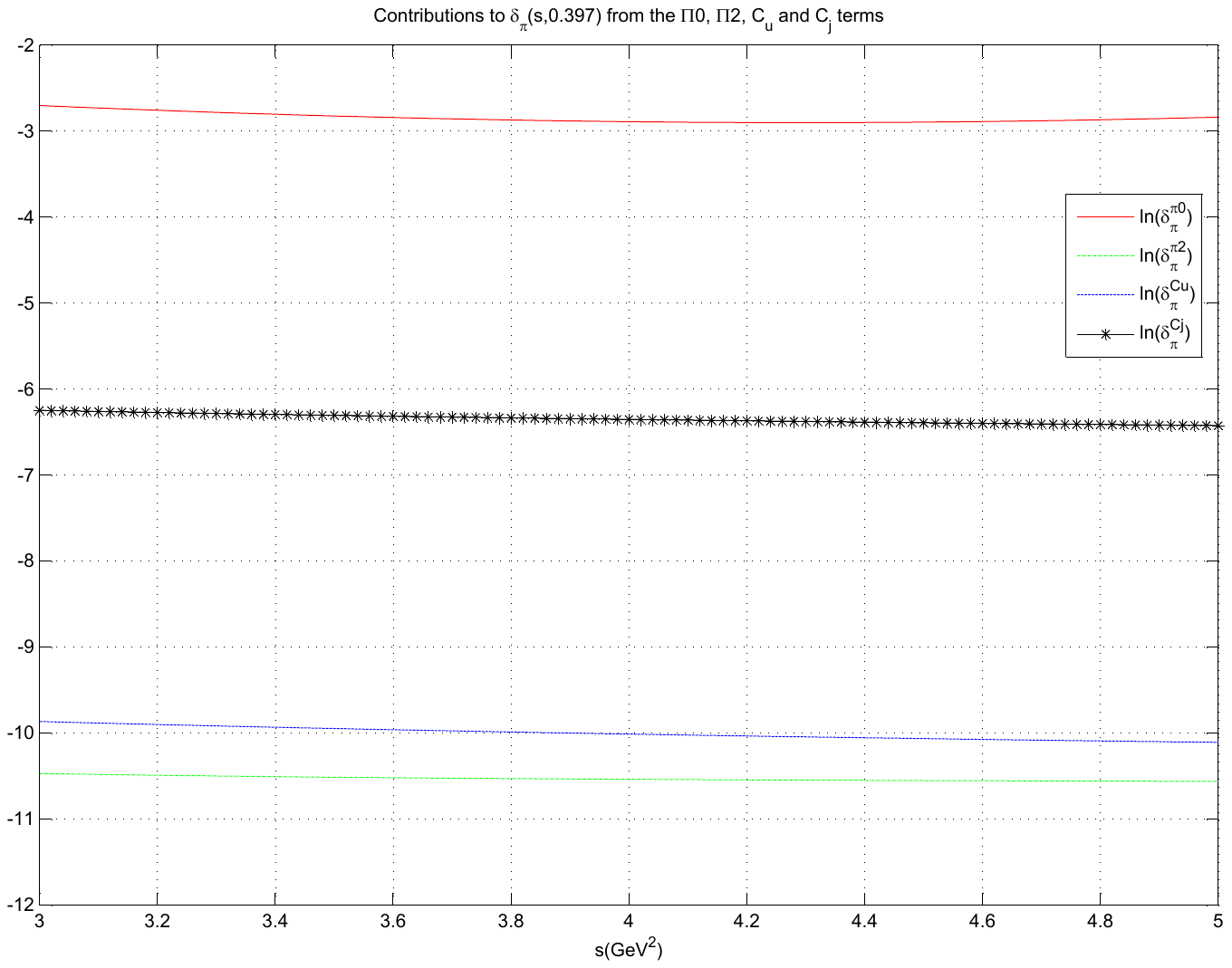}}
	}
	\caption{Plot of the various contributors to $\delta_\pi $ for $\Lambda_{QCD} = 0.397$ GeV. Starting from the top we have $\ln(\delta_{\pi} ^{\pi 0})$, $\ln(\delta_{\pi} ^{C_j})$, $\ln(\delta_{\pi} ^{C_u})$ and $\ln(\delta_{\pi} ^{\pi 2})$ respectively}
	\label{FOPTpioncontT}
\end{figure}

In the case of FOPT the results show good stability, looking at figures \ref{FOPTpsi1} - \ref{FOPTpsi2} in the range $3\le s \le5$ $\times 10^{-4}\rm{GeV}^2$ there is very little change in $\psi_5 (s)$ and $\delta_\pi (s)$. With the inclusion of the resonances the functions still display good stability. From the data used to generate the above plots, the result are tabulated below 
\begin{table}[hb]
	\centering
	\begin{tabular}{ccccc}
\hline
\hline
           &            &  Pion data &            &            \\
\hline
           & \multicolumn{ 2}{c}{No Resonances} & \multicolumn{ 2}{c}{With Resonances} \\
\hline
    $\Lambda_{QCD}$(GeV)       & $\psi_5 (0)(\rm{GeV}^4)\times 10^{-4}$ & $\delta_\pi$ & $\psi_5 (0)(\rm{GeV}^4)\times 10^{-4}$ & $\delta_\pi$ \\
\hline
     0.365 &  3.12-3.14 & 0.059-0.062 &  3.12-3.16 & 0.052-0.054 \\
     0.397 &  3.04-3.09 & 0.072-0.073 &  3.04-3.09 & 0.068-0.071 \\
\hline
\hline
\end{tabular}
\caption{Tabulated values extracted from the plots \ref{FOPTpsi1} - \ref{FOPTpsi2} which are used to estimate $\delta_\pi$.}
\label{piontab}
\end{table}
from values listed in table \ref{piontab}, we estimate the value of $\delta_\pi=0.064 \pm 0.008$.

\clearpage

\subsubsection{$SU(3)\otimes SU(3)$}

\begin{figure}[h]
	\centerline{
	\mbox{\includegraphics[width=9in]{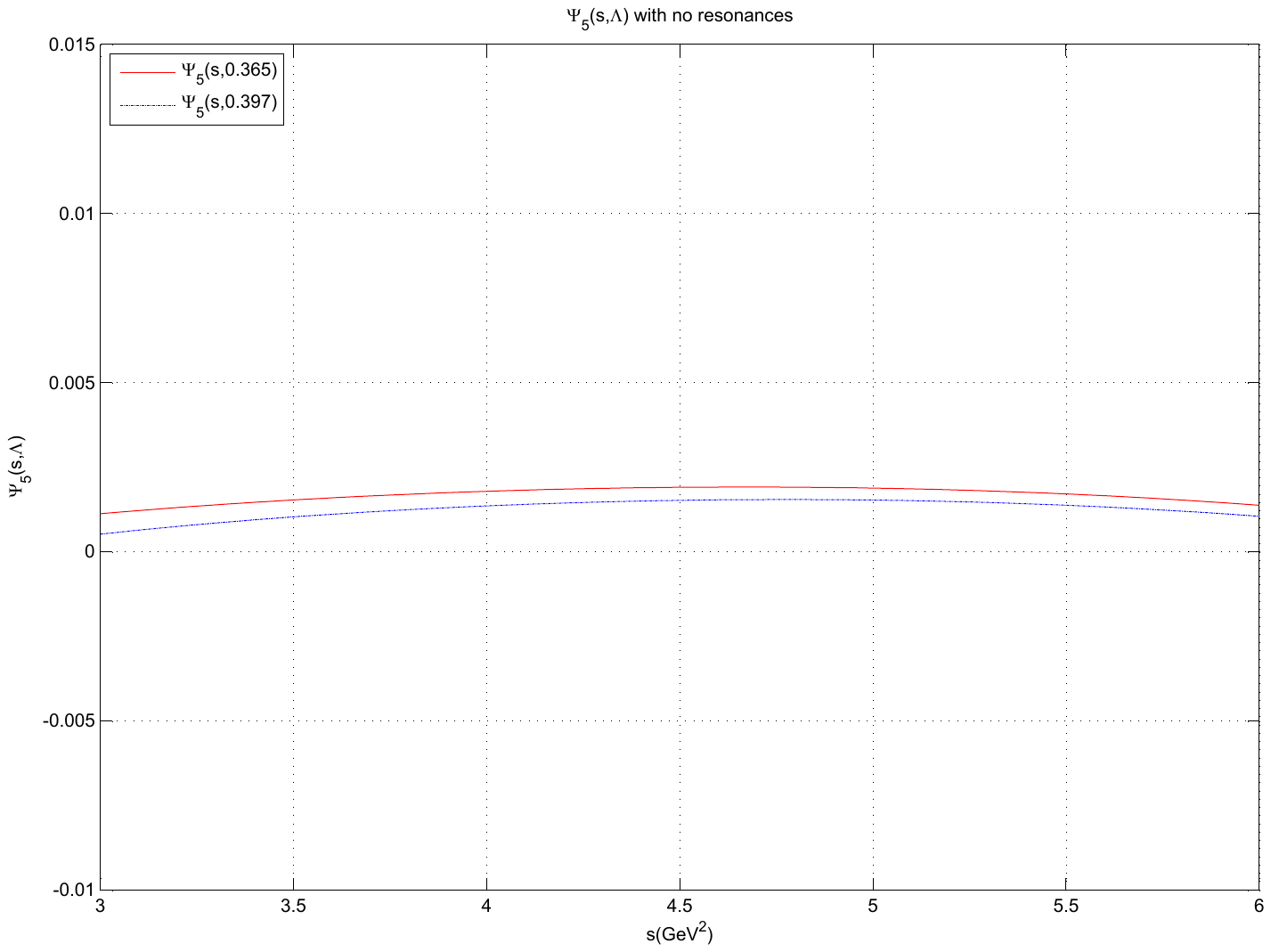}}
	}
	\caption{Results for $\psi_5 (0)$(GeV$^4$) as a function of the energy $s$ in FOPT. We have neglected the two resonances. Starting from the top we have $\psi_5 (s,0.365)$ and $\psi_5 (s,0.397)$ respectively}
	\label{FOPTpsi1K}
\end{figure}

\clearpage

\begin{figure}[h]
	\centerline{
	\mbox{\includegraphics[width=9in]{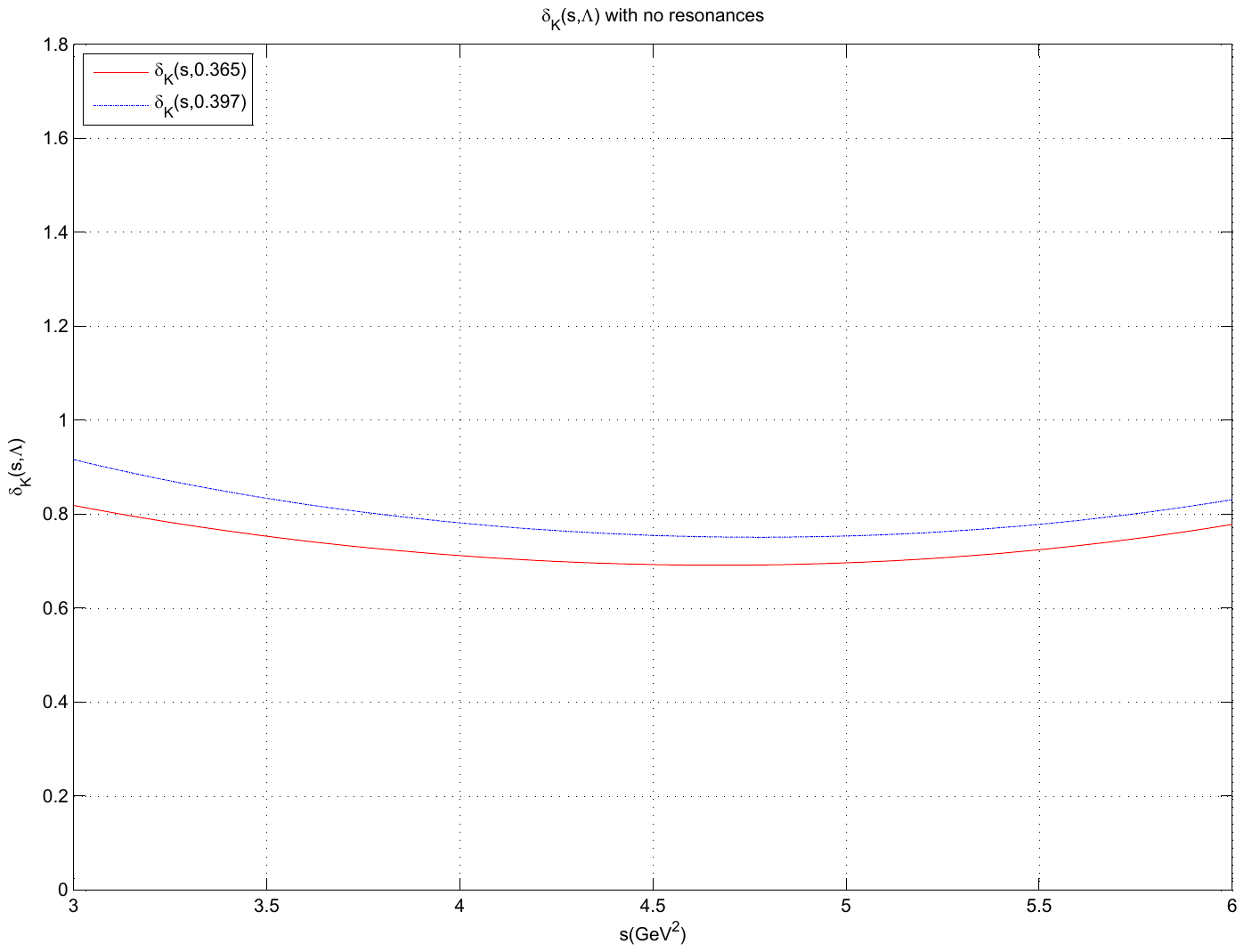}}
	}
	\caption{Results for $\delta_\pi $  as a function of the energy $s$ in FOPT. We have neglected the two resonances. Starting from the top we have $\delta_K (s,0.397)$ and $\delta_K (s,0.365)$ respectively}
	\label{FOPTdeltapi1K}
\end{figure}

\clearpage

\begin{figure}[h]
	\centerline{
	\mbox{\includegraphics[width=9in]{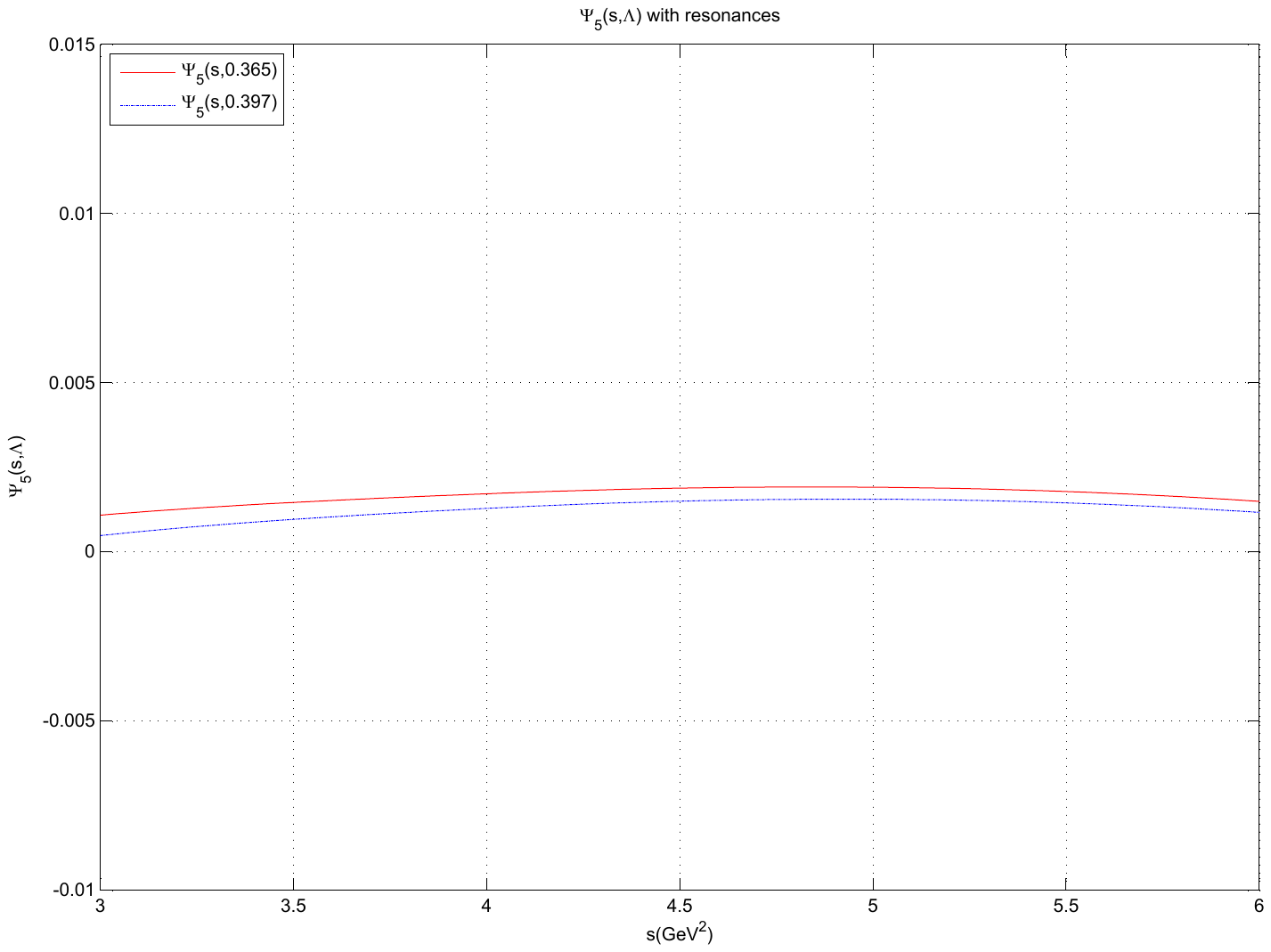}}
	}
	\caption{Results for $\psi_5 (0)$(GeV$^4$) as a function of the energy $s$ in FOPT. We have included the two resonances with the weighting factor $\kappa=0.5$. Starting from the top we have $\psi_5 (s,0.365)$ and $\psi_5 (s,0.397)$ respectively}
	\label{FOPTpsi2K}
\end{figure}

\clearpage

\begin{figure}[h]
	\centerline{
	\mbox{\includegraphics[width=9in]{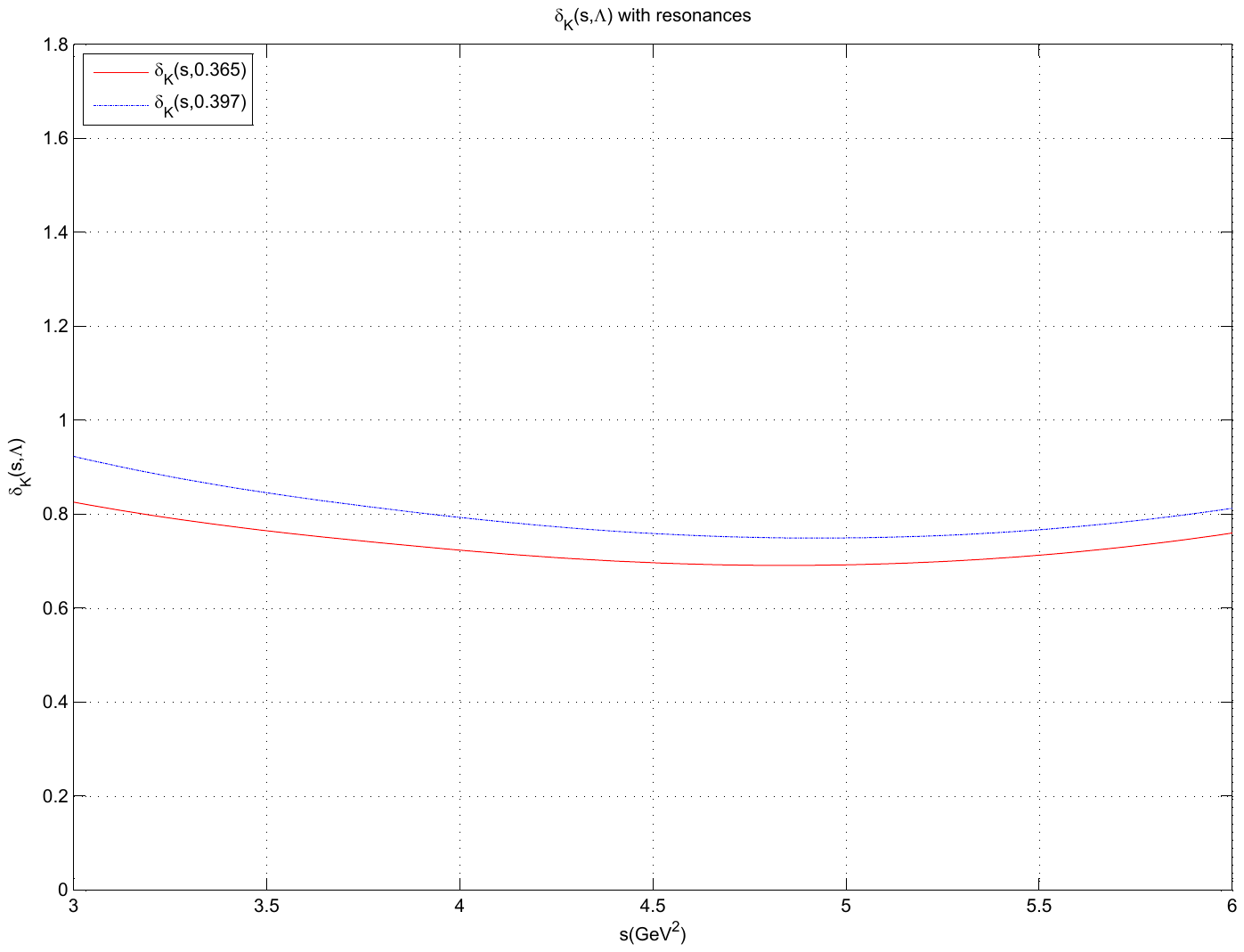}}
	}
	\caption{Results for $\delta_\pi $  as a function of the energy $s$ in FOPT. We have included the two resonances with the weighting factor $\kappa=0.5$. Starting from the top we have $\delta_K (s,0.397)$ and $\delta_K (s,0.365)$ respectively}
	\label{FOPTdeltapi2K}
\end{figure}

\clearpage

\begin{figure}[h]
	\centerline{
	\mbox{\includegraphics[width=9in]{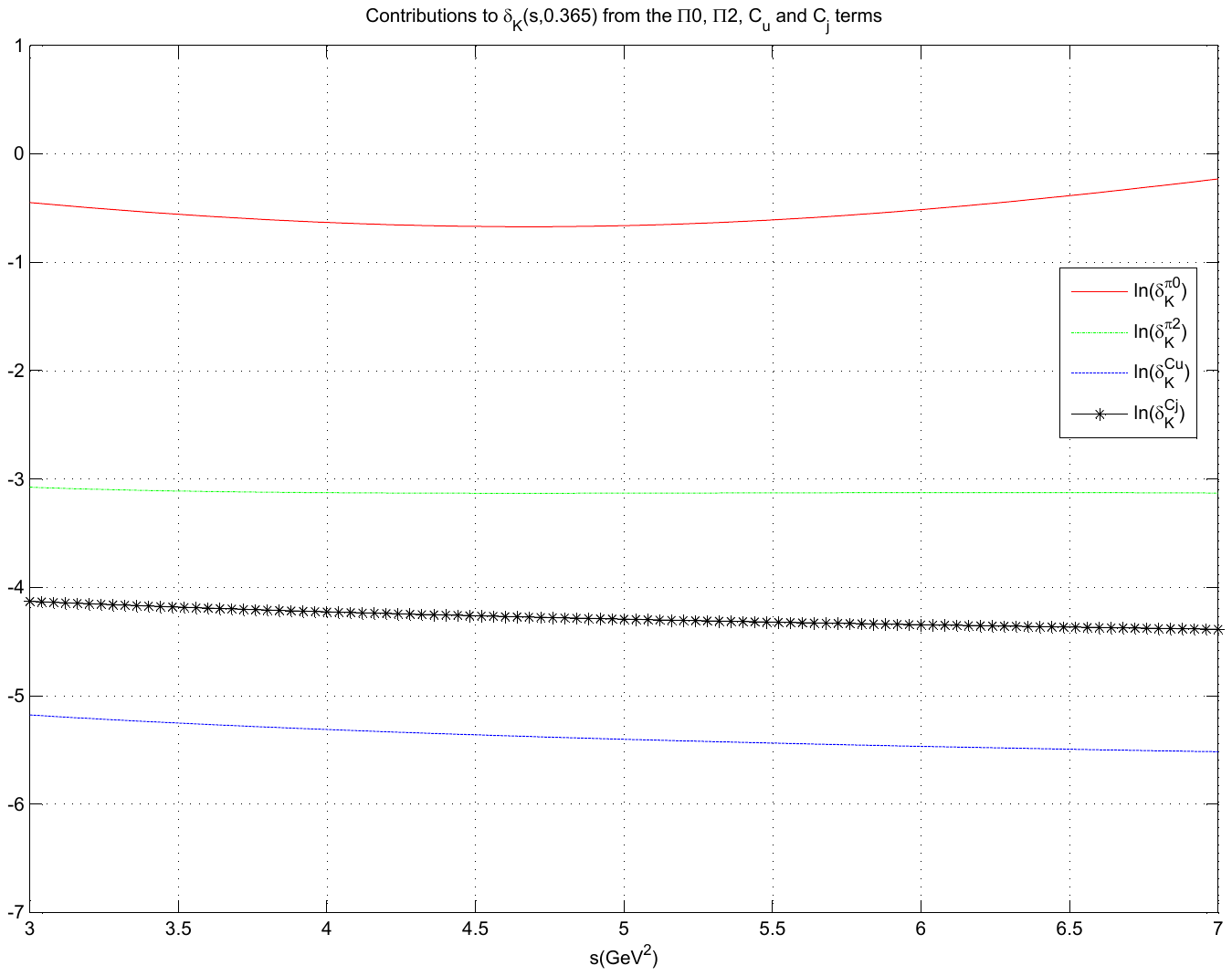}}
	}
	\caption{Plot of the various contributors to $\delta_K $ for $\Lambda_{QCD} = 0.365$ GeV. Starting from the top we have $\ln(\delta_{K} ^{\pi 0})$, $\ln(\delta_{K} ^{\pi 2})$, $\ln(\delta_{K} ^{C_j})$ and $\ln(\delta_{K} ^{C_u})$ respectively}
	\label{FOPTkaoncontL}
\end{figure}

\clearpage

\begin{figure}[h]
	\centerline{
	\mbox{\includegraphics[width=7.0in]{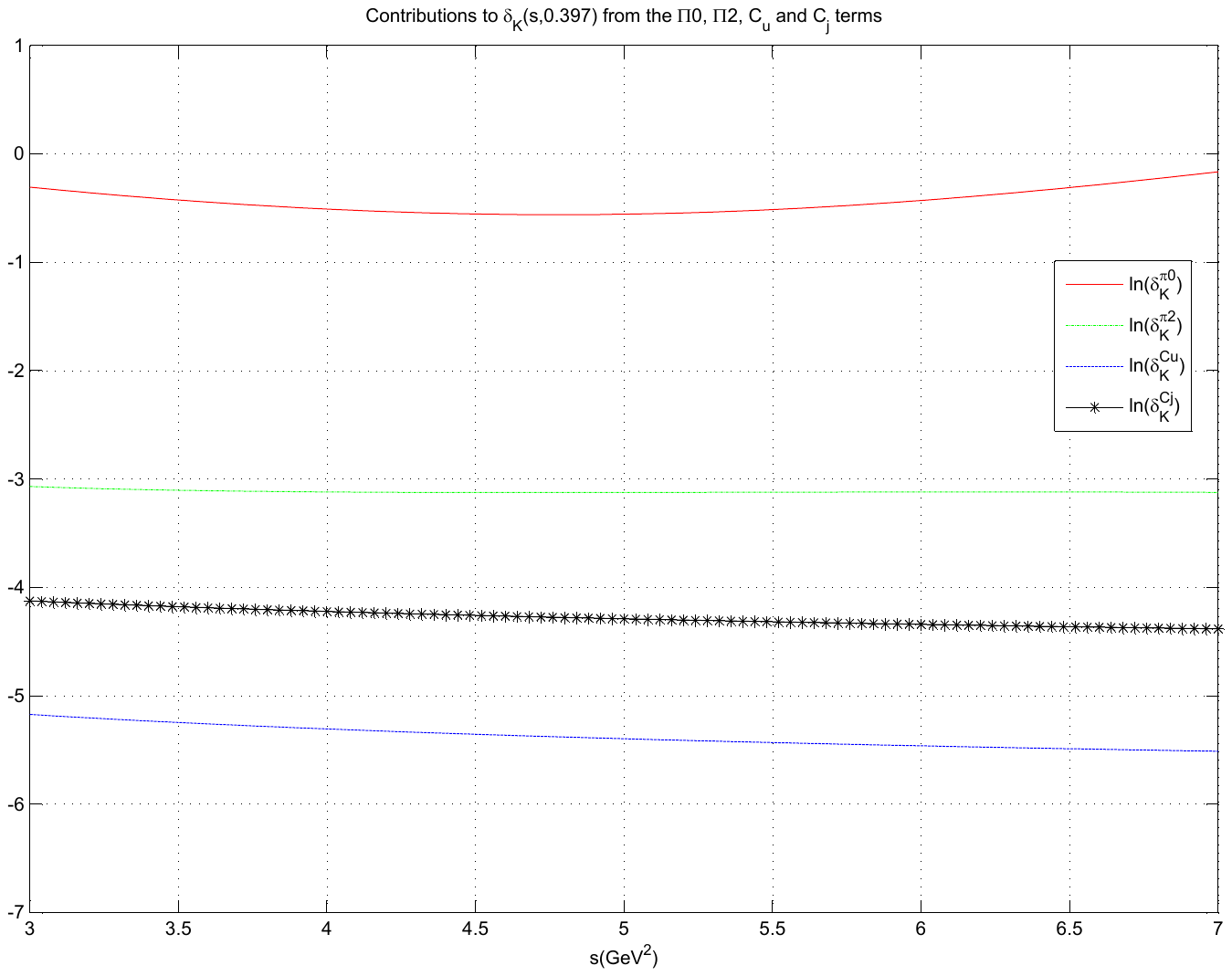}}
	}
	\caption{Plot of the various contributors to $\delta_K $ for $\Lambda_{QCD} = 0.397$ GeV. Starting from the top we have $\ln(\delta_{K} ^{\pi 0})$, $\ln(\delta_{K} ^{\pi 2})$, $\ln(\delta_{K} ^{C_j})$ and $\ln(\delta_{K} ^{C_u})$ respectively}
	\label{FOPTkaoncontT}
\end{figure}

Looking at figures \ref{FOPTpsi1K} - \ref{FOPTdeltapi2K} in the range $4\le s \le5$ $\times 10^{-4}\rm{GeV}^2$ there is very little change in $\psi_5 (s)$ and $\delta_K (s)$. With the inclusion of the resonances the functions still display good stability. From the data used to generate the above plots, the result are tabulated below:

\begin{table}[hb]
	\centering
	\begin{tabular}{ccccc}
\hline
\hline
                     \multicolumn{ 5}{c}{Kaon data} \\
\hline
           & \multicolumn{ 2}{c}{No Resonances} & \multicolumn{ 2}{c}{With Resonances} \\
\hline
$\Lambda_{QCD}(\rm{GeV})$ & $\psi_5 (0)(\rm{GeV}^4)$ & $\delta_K$ & $\psi_5 (0)(\rm{GeV}^4)$ & $\delta_K$ \\
\hline
     0.365 & 0.0017-0.0019 & 0.6929-0.6963 & 0.0018-0.0019 & 0.6921-0.6959 \\
     0.397 & 0.0013-0.0015 & 0.7535-0.7542 & 0.0014-0.0015 & 0.7493-0.7578 \\
\hline
\hline
\end{tabular}
\caption{Tabulated values extracted from the plots \ref{FOPTpsi1K} - \ref{FOPTdeltapi2K} which are used to estimate $\delta_K$.}
\label{kaontab}
\end{table}

From values listed in table \ref{kaontab}, we estimate the value of $\delta_K=0.64 \pm 0.24$.
\clearpage
\subsection{Discussion of Results}
\label{sec:DiscussionOfResults}

We have estimated the values of the $\delta_\pi$ and $\delta_K$ as (taking averages and adding the errors in quadrature)
\begin{eqnarray}  
\delta _\pi   &=& 0.060 \pm 0.014 \label{prepi} \\
  \delta _K  &=& 0.64 \pm 0.24 
\end{eqnarray}

The estimated value of $\delta_\pi$ is larger than the value estimated in \cite{dominguez3} which is based on perturbative QCD up to next-to-leading order. We can use the value for $\delta_\pi$ to find the quark condensate by using the current quark mass of  $0.5\left( {m_u  + m_d } \right) = 4.1 \pm 0.2$ MeV and substituting into (\ref{breakgmor}) to get
\begin{eqnarray}    
\left\langle {0|\bar qq|0} \right\rangle  &=&  - \frac{{f_\pi ^2 M_\pi ^2 \left( {1 - \delta _\pi  } \right)}}{{2 \cdot 0.5\left( {m_u  + m_d } \right)}} \nonumber \\
   &=&  - \left( {266 \pm 5{\text{ MeV}}} \right)^3  
\end{eqnarray}

Comparing to the value calculated in \cite{Jamin0} were
\begin{eqnarray}    
\left\langle {0|\bar qq|0} \right\rangle  =  - \left( {275 \pm 15{\text{ MeV}}} \right)^3  \nonumber
\end{eqnarray}

We see that the two values are in agreement within experimental uncertainty. $\delta_\pi$ is also related to the low energy constants of chiral perturbation theory by (\ref{chptcons})
\begin{eqnarray}
\delta _\pi   = 4\frac{{M_\pi ^2 }}
{{f_\pi ^2 }}\left( {2L_8^r  - H_2^r } \right)
\label{chptcons}
\end{eqnarray}

Using (\ref{chptcons}) and $L_8^r  = \left( {0.88 \pm 0.24} \right) \times 10^{ - 3} $ \cite{Jamin0} we obtain
\begin{eqnarray}  
H_2^r  &=& 2L_8^r  - \frac{{f_\pi ^2 }}{{4M_\pi ^2 }}\delta _\pi   \nonumber \\
   &=&  - \left( {4.9 \pm 1.8} \right) \times 10^{ - 3}  
   \label{calcH}
\end{eqnarray}

which is in agreement with the value quoted in \cite{Jamin0}. Alternative comparisons come from Lattice QCD \cite{lqcdpre} which predicts a value, using $L_8^r  = \left( {0.58 \pm 0.09} \right) \times 10^{ - 3}$ of
\begin{eqnarray} 
H_2^r  =  - \left( {5.5 \pm 2.0} \right) \times 10^{ - 3} 
\end{eqnarray}

Using $L_8^r  = \left( {0.62 \pm 0.20} \right) \times 10^{ - 3}$ \cite{Amoros} we calculate a value of $H_2^r  =  - \left( {5.4 \pm 1.2} \right) \times 10^{ - 3} $, which is in agreement with (\ref{calcH}).
The unphysical constant $H_2^r$ enters the next-to-leading order definition of the quark condensate and leads to some ambiguity in the definition of the condensate. 
\clearpage
\section{Conclusion}
\label{sec:Conclusion}

In this project we have made a direct estimate of $\delta_\pi$ and $\delta_K$ which measures how much the $SU(2)\otimes SU(2)$ and $SU(3)\otimes SU(3)$ symmetry groups are broken with the inclusion of chiral corrections to their respective correlators. We have made use of the methods of Fixed Order and Fixed Renormalization Scale Perturbation theory to calculate the symmetry breaking parameters $\delta_\pi$ and $\delta_K$. We also attempted to use the method of Contour Improved Perturbation theory, but this analysis leads to $\delta _5 \left( z \right)$ being dependent on both $\psi _5 \left( z \right)$ and $\psi^{'} _5 \left( z \right)$ which are not physical quantities as they depend on the renormalization and the regularization schemes used.


We have also performed our analysis using an integration kernel which is a complete second order polynomial which was constrained, with an appropriate choice of its coefficients, to vanish at the known two resonance points in the hadronic region. This leads to drastic reduction of systematic uncertainties that have plagued previous calculations of this nature. For $\delta_\pi$ and $\delta_K$ we predict values of
\begin{eqnarray}  
\delta _\pi   &=& 0.060 \pm 0.014  \nonumber \\
  \delta _K  &=& 0.64 \pm 0.24 \nonumber
\end{eqnarray}

and as a result of our new values for $\delta_\pi$ and $\delta_K$ we calculated the light quark condensate to be
\begin{eqnarray}    
\left\langle {0|\bar qq|0} \right\rangle  =  - \left( {266 \pm 5{\text{ MeV}}} \right)^3  \nonumber
\end{eqnarray}

The chiral perturbation theory low energy constant
\begin{eqnarray} 
H_2^r  =  - \left( {4.9 \pm 1.8} \right) \times 10^{ - 3} \nonumber
\end{eqnarray}

We have compared these predicted values of the symmetry breaking parameters, the quark condensate and the low energy constant to those found in the literature and the agreement is to within one standard deviation of the mean.
\section*{Acknowledgment}

I would like to thank my supervisor Professor Cesareo Dominguez who patiently guided and helped me during the past two years in completing this project. Prof. Dominguez has treated me kindly letting my mind run wild and only when I had strayed too far, bringing me back on the right track! I have matured as a theoretical physicist under his constant guidance. I would also like to thank our collaborators, J.M. Bordes and J.A. Pe$\widetilde{\text{n}}$arrocha from Spain, and Professor Karl Schilcher from Germany for all the discussions and help they had given me for this project. Furthermore, I would also like to thank Yingwen Zhang for helpful discussions on programming techniques and optimization of numerical routines and Sebastian Bodenstein for technical discussions about branch cuts and complex analysis in general and the Renormalization Group.

Finally, I would like to thank the National Institute of Theoretical Physics(NITheP) which provided financial support over the past two years for this project.
\appendix

\clearpage
\section{Miscellaneous Material for the Gell-Mann-Oakes-Renner Relation}
\label{sec:MiscellaneousMaterialForTheGellMannOakesRennerRelation}
\numberwithin{equation}{subsection}

\subsection{A Step Towards Covariance}
\label{AStepTowardsCovariance}

The Heaviside/Step function is represented piecewise as the following function
\begin{eqnarray}
\Theta \left( t \right) = \left\{ {\begin{array}{*{20}c}
   1 & {,t \geq 0}  \\
   0 & {,t \le 0}  \\
 \end{array} } \right.
\end{eqnarray}
were $t \in \mathbb{R}$. In the complex plane the Heaviside function is represented more compactly as a contour integral with the contour $C=C'\cup \mathbb{R}$ where $C'$ is a positively oriented semicircle and $\mathbb{R}$ is the real axis from $+\infty$ to $-\infty$
\begin{eqnarray}
  \Theta \left( t \right) &=& \frac{1}{{2\pi i}}\mathop{\oint \mkern-0.8mu }\limits_C {\frac{{e^{-izt} }}{{z + i\varepsilon }}dz}  \nonumber \\
   &=& \frac{1}{{2\pi i}}\int\limits_\infty ^{ - \infty } {\frac{{e^{-izt} }}{{z + i\varepsilon }}dz}  +\frac{1}{{2\pi i}}\int\limits_{C'} {\frac{{e^{-izt} }}{{z + i\varepsilon }}dz} \nonumber\\
	 &=&  - \frac{1}{{2\pi i}}\int\limits_{ - \infty }^\infty  {\frac{{e^{-izt} }}{{z + i\varepsilon }}dz}  + \underline {\frac{1}{{2\pi i}}\int\limits_{C'} {\frac{{e^{-izt} }}{{z + i\varepsilon }}dz} } \label{jord} \\
	 &=&  - \frac{1}{{2\pi i}}\int\limits_{ - \infty }^\infty  {\frac{{e^{-izt} }}{{z + i\varepsilon }}dz} 
\end{eqnarray}

were $t \in \mathbb{R}$ and $z \in \mathbb{C}$ and we have used Cauchy's Theorem on the analytic function $e^{-izt}$ with a closed contour which is a semi-circle in the lower half plane sent to infinity. The $i\varepsilon$ is added to the denominator to prevent the denominator ever going to zero and the limit is taken at the end of $i\varepsilon \to 0$. This is not a non-rigorous procedure and we can perform the integral without the $i\varepsilon$, but in this case we would have to resort to the method of the Principle Values to extract the integral. Both methods lead to the same solution. The Principle Value is though, aesthetically more pleasing a method, quite difficult to use if there are repeated zeros. The second term which is underlined in (\ref{jord}) does not contribute to the integral due to Jordan's Lemma\cite{JordansLemma1}

Consider now
\begin{eqnarray}  
	e^{ - iE\left( {x_0  - y_0 } \right)} \Theta \left( {x_0  - y_0 } \right) &=&  - \frac{1}{{2\pi i}}\int\limits_{ - \infty }^\infty  {\frac{{e^{ - iz\left( {x_0  - y_0 } \right)} e^{ 							- 			iE\left( {x_0  - y_0 } \right)} }}{{z + i\varepsilon }}dz}  \nonumber \\
   &=&  - \frac{1}{{2\pi i}}\int\limits_{ - \infty }^\infty  {\frac{{e^{ - i\left( {x_0  - y_0 } \right)\left( {z + E} \right)} }}{{z + i\varepsilon }}dz}  
\end{eqnarray}
if we make the following change of variables $q_0  = z + E$ we get
\begin{eqnarray} 
e^{ - iE\left( {x_0  - y_0 } \right)} \Theta \left( {x_0  - y_0 } \right) = \frac{1}
{{2\pi i}}\int\limits_{ - \infty }^\infty  {\frac{{e^{ - i\left( {x_0  - y_0 } \right)q_0 } }}
{{E - q_0  - i\varepsilon }}dq_0 } 
\label{x1}
\end{eqnarray}
we perform the same procedure on the remaining term of (\ref{X}) we obtain
\begin{eqnarray} 
e^{ - iE\left( {y_0  - x_0 } \right)} \Theta \left( {y_0  - x_0 } \right) = \frac{1}
{{2\pi i}}\int\limits_{ - \infty }^\infty  {\frac{{e^{ - i\left( {y_0  - x_0 } \right)q_0 } }}
{{E - q_0  - i\varepsilon }}dq_0 } 
\label{x22}
\end{eqnarray}
making a change of variables for (\ref{x22}) with $q_0  \to  - q_o$ which leads to 
\begin{eqnarray} 
e^{ - iE\left( {y_0  - x_0 } \right)} \Theta \left( {y_0  - x_0 } \right) = \frac{1}
{{2\pi i}}\int\limits_{ - \infty }^\infty  {\frac{{e^{ - i\left( {x_0  - y_0 } \right)q_0 } }}
{{E + q_0  - i\varepsilon }}dq_0 } 
\label{x2}
\end{eqnarray}
$X$ can now be evaluated by combining (\ref{x1}) and (\ref{x2})
\begin{eqnarray}   
X &=& \frac{1}{{2\pi i}}\int\limits_{ - \infty }^\infty  {\frac{{e^{ - i\left( {x_0  - y_0 } \right)q_0 } }}{{E - q_0  - i\varepsilon }}dq_0 }  + \frac{1}{{2\pi i}}\int\limits_{ - \infty }^\infty  {\frac{{e^{ - i\left( {x_0  - y_0 } \right)q_0 } }}{{E + q_0  - i\varepsilon }}dq_0 }  \nonumber \\
   &=& \frac{1}{{2\pi i}}\int\limits_{ - \infty }^\infty  {\frac{{e^{ - i\left( {x_0  - y_0 } \right)q_0 } }}{{E - q_0  - i\varepsilon }} + \frac{{e^{ - i\left( {x_0  - y_0 } \right)q_0 } }}
		{{E + q_0  - i\varepsilon }}dq_0 }  \nonumber \\
   &=& \frac{1}{{2\pi i}}\int\limits_{ - \infty }^\infty  {e^{ - i\left( {x_0  - y_0 } \right)q_0 } \left[ {\frac{1}{{E - q_0  - i\varepsilon }} + \frac{1}{{E + q_0  - i\varepsilon }}} \right]dq_0 }  \nonumber \\
   &=& \frac{{2E}}{{2\pi i}}\int\limits_{ - \infty }^\infty  {\frac{{e^{ - i\left( {x_0  - y_0 } \right)q_0 } }}{{E^2  - q_0 ^2  - i\varepsilon '}}dq_0 } 
\end{eqnarray}
we note that the denominator
$$
E^2  - q_0 ^2  - i\varepsilon ' = m_\pi ^2  + \underline q ^2  - q_0 ^2  - i\varepsilon ' = m_\pi ^2  - \left( { - \underline q ^2  + q_0 ^2 } \right) - i\varepsilon ' = m_\pi ^2  - q_\mu  q^\mu   - i\varepsilon ' = m_\pi ^2  - q^2  - i\varepsilon '$$
leading to 
\begin{eqnarray}   
X = \frac{{2E}}{{2\pi i}}\int\limits_{ - \infty }^\infty  {\frac{{e^{ - i\left( {x_0  - y_0 } \right)q_0 } }}{{m_\pi ^2  - q^2  - i\varepsilon '}}dq_0 } 
\label{finX}
\end{eqnarray}

\subsection{Time Ordered Products}
\label{TimeOrderedProducts}

\begin{eqnarray}
T\left[ {A_\mu  \left( x \right)A_\nu ^\dag  \left( 0 \right)} \right] = \left[ {A_\mu  \left( x \right);A_\nu ^\dag  \left( 0 \right)} \right]\Theta \left( {x_0 } \right) + A_\nu ^\dag  \left( 0 \right)A_\mu  \left( x \right)
\label{top3App}
\end{eqnarray}

we prove this as follows by starting with the right-hand-side(RHS)
\begin{eqnarray}
  \rm{RHS} &=& \left[ {A_\mu  \left( x \right);A_\nu ^\dag  \left( 0 \right)} \right]\Theta \left( {x_0 } \right) + A_\nu ^\dag  \left( 0 \right)A_\mu  \left( x \right) \nonumber \\
      &=& A_\mu  \left( x \right)A_\nu ^\dag  \left( 0 \right)\Theta \left( {x_0 } \right) - A_\nu ^\dag  \left( 0 \right)A_\mu  \left( x \right)\Theta \left( {x_0 } \right) + A_\nu ^\dag  \left( 0 						\right)A_\mu  \left( x \right) \nonumber \\
   		&=& A_\mu  \left( x \right)A_\nu ^\dag  \left( 0 \right)\Theta \left( {x_0 } \right) - A_\nu ^\dag  \left( 0 \right)A_\mu  \left( x \right)\left[ {\Theta \left( {x_0 } \right) - 1} \right]  
\label{rhs1}
\end{eqnarray}
but noting that 
$$
\Theta \left( {x_0 } \right) - 1 =  - \Theta \left( { - x_0 } \right)
$$
and substituting back into (\ref{rhs1}) which leads to 
\begin{eqnarray}  
\rm{RHS} &=& A_\mu  \left( x \right)A_\nu ^\dag  \left( 0 \right)\Theta \left( {x_0 } \right) + A_\nu ^\dag  \left( 0 \right)A_\mu  \left( x \right)\Theta \left( { - x_0 } \right) \nonumber \\
   &=& T\left[ {A_\mu  \left( x \right)A_\nu ^\dag  \left( 0 \right)} \right] \nonumber \\
   &=& \rm{LHS} \nonumber
\end{eqnarray}

\subsection{Preparatory Work I}
\label{PreparatoryWorkI}

We will now do some preparatory work on (\ref{pinu}) before proceeding.  In the Heisenberg representation operators can be propagated through spacetime via a product of unitary transformations. In our case
\begin{eqnarray}
  {\partial ^\mu }{A_\mu }\left( x \right) &=& {e^{ip \cdot x}}{\partial ^\mu }{A_\mu }\left( 0 \right){e^{ - ip \cdot x}} \label{heisen1a} \\
  A_\nu ^\dag \left( 0 \right) &=& {e^{ip \cdot x}}A_\nu ^\dag \left( { - x} \right){e^{ - ip \cdot x}} 
\label{heisen1b}
\end{eqnarray}

here we have taken the operators ${\partial ^\mu }{A_\mu }\left( 0 \right)$ and $A_\nu ^\dag \left( { - x} \right)$ and propagated them from the spacetime points $0$ and $-x$ to the spacetime points $x$ and $0$ respectively. Substituting (\ref{heisen1a}) and (\ref{heisen1b}) into the time-order product (\ref{pinu})
\begin{align}
  T\left[ {{\partial ^\mu }{A_\mu }\left( x \right)A_\nu ^\dag \left( 0 \right)} \right] =& {e^{ip \cdot x}}{\partial ^\mu }{A_\mu }\left( 0 \right){e^{ - ip \cdot x}}{e^{ip \cdot x}}A_\nu ^\dag \left( { - x} \right){e^{ - ip \cdot x}}\Theta \left( {{x_0}} \right) +\nonumber \\
  &+ {e^{ip \cdot x}}A_\nu ^\dag \left( { - x} \right){e^{ - ip \cdot x}}{e^{ip \cdot x}}{\partial ^\mu }{A_\mu }\left( 0 \right){e^{ - ip \cdot x}}\Theta \left( { - {x_0}} \right) \nonumber \\
&= {e^{ip \cdot x}}\left[ {{\partial ^\mu }{A_\mu }\left( 0 \right)A_\nu ^\dag \left( { - x} \right)\Theta \left( {{x_0}} \right) + A_\nu ^\dag \left( { - x} \right){\partial ^\mu }{A_\mu }\left( 0 \right)\Theta \left( { - {x_0}} \right)} \right]{e^{ - ip \cdot x}} 
\label{top4}	
\end{align}

substituting (\ref{top4}) into (\ref{pinu}) we obtain
\begin{eqnarray}
\Pi _{_\nu }^5 \equiv  - \int {{e^{iq\cdot x}}{e^{ip \cdot x}}\left\langle {0\left| {\left[ {{\partial ^\mu }{A_\mu }\left( 0 \right)A_\nu ^\dag \left( { - x} \right)\Theta \left( {{x_0}} \right) + A_\nu ^\dag \left( { - x} \right){\partial ^\mu }{A_\mu }\left( 0 \right)\Theta \left( { - {x_0}} \right)} \right]} \right|0} \right\rangle {e^{ - ip \cdot x}}{d^4}x} \label{pinu2}
\end{eqnarray}

we now make a change of variables with $x \mapsto x =  - y$ and $d^4 x = -d^4 y$ and then 
\begin{eqnarray}
  \Pi _{_\nu }^5 &\equiv& \int {{e^{ - iq\cdot y}}{e^{ - ip \cdot y}}\left\langle {0\left| {\left[ {{\partial ^\mu }{A_\mu }\left( 0 \right)A_\nu ^\dag \left( y \right)\Theta \left( { - {y_0}} \right) + A_\nu ^\dag \left( y \right){\partial ^\mu }{A_\mu }\left( 0 \right)\Theta \left( {{y_0}} \right)} \right]} \right|0} \right\rangle {e^{ip \cdot y}}{d^4}y}  \nonumber \\
   &=& \int {{e^{ - iq\cdot y}}\left\langle {0\left| {T\left[ {A_\nu ^\dag \left( y \right){\partial ^\mu }{A_\mu }\left( 0 \right)} \right]} \right|0} \right\rangle {d^4}y}  
\label{PiNu}
\end{eqnarray}

and now just renaming $y \mapsto y = x$ we obtain 
\begin{eqnarray}
\Pi _{_\nu }^5 = \int {{e^{ - iq\cdot x}}\left\langle {0\left| {T\left[ {A_\nu ^\dag \left( x \right){\partial ^\mu }{A_\mu }\left( 0 \right)} \right]} \right|0} \right\rangle {d^4}x}
\label{Pinu3A}
\end{eqnarray}

\subsection{Preparatory Work II}
\label{PreparatoryWorkII}

We first express the time-ordered product as
\begin{eqnarray}
T\left[ {\partial ^\nu  A_\nu ^\dag  \left( x \right)\partial ^\mu  A_\mu  \left( 0 \right)} \right] = \partial ^\nu  A_\nu ^\dag  \left( x \right)\partial ^\mu  A_\mu  \left( 0 \right)\Theta \left( {x_0 } \right) + \partial ^\mu  A_\mu  \left( 0 \right)\partial ^\nu  A_\nu ^\dag  \left( x \right)\Theta \left( { - x_0 } \right)
\label{top7}
\end{eqnarray}

and as previously stated operators expressed in the Heisenberg representation can be propagated through spacetime via a product of unitary transformations and in this case
\begin{eqnarray}  
\partial ^\nu  A_\nu ^\dag  \left( x \right) &=& e^{ip \cdot x} \partial ^\nu  A_\nu ^\dag  \left( 0 \right)e^{ - ip \cdot x}  \label{heisen2a} \\
  \partial ^\mu  A_\mu  \left( 0 \right) &=& e^{ip \cdot x} \partial ^\mu  A_\mu  \left( { - x} \right)e^{ - ip \cdot x}  
\label{heisen2b}
\end{eqnarray}

now substituting (\ref{heisen2a}) and (\ref{heisen2b}) into the time-ordered product (\ref{top7}) we obtain
\begin{eqnarray}
T\left[ {\partial ^\nu  A_\nu ^\dag  \left( x \right)\partial ^\mu  A_\mu  \left( 0 \right)} \right] = e^{ip \cdot x} \left[ {\partial ^\nu  A_\nu ^\dag  \left( 0 \right)\partial ^\mu  A_\mu  \left( { - x} \right)\Theta \left( {x_0 } \right) + \partial ^\mu  A_\mu  \left( { - x} \right)\partial ^\nu  A_\nu ^\dag  \left( 0 \right)\Theta \left( { - x_0 } \right)} \right]e^{ - ip \cdot x} 
\label{top8}
\end{eqnarray}

we are now ready to cast the second term of (\ref{qnupinu3})
\[
\begin{gathered}
   - i\int {e^{ - iq\cdot x} \left\langle {0\left| {T\left[ {\partial ^\nu  A_\nu ^\dag  \left( x \right)\partial ^\mu  A_\mu  \left( 0 \right)} \right]} \right|0} \right\rangle d^4 x}  \hfill \\
   =  - i\int {e^{ - iq\cdot x} e^{ip \cdot x} \left\langle {0\left| {\left[ {\partial ^\nu  A_\nu ^\dag  \left( 0 \right)\partial ^\mu  A_\mu  \left( { - x} \right)\Theta \left( {x_0 } \right) + \partial ^\mu  A_\mu  \left( { - x} \right)\partial ^\nu  A_\nu ^\dag  \left( 0 \right)\Theta \left( { - x_0 } \right)} \right]} \right|0} \right\rangle e^{ - ip \cdot x} d^4 x}  \hfill \\ 
\end{gathered} 
\]

making the change of variables $x \mapsto x =  - y$ so $d^4 x = -d^4 y$ leads to 
\[
\begin{gathered}
   - i\int {e^{ - iq\cdot x} \left\langle {0\left| {T\left[ {\partial ^\nu  A_\nu ^\dag  \left( x \right)\partial ^\mu  A_\mu  \left( 0 \right)} \right]} \right|0} \right\rangle d^4 x}  \hfill \\
   = i\int {e^{iq\cdot y} e^{ - ip \cdot y} \left\langle {0\left| {\left[ {\partial ^\nu  A_\nu ^\dag  \left( 0 \right)\partial ^\mu  A_\mu  \left( y \right)\Theta \left( { - y_0 } \right) + \partial ^\mu  A_\mu  \left( y \right)\partial ^\nu  A_\nu ^\dag  \left( 0 \right)\Theta \left( {y_0 } \right)} \right]} \right|0} \right\rangle e^{ip \cdot y} d^4 y}  \hfill \\ 
\end{gathered} 
\]

but we now have a new time-ordered product with
\begin{eqnarray}
\partial ^\nu  A_\nu ^\dag  \left( 0 \right)\partial ^\mu  A_\mu  \left( y \right)\Theta \left( { - y_0 } \right) + \partial ^\mu  A_\mu  \left( y \right)\partial ^\nu  A_\nu ^\dag  \left( 0 \right)\Theta \left( {y_0 } \right) = T\left[ {\partial ^\mu  A_\mu  \left( y \right)\partial ^\nu  A_\nu ^\dag  \left( 0 \right)} \right]
\label{top9}
\end{eqnarray}

and
\begin{eqnarray}
 - i\int {e^{ - iq\cdot x} \left\langle {0\left| {T\left[ {\partial ^\nu  A_\nu ^\dag  \left( x \right)\partial ^\mu  A_\mu  \left( 0 \right)} \right]} \right|0} \right\rangle d^4 x}  = i\int {e^{iq\cdot y} \left\langle {0\left| {T\left[ {\partial ^\mu  A_\mu  \left( y \right)\partial ^\nu  A_\nu ^\dag  \left( 0 \right)} \right]} \right|0} \right\rangle d^4 y} 
\label{qcd1}
\end{eqnarray}

we now make a final change of variables of $y \mapsto y = x$ and substituting into (\ref{qcd1})and arrive at
\begin{eqnarray}
 - i\int {e^{ - iq\cdot x} \left\langle {0\left| {T\left[ {\partial ^\nu  A_\nu ^\dag  \left( x \right)\partial ^\mu  A_\mu  \left( 0 \right)} \right]} \right|0} \right\rangle d^4 x}  = i\int {e^{iq\cdot x} \left\langle {0\left| {T\left[ {\partial ^\mu  A_\mu  \left( x \right)\partial ^\nu  A_\nu ^\dag  \left( 0 \right)} \right]} \right|0} \right\rangle d^4 x} 
\label{qcd2A}
\end{eqnarray}

\subsection{Operator Identity}
\label{OperatorIdentity}

We give a quick proof of (\ref{comidentity}) starting from the right-hand-side 
\begin{eqnarray}    
\rm{RHS} &=& A^\dag  \left\{ {B;C^\dag  } \right\}D - C^\dag  \left\{ {D;A^\dag  } \right\}B - \left\{ {A^\dag  ;C^\dag  } \right\}BD + C^\dag  A^\dag  \left\{ {B;D} \right\} \nonumber \\
   &=& A^\dag  \left( {BC^\dag   + C^\dag  B} \right)D - C^\dag  \left( {DA^\dag   + A^\dag  D} \right)B - \left( {A^\dag  C^\dag   + C^\dag  A^\dag  } \right)BD + C^\dag  A^\dag  \left( {BD + DB} \right) \nonumber \\
   &=& A^\dag  BC^\dag  D + A^\dag  C^\dag  BD - C^\dag  DA^\dag  B - C^\dag  A^\dag  DB - A^\dag  C^\dag  BD - C^\dag  A^\dag  BD + C^\dag  A^\dag  BD + C^\dag  A^\dag  DB \nonumber \\
   &=& A^\dag  BC^\dag  D - C^\dag  DA^\dag  B \nonumber \\
   &=& \left[ {A^\dag  B;C^\dag  D} \right] \nonumber \\
   &=& \rm{LHS} \nonumber
\end{eqnarray}
\clearpage
\section{Perturbative QCD Calculations}
\label{PerturbativeQCDCalculations}
\numberwithin{equation}{section}

In this section we will explicitly derive $\left. {\delta _5 \left( R \right)} \right|_{PQCD}$ from a clean slate. The pseudoscalar correlator(\ref{psi}) and the kernel(\ref{kernel}) are shown below
\begin{eqnarray}
\left. {\psi _5 \left( {q^2 } \right)} \right|_{PQCD}  = m_q^2 \left( {\left| q \right|^2 } \right)\left[ { - q^2 \Pi _0 \left( {q^2 } \right) + m_q \left( {\left| q \right|^2 } \right)\Pi _2 \left( {q^2 } \right) + \frac{{C_u }}
{{ - q^2 }}\left. {\left\langle {m_s \bar uu} \right\rangle } \right|_{\mu _0 }  + \sum\limits_{j = 1}^3 {\frac{{C_j }}
{{ - q^2 }}\left\langle {\hat O_j } \right\rangle }  +  \ldots } \right]
\label{psi}
\end{eqnarray}
\begin{eqnarray}
\Delta \left( {{q^2}} \right) = 1 - {a_0}{q^2} - {a_1}{q^4}
\label{kernel}
\end{eqnarray}

were $\Pi _0 \left( {q^2 } \right)$ is used depending on which loop order we want. The remaining terms in the correlator are the correction terms and are usually small relative to the terms in $\Pi _0 \left( {q^2 } \right)$. We attempt to calculate $\delta_5(q^2)$ which is
\begin{eqnarray}
{\delta _5}\left( {{q^2}} \right) = \frac{1}{{2\pi i}}\oint\limits_{} {\frac{1}{{{q^2}}}\Delta \left( {{q^2}} \right){\psi _5}\left( {{q^2}} \right)d{q^2}}
\label{delta5}
\end{eqnarray}

Since $q^2$ is a complex variable, we shall for convenience rename $q^2\rightarrow z$, articles published on the method of the QCD Sum Rules tend to use $s$ but we shall use the letter $z$ as it is traditionally a complex variable.
\subsection{$\Pi_0(z)$ Calculations}
\label{Pi0ZCALCULATIONS}
\numberwithin{equation}{subsection}

\subsubsection{One Loop Calculation}
\label{ONELOOPCALCULATION}
For the one loop
\begin{eqnarray}
 \Pi _0 \left( {q^2 } \right) &=& \frac{1}{{16\pi ^2 }}\left[ { - 12 + 6\ln \left( {\frac{{ - q^2 }}{{\mu ^2 }}} \right) + O\left( {\alpha _s } \right)} \right] 
\end{eqnarray}

\begin{eqnarray}
\psi _5 \left( {q^2 } \right) =  - \frac{{ m _q^2 \left( {\left| q \right|^2 } \right)}}{{16\pi ^2 }}q^2 \left[ { - 12 + 6\ln \left( {\frac{{ - q^2 }}{{\mu ^2 }}} \right)} \right]
\end{eqnarray}

Along the contour the radius is a constant, thus $|q|^2=|z|=R$, so we can take the $ m _q^2(|q|^2)$ term out of the integral. We now have to integrate the following
\begin{eqnarray}
\left. {\delta _5 } \right|_{1\text{loop}}  &=& \frac{{m _q^2 \left( R \right)}}{{16\pi ^2 }}\frac{1}{{2\pi i}}\oint\limits_{C(R)} {\left( {1 - a_0 z - a_1 z^2 } \right)\left( { - z} \right)\left[ { - 12 + 6\ln \left( {\frac{{ - z}}{{\mu ^2 }}} \right)} \right]\frac{1}{z}dz} \cr
 &=& \frac{{m _q^2 \left( R \right)}}{{16\pi ^2 }}\frac{1}{{2\pi i}}\oint\limits_{C(R)} {\left( {1 - a_0 z - a_1 z^2 } \right)\left[ {12 - 6\ln \left( {\frac{{ - z}}{{\mu ^2 }}} \right)} \right]dz} \cr \nonumber
\end{eqnarray}

making the change of variables $z=\mu^2 w$ and $dz=\mu^2 dw$
\begin{eqnarray}
\left. {\delta _5 } \right|_{1\text{loop}}  &=& \frac{{m _q^2 \left( R \right)}}{{16\pi ^2 }}\mu ^2 \frac{1}{{2\pi i}}\oint\limits_{C(r)} {\left( {1 - a_0 \mu ^2 w - a_1 \mu ^4 w^2 } \right)\left[ {12 - 6\ln \left( { - w} \right)} \right]dw} \cr
 &=& \frac{{m _q^2 \left( R \right)}}{{16\pi ^2 }}\left( { - 6\mu ^2 } \right)\frac{1}{{2\pi i}}\oint\limits_{C(r)} {\left( {1 - a_0 \mu ^2 w - a_1 \mu ^4 w^2 } \right)\ln \left( { - w} \right)dw} \cr \nonumber
\end{eqnarray}

Using the integrals generated by using (\ref{I}) we can integrate the above to
\begin{eqnarray}
\left. {\delta _5 } \right|_{1\text{loop}}  = \frac{{m _q^2 \left( R \right)}}{{16\pi ^2 }}\left( { - 6\mu ^2 } \right)\left[ {r - a_0 \mu ^2 \frac{1}{2}r^2  - a_1 \mu ^4 \frac{1}{3}r^3 } \right]
\end{eqnarray}

since the change of variables were $z=\mu^2 w$ $\Rightarrow$ $|z|=\mu^2 |w|$ thus $R=\mu^2 r$, substituting back,then setting $\mu^2 =R$ and simplifying we obtain
\begin{eqnarray}
\left. {\delta _5 } \right|_{1\text{loop}}  =  - \frac{{m _q^2 \left( R \right)}}{{16\pi ^2 }}\left[ {6R - 3a_0 R^2  - 2a_1 R^3 } \right]
\end{eqnarray}

\subsubsection{Two Loop Calculation}
\label{TWO LOOP CALCULATION}
\numberwithin{equation}{subsection}
For the two loop
\begin{eqnarray}
\Pi _0 \left( {q^2 } \right) = \frac{1}{{16\pi ^2 }}\left[ { \ldots  + \left( {\frac{{\alpha _s }}{\pi }} \right)\left( {24\zeta \left( 3 \right) - \frac{{131}}{2} + 34\ln \left( {\frac{{ - q^2 }}{{\mu ^2 }}} \right) - 6\ln ^2 \left( {\frac{{ - q^2 }}{{\mu ^2 }}} \right)} \right) + O\left( {\alpha _s^2 } \right)} \right]
\label{first}
\end{eqnarray}

now substituting (\ref{first}) into (\ref{psi}) we arrive at
\begin{eqnarray}
\left. {\delta _5 } \right|_{2\text{loop}}  =  - \frac{{m _q^2 \left( R \right)}}{{16\pi ^2 }}\left[ {\frac{{\alpha _s }}{\pi }} \right]\frac{1}{{2\pi i}}\oint\limits_{C(R)} {\left( {1 - a_0 z - a_1 z^2 } \right)\left[ {\left( {24\zeta \left( 3 \right) - \frac{{131}}{2}} \right) + 34\ln \left( {\frac{{ - z}}{{\mu ^2 }}} \right) - 6\ln ^2 \left( {\frac{{ - z}}{{\mu ^2 }}} \right)} \right]dz} 
\end{eqnarray}

using the same change of variables as shown in the one-loop calculations and simplify we obtain
\begin{align}
\left. {\delta _5 } \right|_{2\text{loop}} =  - \frac{{m _q^2 \left( R \right)}}{{16\pi ^2 }}\left[ {\frac{{\alpha _s }}{\pi }} \right]\Biggl\{ &34\mu ^2 \frac{1}{{2\pi i}}\oint\limits_{C(r)} {\left( {1 - a_0 \mu ^2 w - a_1 \mu ^4 w^2 } \right)\ln \left( { - w} \right)dw} +\nonumber\\
 &-  \frac{6\mu ^2}{{2\pi i}}\oint\limits_{C(r)} {\left( {1 - a_0 \mu ^2 w - a_1 \mu ^4 w^2 } \right)\ln ^2 \left( {-w} \right)dw}  \Biggr\}
\end{align}

Using the integrals generated by using (\ref{I}) we can integrate the above to
\begin{align}
\left. {\delta _5 } \right|_{2\text{loop}} = - \frac{{m _q^2 \left( R \right)}}{{16\pi ^2 }}\left[ {\frac{{\alpha _s }}{\pi }} \right] &\Biggl\{ 34\mu ^2 \left[ {r - a_0 \mu ^2 \frac{{r^2 }}{2} - a_1 \mu ^4 \frac{{r^3 }}{3}} \right] +\nonumber\\
&- 6\mu ^2 \left[ {2r\left( {\ln r - 1} \right) - a_0 \mu ^2 r^2 \left( {\ln r - \frac{1}{2}} \right) - a_1 \mu ^4 \frac{{r^3 }}{9}\left( {6\ln r - 2} \right)} \right] \Biggr\}
\end{align}

since the change of variables were $z=\mu^2 w$ $\Rightarrow$ $|z|=\mu^2 |w|$ thus $R=\mu^2 r$, substituting back,then setting $\mu^2 =R$ and simplifying we obtain
\begin{eqnarray}
\left. {\delta _5 } \right|_{2\text{loop}}  =  - \frac{{m _q^2 \left( R \right)}}{{16\pi ^2 }}\left[ {\frac{{\alpha _s }}{\pi }} \right]\left[ {46R - 20a_0 R^2  - \frac{{38}}{3}a_1 R^3 } \right]
\end{eqnarray}

\subsubsection{Three Loop Calculation}
\label{THREE LOOP CALCULATION}
\numberwithin{equation}{subsection}
For the three loop
\begin{eqnarray}
\Pi _0 \left( {q^2 } \right) = \frac{1}{{16\pi ^2 }}\left[ { \ldots  + \left( {\frac{{\alpha _s }}{\pi }} \right)^2 \left( {{\rm A_3}\ln \left( {\frac{{ - q^2 }}{{\mu ^2 }}} \right) + {\rm B_3}\ln ^2 \left( {\frac{{ - q^2 }}{{\mu ^2 }}} \right) + {\rm C_3}\ln ^3 \left( {\frac{{ - q^2 }}{{\mu ^2 }}} \right)} \right) + O\left( {\alpha _s^3 } \right)} \right]
\label{second}
\end{eqnarray}

were 
\begin{eqnarray} 
 {\rm A_3} &=& 4n_f \zeta \left( 3 \right) - \frac{{65}}{4}n_f  - 117\zeta \left( 3 \right) + \frac{{10801}}{{24}} + 34 \label{A} \\ 
 {\rm B_3} &=& \frac{{11}}{3}n_f  - 106 \label{B} \\ 
 {\rm C_3} &=& \frac{{19}}{2} - \frac{1}{3}n_f \label{C}
\end{eqnarray}

now substituting (\ref{second}) into (\ref{psi}) and making the change of variables we arrive at
\begin{align}
\left. {\delta _5 } \right|_{3\text{loop}} \left[ { - \frac{{ m_q^2 \left( R \right)}}
	{{16\pi ^2 }}\left[ {\frac{{\alpha _s }}
		{\pi }} \right]^2 } \right]^{ - 1} \!\!\!\!\!\! =&  \frac{{\text{A}_3}}
{{2\pi i}}\oint\limits_{C(r)} {\left( {1 - a_0 \mu ^2 w - a_1 \mu ^4 w^2 } \right)\ln \left( { - w} \right)dw} +\nonumber\\
&+ \frac{{\text{B}_3}}
{{2\pi i}}\oint\limits_{C(r)} {\left( {1 - a_0 \mu ^2 w - a_1 \mu ^4 w^2 } \right)\ln ^2 \left( { - w} \right)dw}\nonumber\\ 
&+ \frac{{\text{C}_3}}
{{2\pi i}}\oint\limits_{C(r)} {\left( {1 - a_0 \mu ^2 w - a_1 \mu ^4 w^2 } \right)\ln ^3 \left( { - w} \right)dw} 	
\end{align}

Using the integrals generated by using (\ref{I}) we can integrate the above to
\begin{align}
&\left. {\delta _5 } \right|_{3\text{loop}} \left[ { - \frac{{m_q^2 \left( R \right)}}
	{{16\pi ^2 }}\left[ {\frac{{\alpha _s }}
		{\pi }} \right]^2 } \right]^{ - 1}  \nonumber\\
 =& {\text{A}_3}\left[ {r - a_0 \mu ^2 \frac{{r^2 }}
	{2} - a_1 \mu ^4 \frac{{r^3 }}
	{3}} \right] + {\text{B}_3}\left[ {2r\left( {\ln r - 1} \right) - a_0 \mu ^2 r^2 \left( {\ln r - \frac{1}
		{2}} \right) - a_1 \mu ^4 \frac{{r^3 }}
	{9}\left( {6\ln r - 2} \right)} \right] \nonumber\\ 
+& {\text{C}_3}\left[ {r\left( {3\ln ^2 r - 6\ln r + 6 - \pi ^2 } \right) - a_0 \mu ^2 r^2 \left( {\frac{3}
		{2}\ln ^2 r - \frac{3}
		{2}\ln r + \frac{3}
		{4} - \frac{{\pi ^2 }}
		{2}} \right) - a_1 \mu ^4 r^3 \left( {\ln ^3 r - 2\ln r + \frac{2}
		{9} - \frac{{\pi ^2 }}
		{3}} \right)} \right]\nonumber\\	
\end{align}
since the change of variables were $z=\mu^2 w$ $\Rightarrow$ $|z|=\mu^2 |w|$ thus $R=\mu^2 r$, substituting back,then setting $\mu^2 =R$ and simplifying we obtain
\begin{align}
\left. {\delta _5 } \right|_{3\text{loop}} \left[ { - \frac{{m_q^2 \left( R \right)}}
	{{16\pi ^2 }}\left[ {\frac{{\alpha _s }}
		{\pi }} \right]^2 } \right]^{ - 1}  = \left[ {A_3 - 2B_3 + C_3 \left( {6 - \pi ^2 } \right)} \right]R + \frac{{a_0 }}
{2}\left[ {B_3 - A_3 - 2C_3 \left( {\frac{3}
		{4} - \frac{{\pi ^2 }}
		{2}} \right)} \right]R^2 +\nonumber\\ 
+ \frac{{a_1 }}
{3}\left[ {\frac{{2B_3}}
	{3} - A_3 - C_3 \left( {\frac{{2 - 3\pi ^2 }}
		{3}} \right)} \right]R^3 	
\end{align}

using the A$_3$, B$_3$ and C$_3$ defined above in (\ref{A}),(\ref{B}),(\ref{C}) we find the coefficients of $R$,$R^2$ and $R^3$ are
\begin{align}
A_3 - 2B_3 + C_3 \left( {6 - \pi ^2 } \right) &= \left( {4n_f  - 117} \right)\zeta \left( 3 \right) + \frac{{15889 - 566n_f }}{{24}} + \left( {2n_f  - 57} \right)\left( {\frac{{\pi ^2 }}{6} - 1} \right) \\ 
\frac{{a_0 }}{2}\left[ {B_3 - A_3 - 2C_3 \left( {\frac{3}{4} - \frac{{\pi ^2 }}{2}} \right)} \right] &= \frac{{a_0 }}{2}\left[ {\left( {117 - 4n_f } \right)\zeta \left( 3 \right) + \frac{{478n_f  - 13345}}{{24}} + \left( {57 - 2n_f } \right)\left( {\frac{{\pi ^2 }}{6} - \frac{1}{4}} \right)} \right] \\ 
\frac{{a_1 }}{3}\left[ {\frac{{2B_3}}{3} - A_3 - C_3 \left( {\frac{{2 - 3\pi ^2 }}{3}} \right)} \right] &= \frac{{a_1 }}{3}\left[ {\left( {117 - 4n_f } \right)\zeta \left( 3 \right) + \frac{{1346n_f  - 37491}}{{72}} + \left( {57 - 2n_f } \right)\left( {\frac{{\pi ^2 }}{6} - \frac{1}{9}} \right)} \right] 
\end{align}
\subsubsection{Four Loop Calculation}
\label{FOUR LOOP CALCULATION}
\numberwithin{equation}{subsection}
For the three loop
\begin{align}
\Pi \left( {q^2 } \right) = \frac{1}{{16\pi ^2 }}\left[ { \ldots  + \left( {\frac{{\alpha _s }}{\pi }} \right)^3 \left( {A\ln \left( {\frac{{ - q^2 }}{{\mu ^2 }}} \right) + B\ln ^2 \left( {\frac{{ - q^2 }}{{\mu ^2 }}} \right) + C\ln ^3 \left( {\frac{{ - q^2 }}{{\mu ^2 }}} \right) + D\ln ^4 \left( {\frac{{ - q^2 }}{{\mu ^2 }}} \right)} \right) + O\left( {\alpha _{_s }^4 } \right)} \right]
\label{fourth}	
\end{align}

were
\begin{eqnarray}
 A &=& A_1 \label{AA} \\ 
 B &=&  - 6\left[ {\frac{{4781}}{{18}} - \frac{{475}}{8}\zeta \left( 3 \right)} \right] \label{BB} \\ 
 C &=& 229 \label{CC} \\ 
 D &=&  - \frac{{221}}{{16}} \label{DD}
\end{eqnarray}

now substituting (\ref{fourth}) into (\ref{psi}) and making the change of variables we arrive at
\begin{align}
&\left. \delta  \right|_{4\text{loop}} \left[ { - \mu ^2 \frac{{m_q^2 \left( R \right)}}
	{{16\pi ^2 }}\left[ {\frac{{\alpha _s }}
		{\pi }} \right]^3 } \right]^{ - 1}  \nonumber\\
=& \frac{A}
{{2\pi i}}\oint\limits_{C(r)} {\left( {1 - a_0 \mu ^2 w - a_1 \mu ^4 w^2 } \right)\ln \left( { - w} \right)dw}  + \frac{B}
{{2\pi i}}\oint\limits_{C(r)} {\left( {1 - a_0 \mu ^2 w - a_1 \mu ^4 w^2 } \right)\ln ^2 \left( { - w} \right)dw}\nonumber\\ 
+& \frac{C}
{{2\pi i}}\oint\limits_{C(r)} {\left( {1 - a_0 \mu ^2 w - a_1 \mu ^4 w^2 } \right)\ln ^3 \left( { - w} \right)dw}  + \frac{D}
{{2\pi i}}\oint\limits_{C(r)} {\left( {1 - a_0 \mu ^2 w - a_1 \mu ^4 w^2 } \right)\ln ^4 \left( { - w} \right)dw} \nonumber	
\end{align}

Using the integrals generated by using (\ref{I}) we can integrate the above to
\begin{align}
&\left. \delta  \right|_{4\text{loop}} \left[ { - \mu ^2 \frac{{m_q^2 \left( R \right)}}
	{{16\pi ^2 }}\left[ {\frac{{\alpha _s }}
		{\pi }} \right]^3 } \right]^{ - 1}  \nonumber\\
=& A\left[ {r - a_0 \mu ^2 \frac{{r^2 }}
	{2} - a_1 \mu ^4 \frac{{r^3 }}
	{3}} \right] + B\left[ {2r\left( {\ln r - 1} \right) - a_0 \mu ^2 r^2 \left( {\ln r - \frac{1}
		{2}} \right) - a_1 \mu ^4 \frac{{r^3 }}
	{9}\left( {6\ln r - 2} \right)} \right] \nonumber\\ 
+& C\left[ {r\left( {3\ln ^2 r - 6\ln r + 6 - \pi ^2 } \right) - a_0 \mu ^2 r^2 \left( {\frac{3}
		{2}\left( {\ln ^2 r - \ln r} \right) + \frac{3}
		{4} - \frac{{\pi ^2 }}
		{2}} \right) - a_1 \mu ^4 r^3 \left( {\ln ^3 r - 2\ln r + \frac{2}
		{9} - \frac{{\pi ^2 }}
		{3}} \right)} \right] \nonumber\\ 
+& D\left[ {r\left( {4\ln ^3 r - 12\ln ^2 r + 4\left( {6 - \pi ^2 } \right)\ln r + 4\pi ^2  - 24} \right) - a_0 \mu ^2 r^2 \left( {2\ln ^3 r - 3\ln ^2 r + \left( {3 - 2\pi ^2 } \right)\ln r + \pi ^2  - \frac{3}
		{2}} \right)} \right] \nonumber\\ 
+& D\left[ { - a_1 \mu ^4 r^3 \left( {\frac{4}
		{3}\left( {\ln ^3 r - \ln ^2 r + \left( {\frac{2}
				{3} - \pi ^2 } \right)\ln r} \right) + \frac{4}
		{{3^2 }}\left( {\pi ^2  - \frac{2}
			{3}} \right)} \right)} \right] 
\end{align}

since the change of variables were $z=\mu^2 w$ $\Rightarrow$ $|z|=\mu^2 |w|$ thus $R=\mu^2 r$, substituting back,then setting $\mu^2 =R$ and simplifying we obtain
\begin{align}
&\left. \delta  \right|_{4\text{loop}} \left[ {\frac{{ - m_q^2 \left( R \right)}}
	{{16\pi ^2 }}\left[ {\frac{{\alpha _s }}
		{\pi }} \right]^3 } \right]^{ - 1}  \nonumber\\
=& \left[ {A - 2B + \left( {6 - \pi ^2 } \right)C + 4\left( {\pi ^2  - 6} \right)D} \right]R + \frac{{a_0 }}
{2}\left[ { - A + B - 2\left( {\frac{3}
		{4} - \frac{{\pi ^2 }}
		{2}} \right)C - 2\left( {\pi ^2  - \frac{3}
		{2}} \right)D} \right]R^2 \nonumber\\ 
+& \frac{{a_1 }}
{3}\left[ { - A - \frac{2}
	{3}B - \left( {\frac{2}
		{3} - \pi ^2 } \right)C - \frac{4}
	{3}\left( {\pi ^2  - \frac{2}
		{3}} \right)D} \right]R^3	
\end{align}

using the A, B, C and D defined above in (\ref{AA}-\ref{DD}) we find the coefficients of $R$,$R^2$ and $R^3$ are
\begin{eqnarray}
 A - 2B + \left( {6 - \pi ^2 } \right)C + 4\left( {\pi ^2  - 6} \right)D &=& {\rm A}_{\rm 1}  + 12\left[ {\frac{{4781}}{{18}} - \frac{{475}}{8}\zeta \left( 3 \right)} \right] - \frac{{3411}}{2}\left( {\frac{{\pi ^2 }}{6} - 1} \right) \\ 
  - A + B - 2\left( {\frac{3}{4} - \frac{{\pi ^2 }}{2}} \right)C - 2\left( {\pi ^2  - \frac{3}{2}} \right)D &=&  - {\rm A}_{\rm 1}  - 6\left[ {\frac{{4781}}{{18}} - \frac{{475}}{8}\zeta \left( 3 \right)} \right] + \frac{{6159}}{4}\left( {\frac{{\pi ^2 }}{6} - \frac{1}{4}} \right) \\ 
  - A - \frac{2}{3}B - \left( {\frac{2}{3} - \pi ^2 } \right)C - \frac{4}{3}\left( {\pi ^2  - \frac{2}{3}} \right)D &=&  - {\rm A}_{\rm 1}  + 4\left[ {\frac{{4781}}{{18}} - \frac{{475}}{8}\zeta \left( 3 \right)} \right] + \frac{{2969}}{2}\left( {\frac{{\pi ^2 }}{6} - \frac{1}{4}} \right) 
\end{eqnarray}

\subsubsection{Five Loop Calculation}
\label{FIVE LOOP CALCULATION}
\numberwithin{equation}{subsection}
For the three loop
\begin{eqnarray}
\Pi _{\rm 0} \left( {q^2 } \right) = \frac{1}{{16\pi ^2 }}\left[ { \ldots  + \left( {\frac{{\alpha _s }}{\pi }} \right)^4 \sum\limits_{i = 1}^5 {H_i L^i }  + O\left( {\alpha _s^5 } \right)} \right]
\label{fifth}
\end{eqnarray}

where
\[
L^i  = \ln ^i \left( {\frac{{ - q^2 }}{{\mu ^2 }}} \right)
\]

and $$H_i \in \mathbb{R} $$

now substituting (\ref{fifth}) into (\ref{psi}) and making the change of variables we arrive at
\begin{align}
&\left. {\delta _5 } \right|_{5\text{loop}} \left[ { - \mu ^2 \frac{{m_q^2 \left( R \right)}}
	{{16\pi ^2 }}\left[ {\frac{{\alpha _s }}
		{\pi }} \right]^4 } \right]^{ - 1}  \nonumber\\
=& \frac{{H_1 }}
{{2\pi i}}\oint\limits_{C(r)} {\left( {1 - a_0 \mu ^2 w - a_1 \mu ^4 w^2 } \right)\ln \left( { - w} \right)dw}  + \frac{{H_2 }}
{{2\pi i}}\oint\limits_{C(r)} {\left( {1 - a_0 \mu ^2 w - a_1 \mu ^4 w^2 } \right)\ln ^2 \left( { - w} \right)dw} \nonumber\\  
+& \frac{{H_3 }}
{{2\pi i}}\oint\limits_{C(r)} {\left( {1 - a_0 \mu ^2 w - a_1 \mu ^4 w^2 } \right)\ln ^3 \left( { - w} \right)dw}  + \frac{{H_4 }}
{{2\pi i}}\oint\limits_{C(r)} {\left( {1 - a_0 \mu ^2 w - a_1 \mu ^4 w^2 } \right)\ln ^4 \left( { - w} \right)dw} \nonumber\\  
+& \frac{{H_5 }}
{{2\pi i}}\oint\limits_{C(r)} {\left( {1 - a_0 \mu ^2 w - a_1 \mu ^4 w^2 } \right)\ln ^5 \left( { - w} \right)dw} 	
\end{align}

Using the integrals generated by using (\ref{I}) we can integrate the above to
\begin{align}
&\left. {\left[ { - \frac{{m _q^2 \mu ^2 }}{{16\pi ^2 }}\left[ {\frac{{\alpha _s }}{\pi }} \right]^4 } \right]^{ - 1} \delta _5 } \right|_{5\text{loop}}  \nonumber\\
=& H_1 \left( {r - \frac{{a_0 \mu ^2 }}{2}r^2  - \frac{{a_1 \mu ^4 }}{4}r^3 } \right) + H_2 \left( {2r\left( {\ln r - 1} \right) - a_0 \mu ^2 r^2 \left( {\ln r - \frac{1}{2}} \right) - \frac{{a_1 \mu ^4 }}{9}r^3 \left( {6\ln r - 2} \right)} \right) + \nonumber \\ 
+& H_3 \left( {r\left( {3\ln ^2 r - 6\ln r + 6 - \pi ^2 } \right) - a_0 \mu ^2 r^2 \left( {\frac{3}{2}\ln ^2 r - \frac{3}{2}\ln r + \frac{3}{4} - \frac{{\pi ^2 }}{2}} \right)} \right)+\nonumber \\ 
-& a_1 \mu ^4 r^3 H_3 \left( {\ln ^3 r - 2\ln r + \frac{2}{9} - \frac{{\pi ^2 }}{3}} \right) + \nonumber \\ 
+& rH_4 \left( {4\ln ^3 r - 12\ln ^2 r + 4\ln r\left( {6 - \pi ^2 } \right) + 4\left( {\pi ^2  - 6} \right)} \right) \nonumber \\ 
-& a_0 \mu ^2 r^2 H_4 \left( {2\ln ^3 r - 3\ln ^2 r + \ln r\left( {3 - 2\pi ^2 } \right) + \pi ^2  - \frac{3}{2}} \right) + \nonumber \\ 
-& a_1 \mu ^4 r^3 H_4 \left( {\frac{4}{3}\left( {\ln ^3 r - \ln ^2 r + \ln r\left( {\frac{2}{3} - \pi ^2 } \right)} \right) + \frac{4}{9}\left( {\pi ^2  - \frac{2}{3}} \right)}\right)+\nonumber \\ 
+& rH_5 \left( {5\ln ^4 r - 20\ln ^3 r + 10\left( {\ln r - 2} \right)\left( {6 - \pi ^2 } \right)\ln r + \pi ^4  - 20\pi ^2  + 120} \right) + \nonumber \\ 
-& a_0 \mu ^2 r^2 H_5 \left( {\frac{5}{2}\ln ^4 r - 5\ln ^3 r + 5\left( {\ln r - 1} \right)\left( {\frac{3}{2} - \pi ^2 } \right)\ln r + \frac{{\pi ^4 }}{2} - \frac{5}{2}\pi ^2  + \frac{{15}}{4}} \right) + \nonumber \\ 
-& a_1 \mu ^4 r^3 H_5 \left( {\frac{5}{3}\ln ^4 r - \frac{{20}}{9}\ln ^3 r + \frac{{10}}{9}\left( {\ln r - \frac{2}{3}} \right)\left( {\frac{6}{9} - \pi ^2 } \right)\ln r + \frac{{\pi ^4 }}{3} - \frac{{20}}{{27}}\pi ^2  + \frac{{40}}{{81}}} \right) 	
\end{align}

since the change of variables were $z=\mu^2 w$ $\Rightarrow$ $|z|=\mu^2 |w|$ thus $R=\mu^2 r$, substituting back,then setting $\mu^2 =R$ and simplifying we obtain
\begin{eqnarray}
\left. {\delta _5 } \right|_{5\text{loop}}  =  - \frac{{m _q^2 }}{{16\pi ^2 }}\left[ {\frac{{\alpha _s }}{\pi }} \right]^4 \left[ {\alpha R + \beta R^2  + \gamma R^3 } \right]
\end{eqnarray}

where
\begin{eqnarray}
 \alpha  &=& H_1  - 2H_2  + \left( {4H_4  - H_3 } \right)\left( {\pi ^2  - 6} \right) + \left( {\pi ^4  - 20\pi ^2  + 120} \right)H_5  \\ 
 \beta  &=&  - H_1  + H_2  + \left( {H_3  - 2H_4 } \right)\left( {\pi ^2  - \frac{3}{2}} \right) - \left( {\pi ^4  - 5\pi ^2  + \frac{{15}}{2}} \right)H_5  \\ 
 \gamma  &=&  - H_1  + \frac{2}{3}H_2  + \left( {H_3  - \frac{4}{3}H_4 } \right)\left( {\pi ^2  - \frac{2}{3}} \right) - \left( {\pi ^4  - \frac{{20}}{9}\pi ^2  + \frac{{40}}{{27}}} \right)H_5  
\end{eqnarray}
\subsection{$\Pi_2(z)$ Calculations}
\label{sec:Pi2ZCALCULATIONS}

\begin{eqnarray}
\Pi _2 \left( z \right) = \frac{1}
{{16\pi ^2 }}\left\{ { - 12 + 12\ln \left( {\frac{{ - z}}{{\mu ^2 }}} \right) + {\frac{\alpha _s \left( {\left| z \right|} \right) }{\pi }}\left[ { - 1000 + 64\ln \left( {\frac{{ - z}}{{\mu ^2 }}} \right) - 24\left[ {\ln \left( {\frac{{ - z}}{{\mu ^2 }}} \right)} \right]^2  + 48\zeta \left( 3 \right)} \right]} \right\}
\label{pi2}
\end{eqnarray}

Now substituting (\ref{pi2}) into the correlator (\ref{psi}) and focusing on this term in the correlator we get
\begin{eqnarray}
\psi _5 \left( z \right) = m_q^4 \left( {\left| z \right| } \right)\Pi _2 \left( z \right)
\label{pi2true}
\end{eqnarray}

we can now calculate the $\delta_5$ by using (\ref{delta5}). We will break the calculation up into to manageable pieces so 
\begin{align}
		\delta _5 &= \frac{1}{{2\pi i}}\oint\limits_{C\left( R \right)} {\frac{1}{z}\left( {1 - a_0 z - a_1 z^2 } \right)\frac{{m_q^4 \left( {\left| z \right|} \right)}}{{16\pi ^2 }}\left( { - 12 +12\ln 								\left( {\frac{{ - z}}{{\mu ^2 }}} \right)} \right)dz}  \nonumber \\   
&= \frac{1}{{2\pi i}}\oint\limits_{C\left( R \right)} {\frac{1}{z}\left( {1 - a_0 z - a_1 z^2 } \right)\frac{{ - 12m_q^4 \left( {\left| z \right|} \right)}}{{16\pi ^2 }}dz}  + \frac{1}
{{2\pi i}}\oint\limits_{C\left( R \right)} {\frac{1}{z}\left( {1 - a_0 z - a_1 z^2 } \right)\frac{{m_q^4 \left( {\left| z \right|^2 } \right)}}{{16\pi ^2 }}12\ln \left( {\frac{{ - z}}
		{{\mu ^2 }}} \right)dz}  \label{explain} \\
&= \frac{{ - 12m_q^4 \left( R \right)}}{{16\pi ^2 }}\frac{1}{{2\pi i}}\oint\limits_{C\left( R \right)} {\frac{1}{z}\left( {1 - a_0 z - a_1 z^2 } \right)dz} + \underline {\frac{{12m_q^4 									\left( R \right)}}{{16\pi ^2 }}\frac{1}{{2\pi i}}\oint\limits_{C\left( R \right)} {\frac{1}{z}\left( {1 - a_0 z - a_1 z^2 } \right)\ln \left( {\frac{{ - z}}{{\mu ^2 }}} \right)dz} } 
\label{pi2step}	
\end{align}

in (\ref{explain}) we are integrating on a circle of radius $R$ so the only change is in the angle on the loop. Since the mass only depends on the magnitude of the complex variable, we can take the mass term out of the integral since $|z|=R$. Making a change of variables with $z=\mu ^2 w$ $\Rightarrow$ $dz=\mu ^2 dw$ on the underlined term in (\ref{pi2step}) we obtain
\begin{align}
\delta _5  =& \frac{{ - 12m_q^4 \left( R \right)}}{{16\pi ^2 }}\frac{1}{{2\pi i}}\oint\limits_{C\left( R \right)} {\left( {\frac{1}{z} - a_0  - a_1 z} \right)dz} +\nonumber\\ 
&+ \frac{{12m_q^4 \left( R \right)}}
{{16\pi ^2 }}\frac{1}{{2\pi i}}\oint\limits_{C\left( r \right)} {\left( {\frac{1}{w}\ln \left( { - w} \right) - a_0 \mu ^2 \ln \left( { - w} \right) - a_1 \mu ^4 w\ln \left( { - w} \right)} \right)dw} \label{toint} 	
\end{align}

were $C(r)$ is a new contour of integration. Now using the integrals generated by (\ref{I}) we can integrate (\ref{toint}) to
\begin{eqnarray}  
	\delta _5  &=& \frac{{ - 12m_q^4 \left( R \right)}}{{16\pi ^2 }} + \frac{{12m_q^4 \left( R \right)}}{{16\pi ^2 }}\left[ {\ln r - a_0 \mu ^2 r - \frac{1}{2}a_1 \mu ^4 r^2 } \right] \nonumber \\
   &=&  - \frac{{12m_q^4 \left( R \right)}}{{16\pi ^2 }}\left[ {1 - \ln r + a_0 \mu ^2 r + \frac{1}{2}a_1 \mu ^4 r^2 } \right]
	\label{pi2step2}
\end{eqnarray}

since $|z|=\mu ^2 |w|$ $\Rightarrow$ $R=\mu ^2 r$ now substituting for $r$ into (\ref{pi2step2}) we obtain
\begin{eqnarray}  
\delta _5  =  - \frac{{12m_q^4 \left( R \right)}}{{16\pi ^2 }}\left[ {1 - \ln \left( {\frac{R}{{\mu ^2 }}} \right) + a_0 \mu ^2 \left( {\frac{R}{{\mu ^2 }}} \right) + \frac{1}{2}a_1 \mu ^4 \left( {\frac{R}{{\mu ^2 }}} \right)^2 } \right]
\end{eqnarray}

now taking the limit as $\mu ^2 \to R$ we obtain
\begin{eqnarray} 
\delta _5  =  - \frac{{12m_q^4 \left( R \right)}} {{16\pi ^2 }}\left[ {1 + a_0 R + \frac{1}{2}a_1 R^2 } \right]
\label{pi2part1}
\end{eqnarray}

We now turn to the second term of (\ref{pi2})

\begin{eqnarray} 
\Pi _2 \left( z \right) = \frac{1}{{16\pi ^2 }}\left\{ { \cdots  + \frac{{\alpha _s \left( {\left| z \right|} \right)}}{\pi }\left[ { - 1000 + 64\ln \left( {\frac{{ - z}}{{\mu ^2 }}} \right) - 24\left[ {\ln \left( {\frac{{ - z}}{{\mu ^2 }}} \right)} \right]^2  + 48\zeta \left( 3 \right)} \right]} \right\}
\label{pi2part2}
\end{eqnarray}

now substituting (\ref{pi2part2}) into the correlator and then into (\ref{delta5}) we obtain
\begin{align}
	\delta _5  &= \frac{1}{{2\pi i}}\oint\limits_{C\left( R \right)} {\frac{1}{z}\left( {1 -a_0 z -a_1 z^2 } \right)\frac{{m_q^4 \left( {\left| z \right|} \right)}}{{16\pi ^2 }}\frac{{\alpha _s \left( 								{\left| z \right|} \right)}}{\pi }\left[ {\left( {48\zeta \left( 3 \right) - 1000} \right) + 64\ln \left( {\frac{{ - z}}{{\mu ^2 }}} \right) - 24\left[ {\ln \left( {\frac{{ -z}}{{\mu 										^2 }}} \right)} \right]^2 } \right]dz}  \nonumber \\
&= \frac{{m_q^4 \left( R \right)}}{{16\pi ^2 }}\frac{{\alpha _s \left( R \right)}}{\pi }\left[ {48\zeta \left( 3 \right) - 1000} \right]\frac{1}{{2\pi i}}\oint\limits_{C\left( R\right)} 								{\frac{1}{z}\left( {1 - a_0 z - a_1 z^2 } \right)dz}  +\nonumber \\
&+ \underline{64\frac{{m_q^4 \left( R \right)}}{{16\pi ^2 }}\frac{{\alpha _s \left( R \right)}}{\pi }\frac{1}{{2\pi 																i}}\oint\limits_{C\left(R \right)} {\frac{1}{z}\left( {1 - a_0 z - a_1 z^2 } \right)\ln \left( {\frac{{ - z}}{{\mu ^2 }}} \right)dz} }  \nonumber\\
&- \underline{ 24\frac{{m_q^4 \left( R \right)}}{{16\pi ^2 }}\frac{{\alpha _s \left( R\right)}}{\pi }\frac{1}{{2\pi i}}\oint\limits_{C\left( R \right)} {\frac{1}{z}\left( {1 - a_0 z - 										a_1 z^2 }\right)\left[ {\ln \left( {\frac{{ - z}}{{\mu ^2 }}} \right)} \right]^2 dz} } 
\label{pi2part3}	
\end{align}

making a change of variables of $z=\mu^2 w$ for the underlined terms of (\ref{pi2part3}) and using the integrals of (\ref{I}) we obtain
\begin{eqnarray}   
	\delta _5  &=&  \frac{{m_q^4 \left( R \right)}}{{16\pi ^2 }}\frac{{\alpha _s \left( R \right)}}{\pi }\left[ {48\zeta \left( 3 \right) - 1000} \right] + 64\frac{{m_q^4 \left( R \right)}}{{16\pi ^2 										}}\frac{{\alpha _s \left( R \right)}}{\pi }\left[ {\ln r - a_0 \mu ^2 r - \frac{1}{2}a_1 \mu ^4 r^2 } \right] \nonumber \\ 
						&-& 24\frac{{m_q^4 \left( R \right)}}{{16\pi ^2 }}\frac{{\alpha _s \left(R \right)}}{\pi }\left[ {\ln ^2 r - \frac{{\pi ^2 }}{3} - 2a_0 \mu ^2 r\left( {\ln r - 1} \right) - a_1 \mu ^4 r^2 								\left( {\ln r - \frac{1}{2}} \right)} \right]
	\label{pi2part4}
\end{eqnarray}

substituting $R=\mu^2 r$ into (\ref{pi2part4}) for $r$
\begin{align}
	\delta _5  &= \frac{{m_q^4 \left( R \right)}}{{16\pi ^2 }}\frac{{\alpha _s \left( R \right)}}{\pi }\left[ {48\zeta \left( 3 \right) - 1000} \right] + 64\frac{{m_q^4 \left( R \right)}}{{16\pi ^2 											}}\frac{{\alpha _s \left( R \right)}}{\pi }\left[ {\ln \left( {\frac{R}{{\mu ^2 }}} \right) - a_0 \mu ^2 \frac{R}{{\mu ^2 }} - \frac{1}{2}a_1 \mu ^4 \left( {\frac{R}{{\mu ^2 }}} 											\right)^2 } \right] \nonumber \\
&- 24\frac{{m_q^4 \left( R \right)}}{{16\pi ^2 }}\frac{{\alpha _s \left( R \right)}}{\pi }\left[ {\ln ^2 \left( {\frac{R}{{\mu ^2 }}} \right) - \frac{{\pi ^2 }}
	{3} - 2a_0 \mu ^2 \frac{R}{{\mu ^2 }}\left( {\ln \left( {\frac{R}{{\mu ^2 }}} \right) - 1} \right) - a_1 \mu ^4 \left( {\frac{R}{{\mu ^2 }}} \right)^2 \left( {\ln \left( {\frac{R}
			{{\mu ^2 }}} \right) - \frac{1}{2}} \right)} \right]
\label{pi2part5}	
\end{align}

we now set $\mu^2 = R$ to get 
\begin{align*}
\delta _5  =& \frac{{m_q^4 \left( R \right)}}{{16\pi ^2 }}\frac{{\alpha _s \left( R \right)}}{\pi }\left[ {48\zeta \left( 3 \right) - 1000} \right] + 64\frac{{m_q^4 \left( 																			R \right)}}{{16\pi ^2	}}\frac{{\alpha _s \left( R \right)}}{\pi }\left[ { - a_0 R - \frac{1}{2}a_1 R^2 } \right] +\nonumber\\
-& 24\frac{{m_q^4 \left( R \right)}}{{16\pi ^2 																					}}\frac{{\alpha _s \left( R \right)}} {\pi}\left[ { - \frac{{\pi ^2 }}{3} + 2a_0 R + \frac{1}{2}a_1 R^2 } \right] \nonumber \\   
=& \frac{{m_q^4 \left( R \right)}}{{16\pi ^2 }}\frac{{\alpha _s \left( R \right)}}{\pi }\left\{ {\left[ {48\zeta \left( 3 \right) - 1000 + 8\pi ^2 } \right] - 																						112a_0 R - 44a_1 R^2 }\right\} 
\label{pi2part6}	
\end{align*}

now adding the two contributions up we get the total contribution from the $\Pi_2(z)$ term to the symmetry breaking measure $\delta_5$
\begin{eqnarray}   
\left. {\delta _5 } \right|_{\Pi _2 }  =  - \frac{{12m_q^4 \left( R \right)}} {{16\pi ^2 }}\left[ {1 + a_0 R + \frac{1}{2}a_1 R^2 } \right] + \frac{{m_s^4 \left( R \right)}}{{16\pi ^2 }}\frac{{\alpha _s \left( R \right)}}{\pi }\left\{ {\left[ {48\zeta \left( 3 \right) - 1000 + 8\pi ^2 } \right] - 112a_0 R - 44a_1 R^2 } \right\}
	\label{pi2part7}
\end{eqnarray}

\subsection{C$_u$ Calculation}
\label{sec:allinone}

\begin{eqnarray}
  \psi _5 \left( z \right) &=& m_q^2 \left( {\left| z \right|} \right)\left\{ { \cdots  + \frac{{C_u }}{{ - z}}m_q \left( {\left| z \right|} \right)\left\langle {\bar uu} \right\rangle _{\mu _0 }  +  																\cdots } \right\} \nonumber \\
   												 &=& m_q^3 \left( {\left| z \right|} \right)\frac{{C_u }}{{ - z}}\left\langle {\bar uu} \right\rangle _{\mu _0 } 
   												 \label{cu1}
\end{eqnarray}

were $\left\langle {\bar uu} \right\rangle _{\mu _0 }$ is actually $\left\langle {0|\bar uu|0} \right\rangle _{\mu_0}$ which is the quark vacuum expectation value dependent on the renormalization scale $\mu_0$ and $C_u$ is
\[
C_u  = 1 + \left\{ {\frac{{14}}
{3} - 2\ln \left( {\frac{{ - z}}
{{\mu ^2 }}} \right)} \right\}\frac{{\alpha _s \left( {\left| z \right|} \right)}}
{\pi }
\]

substituting $C_u$ into (\ref{cu1})
\begin{eqnarray}
\psi _5 \left( z \right) = - m_q^3 \left( {\left| z \right|} \right)\left\langle {\bar uu} \right\rangle _{\mu _0 } \frac{1}{z} - m_q^3 \left( {\left| z \right|} \right)\left\langle {\bar uu}																		\right\rangle _{\mu _0 } \frac{{\alpha _s \left( {\left| z \right|} \right)}}{\pi }\frac{1}{z}\left\{ {\frac{{14}}{3} - 2\ln \left( {\frac{{ - z}}{{\mu ^2 }}} \right)} 																\right\}
   												\label{cu2}
\end{eqnarray}

we now proceed in calculating $\delta_5$ using (\ref{delta5}) to get
\begin{align}
\delta _5  =& \frac{1}{{2\pi i}}\oint\limits_{C\left( R \right)} {\frac{1}{z}\left( {1 - a_0 z - a_1 z^2 } \right)\left[ { - m_q^3 \left( {\left| z \right|} \right)\left\langle {\bar uu} 													\right\rangle _{\mu _0 } \frac{1}{z} - m_q^3 \left( {\left| z \right|} \right)\left\langle {\bar uu} \right\rangle _{\mu _0 } \frac{{\alpha _s \left( {\left| z \right|} \right)}}{\pi 									}\frac{1}{z}\left\{ {\frac{{14}}{3} - 2\ln \left( {\frac{{ - z}}{{\mu ^2 }}} \right)} \right\}} \right]dz}  \nonumber \\   
=&  - m_q^3 \left( R \right)\left\langle {\bar uu} \right\rangle _{\mu _0 } \frac{1}{{2\pi i}}\oint\limits_{C\left( R \right)} {\left( {\frac{1}{{z^2 }} - a_0 \frac{1}{z} - a_1 } 											\right)dz} +\nonumber\\
-& m_q^3 \left( R \right)\left\langle {\bar uu} \right\rangle _{\mu _0 } \frac{{\alpha _s \left( R \right)}}{\pi }\underline {\frac{1}{{2\pi i}}\oint\limits_{C\left( R \right)} 						{\frac{1}{{z^2 }}\left( {1 - a_0 z - a_1 z^2 } \right)\left[ {\frac{{14}}{3} - 2\ln \left( {\frac{{ - z}}{{\mu ^2 }}} \right)} \right]dz} } \nonumber \\
\label{cu3}
\end{align}
for the underlined term in (\ref{cu3}) changing variables to $z=\mu^2 w$ $\Rightarrow$ $dz=\mu^2 dw$
\begin{eqnarray}  
\delta _5  &=& - m_q^3 \left( R \right)\left\langle {\bar uu} \right\rangle _{\mu _0}\frac{1}{{2\pi i}}\oint\limits_{C\left( R \right)} {\left( {\frac{1}{{z^2 }} - a_0 \frac{1}{z} - a_1 } \right)dz} 									\nonumber \\
  				 &-& m_q^3 \left( R \right)\left\langle {\bar uu} \right\rangle _{\mu _0 } \frac{{\alpha _s \left( R \right)}}{\pi }\frac{1}{{2\pi i}}\oint\limits_{C\left( r \right)} {\frac{1}
						{{\mu ^4 w^2 }}\left( {1 - a_0 \mu ^2 w - a_1 \mu ^4 w^2 } \right)\left[ {\frac{{14}}{3} - 2\ln \left( { - w} \right)} \right]\mu ^2 dw} 
					\label{cu4}
\end{eqnarray}

we can now integrate (\ref{cu4}) with the integrals generated by (\ref{I}) and using $R=\mu^2 r$ for $r$ we obtain
\begin{eqnarray}  
	\delta _5  = m_q^3 \left( R \right)\left\langle {\bar uu} \right\rangle _{\mu _0 } a_0  + \frac{{14}}{3}m_s^3 \left( R \right) \frac{{\alpha _s \left( R \right)}}{\pi 																								}\left\langle {\bar uu} \right\rangle _{\mu _0 } a_0  + 2m_q^3 \left( R\right)\frac{{\alpha _s \left( R \right)}}{\pi }\left\langle {\bar uu} \right\rangle _{\mu _0 } \left[ { - 											\frac{1}{R} - a_0 \ln \left( {\frac{R}{{\mu ^2 }}} \right) - a_1 R} \right]
					\label{cu5}
\end{eqnarray}

we set $\mu^2 =R$ to get
\begin{eqnarray}   
 \left. {\delta _5 } \right|_{C_u }  &=& m_q^3 \left( R \right)\left\langle {\bar uu} \right\rangle _{\mu _0 } a_0  + \frac{{14}}{3}m_q^3 \left( R \right)\frac{{\alpha _s \left( R \right)}}{\pi 																								}\left\langle {\bar uu} \right\rangle _{\mu _0 } a_0  + 2m_q^3 \left( R \right)\frac{{\alpha _s \left( R \right)}}{\pi }\left\langle {\bar uu} \right\rangle 																						_{\mu _0 } \left[ { - \frac{1}{R} - a_1 R} \right] \nonumber \\
 																	   &=& m_q^3 \left( R \right)\left\langle {\bar uu} \right\rangle _{\mu _0 } \left[ {a_0  + \left( {\frac{{14}}{3}a_0  - \frac{2}{R} - 2a_1 R} \right)\frac{{\alpha 																					_s \left( R \right)}}{\pi }} \right]
					\label{cu6}
\end{eqnarray}

\subsection{C$_j$ Calculations}
\label{sec:CJCalculations}

\begin{eqnarray} 
\psi _5 \left( z \right) = m_s^2 \left( {\left| z \right|} \right)\left\{ { \cdots  + \sum\limits_{j = 1}^3 {\frac{{C_j }}{{ - z}}\left\langle {\widehat O_j } \right\rangle }  +  \cdots } \right\}
\label{cj}
\end{eqnarray}

were the coefficients $C_j$ and the operators $\widehat O_j$ are given by
\begin{eqnarray}  
  C_1  						&=& \frac{1}{8} \nonumber \\
   \widehat O_1   &=& \frac{{\alpha _s }}{\pi }G_{\mu \nu } G^{\mu \nu }  \nonumber \\
  C_2  						&=& \frac{1}{2}\left\{ {1 + \left[ {\frac{{11}}{3} - 2\ln \left( {\frac{{ - z}}{{\mu ^2 }}} \right)} \right]\frac{{\alpha _s \left( {\left| z \right|} \right)}}{\pi }} \right\} 														\nonumber \\
  \widehat O_2    &=& m_q \bar qq \nonumber \\
  C_3  						&=& \frac{3}{{16\pi ^2 }} \nonumber \\ 
	\widehat O_3  	&=& m_q^4 \nonumber
\end{eqnarray}

substituting (\ref{cj}) into (\ref{delta5}) we obtain
\begin{eqnarray}  
\delta _5  = \frac{1}{{2\pi i}}\oint\limits_{C\left( R \right)} {\frac{1}{z}\left( {1 - a_0 z - a_1 z^2 } \right)m_q^2 \left( {\left| z \right|} \right)\sum\limits_{j = 1}^3 {\frac{{C_j }}
{{ - z}}\left\langle {\widehat O_j } \right\rangle } dz} 
\label{cj2}
\end{eqnarray}

now for $j=1$
\begin{eqnarray}  
\left. {\delta _5 } \right|_{C_1 } &=& \frac{1}{{2\pi i}}\oint\limits_{C\left( R \right)} { - \frac{1}{8}m_q^2 \left( {\left| z \right|} \right)\left\langle {\frac{{\alpha _s }}{\pi }G_{\mu \nu } 																						G^{\mu \nu } } \right\rangle \frac{1}{{z^2 }}\left( {1 - a_0 z - a_1 z^2 } \right)dz}  \nonumber \\
																	 &=&  - \frac{1}{8}m_q^2 \left( R \right)\left\langle {\frac{{\alpha _s }}{\pi }G_{\mu \nu } G^{\mu \nu } } \right\rangle \frac{1}{{2\pi i}}\oint\limits_{C\left( R 																					\right)} {\frac{1}{{z^2 }}\left( {1 - a_0 z - a_1 z^2 } \right)dz}  \nonumber \\
																   &=& \frac{1}{8}m_q^2 \left( R \right)\left\langle {\frac{{\alpha _s }}{\pi }G_{\mu \nu } G^{\mu \nu } } \right\rangle a_0  
																			\label{cjeq1}
\end{eqnarray}

now for $j=2$
\begin{eqnarray}    
\left. {\delta _5 } \right|_{C_2 } &=& \frac{1}{{2\pi i}}\oint\limits_{C\left( R \right)} { -m_q^2 \left( {\left| z \right|} \right)\left\langle {m_q \bar qq} \right\rangle \frac{1}{{z^2 }}\left( {1 																				- a_0 z - a_1 z^2 } \right)\left\{ {\frac{1}{2}\left[ {1 + \frac{{11}}{3}\frac{{\alpha _s \left( {\left| z \right|} \right)}}{\pi }} \right] - \ln \left({\frac{{ 																			- z}}{{\mu ^2 }}} \right)\frac{{\alpha _s \left( {\left| z \right|} \right)}}{\pi }} \right\}dz}  \nonumber \\
																   &=&  - \frac{1}{2}m_q^2 \left( R \right)\left\langle {m_q \bar qq} \right\rangle \left[ {1 + \frac{{11}}{3}\frac{{\alpha _s \left( R \right)}}{\pi }} \right]\frac{1}
																			{{2\pi i}}\oint\limits_{C\left( R \right)} {\frac{1}{{z^2 }}\left( {1 - a_0 z - a_1 z^2 } \right)dz}  \nonumber \\
																	&+&  m_q^2 \left( R \right)\left\langle {m_q \bar qq}\right\rangle \frac{{\alpha _s \left( R \right)}}{\pi }\underline {\frac{1}{{2\pi i}}\oint\limits_{C\left( R 																					\right)} {\frac{1}{{z^2 }}\left( {1 - a_0 z - a_1 z^2 } \right)\ln \left( {\frac{{ - z}}{{\mu ^2 }}} \right)dz} } 
																			\label{cj3}
\end{eqnarray}

we make a change of variables for the underlined term in (\ref{cj3}) to $z=\mu^2 w$ $\Rightarrow$ $dz=\mu^2 dw$
\begin{align}
\left. {\delta _5 } \right|_{C_2 }  &=  - \frac{1}{2}m_q^2 \left( R \right)\left\langle {m_q \bar qq} \right\rangle \left[ {1 + \frac{{11}}{3}\frac{{\alpha _s \left( R \right)}}{\pi }} \right]\left( { - a_0 } \right) +\nonumber\\
&+ m_q^2 \left( R \right)\left\langle {m_q \bar qq} \right\rangle \frac{{\alpha _s \left( R \right)}}{\pi }\frac{1}{{2\pi i}}\oint\limits_{C\left( r \right)} {\frac{1}{{\mu ^4 w^2 }}\left( {1 - a_0 \mu ^2 w - a_1 \mu ^4 w^2 } \right)\ln \left( { - w} \right)\mu ^2 dw} \nonumber \\
\label{cj4}	
\end{align}

we can now integrate (\ref{cj4}) by using (\ref{I}) and the using $R=\mu^2 r$ for $r$ to get
\begin{eqnarray}
\left. {\delta _5 } \right|_{C_2 }  = \frac{1}{2}m_q^2 \left( R \right)\left\langle {m_q \bar qq} \right\rangle \left[ {1 + \frac{{11}}{3}\frac{{\alpha _s \left( R \right)}}{\pi }} \right]a_0  + m_q^2 \left( R \right)\left\langle {m_q \bar qq} \right\rangle \frac{{\alpha _s \left( R \right)}}{\pi }\left[ { - \frac{1}{R} - a_0 \ln \left( {\frac{R}{{\mu ^2 }}} \right) - a_1 R} \right]
																			\label{cj5}
\end{eqnarray}

and now setting $\mu^2 = R$ we obtain
\begin{eqnarray}
\left. {\delta _5 } \right|_{C_2 }  = m_q^2 \left( R \right)\left\langle {m_q \bar qq} \right\rangle \left[ {\frac{1}{2}a_0  + \left( {\frac{{11}}{6}a_0  - \frac{1}{R} - a_1 R} \right)\frac{{\alpha 																			_s \left( R \right)}}{\pi }} \right]
																			\label{cjeq2}
\end{eqnarray}

now for $j=3$
\begin{eqnarray}  
  \left. {\delta _5 } \right|_{C_3 } &=& \frac{1}{{2\pi i}}\oint\limits_{C\left( R \right)} { - m_q^2 \left( {\left| z \right|} \right)\left\langle {m_q^4 } \right\rangle \frac{3}{{16\pi ^2 }}\frac{1}
																				{{z^2 }}\left( {1 - a_0 z - a_1 z^2 } \right)dz}  \nonumber \\
   																	 &=&  - m_q^2 \left( R \right)\left\langle {m_q^4 } \right\rangle \frac{3}{{16\pi ^2 }}\frac{1}{{2\pi i}}\oint\limits_{C\left( R \right)} {\frac{1}{{z^2 }}\left( 																					{1 - a_0 z - a_1 z^2 } \right)dz}  \nonumber \\
   																	 &=& \frac{3}{{16\pi ^2 }}m_q^2 \left( R \right)\left\langle {m_q^4 } \right\rangle a_0
																			\label{cjeq3}
\end{eqnarray}

we can now add up the individual $\delta_j$'s to get the whole contribution 
\begin{align}
\left. {\delta _5 } \right|_{Cj}  =& \left. {\delta _5 } \right|_{C_1 }  + \left. {\delta _5 } \right|_{C_2 }  + \left. {\delta _5 } \right|_{C_3 }  \nonumber \\
=& \frac{1}{8}a_0 \left\langle {\frac{{\alpha _s }}{\pi }G_{\mu \nu } G^{\mu \nu } } \right\rangle m_q^2 \left( R \right) + \left\langle {m_q \bar qq} \right\rangle \left[ {\frac{1}
	{2}a_0  + \left( {\frac{{11}}{6}a_0  - \frac{1}{R} - a_1 R} \right)\frac{{\alpha _s \left( R \right)}}{\pi }} \right]m_s^2 \left( R \right) + \frac{3}{{16\pi ^2 }}a_0 \left\langle {m_q^4 } \right\rangle m_q^2 \left( R \right) \nonumber \\
\label{cjtot}	
\end{align}
\numberwithin{equation}{section}
\section{Renormalization Group Equations for the coupling and mass}
\label{RenormalizationGroupEquationsForTheCouplingAndMass}

The evolution of the strong coupling is governed by the Renormalization Group Equation \cite{pascual}\cite{hey}:
\begin{eqnarray}   
		z\frac{{da_s \left( { - z} \right)}}{{dz}} = \beta \left( {a_s } \right) =  - \sum\limits_{n = 0} {\beta _n a_s^{n + 2} \left( { - z} \right)} 
\label{rgeas}
\end{eqnarray}

were $\beta \left( {a_s } \right)$ is the QCD beta function which is in turn expand as a power series in $a_s(-z)$. The $\beta_n$'s are the coefficients of the expansion and these coefficients depend on the number of quarks. The $a_s(z)$ is defined as
\[
a_s \left( z \right) \equiv \frac{{\alpha _s \left( z \right)}}
{\pi }
\]

we make a change of variables to remove the negative sign on the inside of $a_s(-z)$ with $z \mapsto z =  - y$ $\Rightarrow$ $dz =  - dy$ and 
\begin{eqnarray} 
y\frac{{da_s \left( y \right)}}{{dy}} =  - \sum\limits_{n = 0} {\beta _n a_s^{n + 2} \left( y \right)} 
\label{rgeasy}
\end{eqnarray}

In the complex plane differential equation (\ref{rgeasy}) behaves in a simpler manner, to go to the complex plane we make a change of variables with $y  \mapsto y = y_0 e^{ix}$ $\Rightarrow$ 
$dy = y_0 ie^{ix} dx$ ,  here $x$ is $x \in \left( { - \pi ;\pi } \right)$ .With this change of variables (\ref{rgeasy}) becomes
\begin{eqnarray} 
\frac{{da_s \left( x \right)}}{{dx}} =  - i\sum\limits_{n = 0} {\beta _n a_s^{n + 2} \left( x \right)} 
\label{rgeasx}
\end{eqnarray}
 
The differential equation can be solved numerically via the use of the Modified Euler method with an initial condition set by $a_s \left( 0 \right) = a_s \left( { - z_0 } \right)$.

The mass is also governed by a Renormalization Group Equation:
\begin{eqnarray} 
\frac{z}{{m\left( { - z} \right)}}\frac{{dm\left( { - z} \right)}}{{dz}} = \gamma \left( {a_s } \right) =  - \sum\limits_{n = 0} {\gamma _n a_s^{n + 1} \left( { - z} \right)} 
\label{rgemz}
\end{eqnarray}

we can remove the negative sign again by using the above change of variables for $z$, (\ref{rgemz}) changes to 
\[
\frac{y}
{{m\left( y \right)}}\frac{{dm\left( y \right)}}
{{dy}} =  - \sum\limits_{n = 0} {\gamma _n a_s^{n + 1} \left( y \right)} 
\]

and again changing variables for $y$ as shown above we finally obtain the form of the differential equation in the complex plane
\begin{eqnarray} 
\frac{1}{{m\left( x \right)}}\frac{{dm\left( x \right)}}{{dx}} =  - i\sum\limits_{n = 0} {\gamma _n a_s^{n + 1} \left( x \right)} 
\label{rgemxde}
\end{eqnarray}

Surprisingly (\ref{rgemxde}) is a separable differential equation and we can integrate the left hand side  
\begin{eqnarray} 
\int\limits_{m\left( 0 \right)}^{m\left( x \right)} {\frac{{dm\left( x \right)}}
{{m\left( x \right)}}}  =  - i\int\limits_0^x {\sum\limits_{n = 0} {\gamma _n a_s^{n + 1} \left( {x'} \right)} dx'} 
\label{rgemxde2}
\end{eqnarray}

which leads to 
\begin{eqnarray} 
m\left( x \right) = m\left( 0 \right)\exp \left[ { - i\int\limits_0^x {\sum\limits_{n = 0} {\gamma _n a_s^{n + 1} \left( {x'} \right)} dx'} } \right]
\label{rgemxx}
\end{eqnarray}

were $m\left( 0 \right) \equiv m\left( {z_0 } \right)$


\section{A comment about the CIPT}
\label{ACommentAboutTheCIPTIntegrals}

Using the method of Contour Improved Perturbation Theory we find that $\delta_5$ is
\begin{eqnarray}
{\left. {{\delta _5}\left( {{z_0}} \right)} \right|_{PQCD}}  = {\psi _5}\left( {{z_0}} \right) + {z_0}\psi _5 ^{'} \left( {{z_0}} \right) - \frac{1}{{16{\pi ^2}}}\frac{{{{\tilde m}^2}\left( {{z_0}} \right)}}{{2\pi }}\sum\limits_{j = 0}^4 {{K_j}\left[ { - i{z_0}I_j^{\left( 1 \right)} + \sum\limits_{n = 1}^3 {{b_n}{{\left( { - {z_0}} \right)}^n}I_{j;n}^{\left( 2 \right)}}  - F\left( {{z_0}} \right)I_j^{\left( 3 \right)}} \right]}
\label{ciptd5}
\end{eqnarray}


Looking at \ref{ciptd5} we see that there is a dependence on the $\psi _5\left( z_0 \right)$ and $\psi _5^{'}\left( z_0 \right)$ which are functions that are dependent on the renormalization and the regularization schemes! The $b_n$'s are some real coefficients and the $F(z_0)$ is some real function. The $K_j$'s are given in \cite{CADMass}. This makes the $\delta_5$ an unphysical quantity! There are other problems too with this method about the reality of $\delta_5$ which is discussed below. Looking at the RGE for the strong coupling
\begin{eqnarray}
\frac{d}{{dx}}a_s \left( x \right) =  - i\sum\limits_{n = 0} {\beta _n a_s^{n + 2} \left( x \right)} 
\label{rgea}
\end{eqnarray}

with $a_s \left( x \right)$ a solution to the above differential equation. Since this is an equality we can take the complex conjugate on both sides of (\ref{rgea}) and the equality will still hold true i.e.
\begin{eqnarray}  
\overline {\frac{d}{{dx}}a_s \left( x \right)}   &=& \overline { - i\sum\limits_{n = 0} {\beta _n a_s^{n + 2} \left( x \right)} }  \nonumber \\
\frac{d}{{dx}}\overline a_s \left( x \right) &=& i\sum\limits_{n = 0} {\beta _n \overline a_s^{n + 2} \left( x \right)}  
\end{eqnarray}

and making the change of variables $  x \mapsto x =  - v $ $\Rightarrow$ $  dx =  - dv $
\begin{eqnarray} 
\frac{d}{{dv}} \overline a_s \left( { - v} \right) =  - i\sum\limits_{n = 0} {\beta _n \overline a_s^{n + 2} \left( { - v} \right)} 
\end{eqnarray}

and now changing back to $x$'s with $v \mapsto v = x$ $\Rightarrow$ $  dv = dx $ we get 
\begin{eqnarray} 
\frac{d}{{dx}} \overline a_s \left( { - x} \right) =  - i\sum\limits_{n = 0} {\beta _{n }  \overline a_s^{n + 2} \left( { - x} \right)} 
\label{invol}
\end{eqnarray}

Now we see that we have recovered the original differential equation (\ref{rgea}) in (\ref{invol}) by the process of complex conjugation and changing variables. This yields for us a property that the coupling must satisfy i.e.
\begin{eqnarray}
a_s \left( x \right) = \overline a_s \left( { - x} \right)
\label{invol2}
\end{eqnarray}

Applying the complex conjugate on both sides of (\ref{invol2}) leads to a more useful form of this property\footnote{I have searched the literature and could find no mention of this property in the past. This property was discovered after debating about the reality of the integral (\ref{i1}) with Prof. C.A. Dominguez. While attempting to prove that the integral was indeed complex I stumbled upon this property.  }
\begin{eqnarray}
a_s \left( { - x} \right) = \overline a_s \left( x \right)
\label{invol3}
\end{eqnarray}

\begin{figure}[h]
	\centerline{
	\mbox{\includegraphics[width=7in]{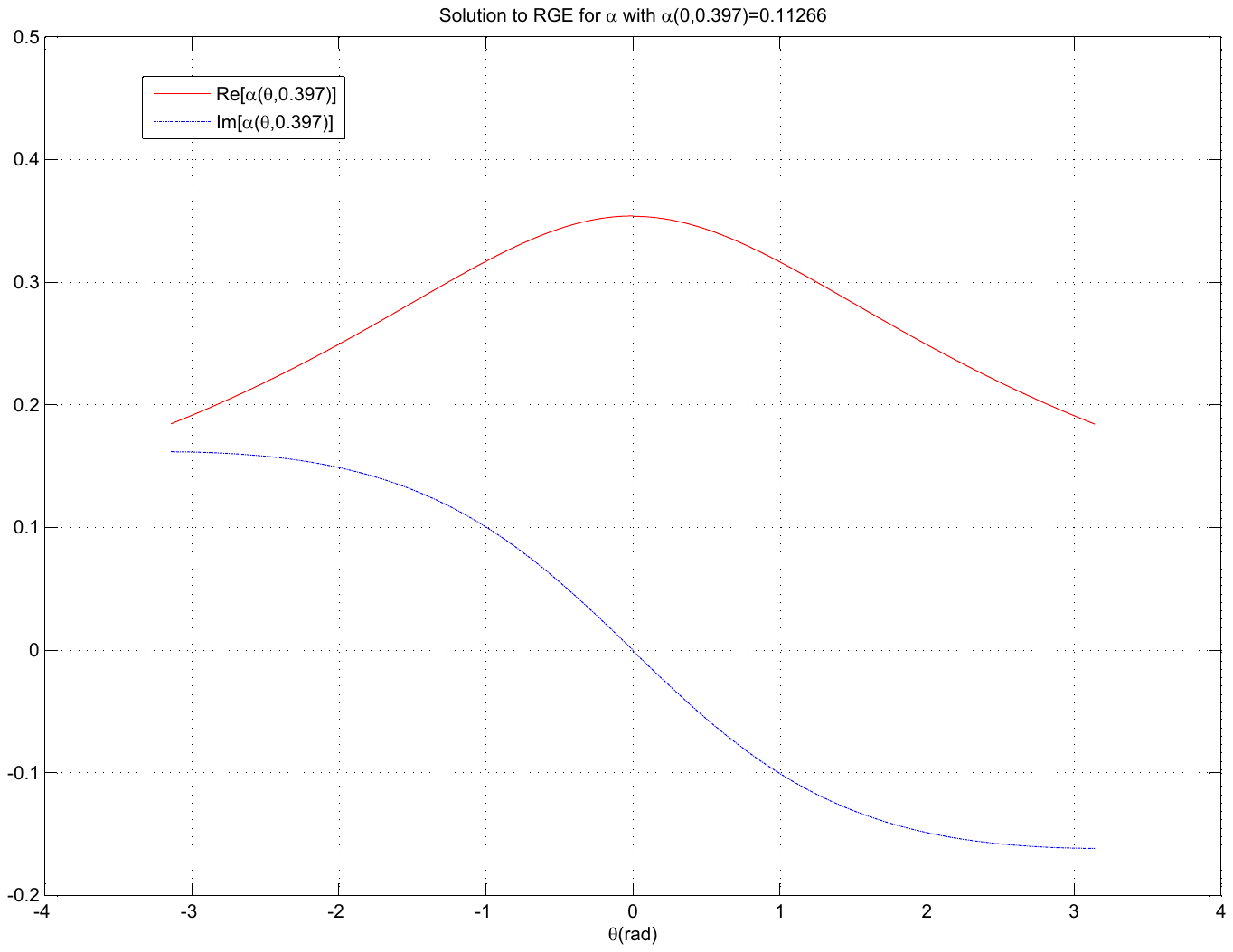}}
	}
	\caption{Numerical solution to the RGE for $\alpha_s(0,0.397)=0.11266$, with the top curve the real part and the bottom the imaginary part of $\alpha_s $}
	\label{RGESolution1}
\end{figure}

\ref{invol3} is the defining condition for the Kramers-Kronig Relation \cite{JordansLemma1}. From the Kramers-Kronig relation we know that the real part of the function must be even and the imaginary part must be odd. We solved numerically the RGE for the coupling using as initial conditions the values of $\alpha_s (z, \Lambda )$ from (\ref{formulaalpha}), a specific initial case was chosen as shown in figure \ref{RGESolution1}
\clearpage

The above numerical solution matches that given in \cite{Daviera,Daviera2}. From figure \ref{RGESolution1} we can see that the property given by (\ref{invol3}) is valid. Now using (\ref{invol3}) we can prove some interesting results about the integrals 
\begin{eqnarray}   
f_1 &=& \frac{{z_0 \widetilde m^2 \left( {z_0 } \right)}}{{2\pi i}}\int\limits_{ - \pi }^\pi  {\left( {\pi  + x} \right)a_s^j \left( x \right)\exp \left[ {ix - 2i\sum\limits_{l = 0}^3 {\gamma _l 						\int\limits_0^x {a_s^{l + 1} \left( {x'} \right)} dx'} } \right]dx}  \label{i1} \\
f_2 &=& \frac{{\left( { - z_0 } \right)^n \widetilde m^2 \left( {z_0 } \right)}}{{2\pi }}\int\limits_{ - \pi }^\pi  {a_s^j \left( x \right)\exp \left[ {inx - 2i\sum\limits_{l = 0}^3 {\gamma _l 							\int\limits_0^x {a_s^{l + 1} \left( {x'} \right)} dx'} } \right]dx}  \hfill \\
f_3 &=& \frac{{\widetilde m^2 \left( {z_0 } \right)}}{{2\pi }}\int\limits_{ - \pi }^\pi  {a_s^j \left( x \right)\exp \left[ { - 2i\sum\limits_{l = 0}^3 {\gamma _l \int\limits_0^x {a_s^{l + 1} \left( {x'} \right)} dx'} } \right]dx} 
\end{eqnarray}

Consider the third integral 
\begin{eqnarray}  
I_j^{\left( 3 \right)} &=& \int\limits_{-\pi }^\pi {a_s^j \left( x \right)\exp \left[ { - 2i\sum\limits_{l = 0}^3 {\gamma _l \int\limits_0^x {a_s^{l + 1} \left( y \right)} dy} } \right]dx}  \nonumber \\
  &=& \underline {\int\limits_{ - \pi }^0 {a_s^j \left( x \right)\exp \left[ { - 2i\sum\limits_{l = 0}^3 {\gamma _l \int\limits_0^x {a_s^{l + 1} \left( y \right)} dy} } \right]dx} }  + \int\limits_0^\pi  {a_s^j \left( x \right)\exp \left[ { - 2i\sum\limits_{l = 0}^3 {\gamma _l \int\limits_0^x {a_s^{l + 1} \left( y \right)} dy} } \right]dx}  
\label{qcdin3t1}
\end{eqnarray}

for the underlined term we make a change of variables $x \mapsto x =  - v$ $\Rightarrow$ $dx =  - dv $ and we obtain
\begin{eqnarray} 
I_j^{\left( 3 \right)}  = \int\limits_0^\pi  {a_s^j \left( { - v} \right)\exp \left[ { - 2i\sum\limits_{l = 0}^3 {\gamma _l \underline {\int\limits_0^{ - v} {a_s^{l + 1} \left( y \right)} dy} } } \right]dv}  + \int\limits_0^\pi  {a_s^j \left( x \right)\exp \left[ { - 2i\sum\limits_{l = 0}^3 {\gamma _l \int\limits_0^x {a_s^{l + 1} \left( y \right)} dy} } \right]dx} 
\label{qcdin3t2}
\end{eqnarray}

we make another change of variables with $y \mapsto y =  - u $ $\Rightarrow$ $dy =  - du $  and renaming $v$ as $x$ which leads to 
\begin{eqnarray}   
I_j^{\left( 3 \right)}  &=& \int\limits_0^\pi  {a_s^j \left( { - x} \right)\exp \left[ {2i\sum\limits_{l = 0}^3 {\gamma _l \int\limits_0^x {a_s^{l + 1} \left( { - u} \right)} du} } \right]dx}  + \int\limits_0^\pi  {a_s^j \left( x \right)\exp \left[ { - 2i\sum\limits_{l = 0}^3 {\gamma _l \int\limits_0^x {a_s^{l + 1} \left( y \right)} dy} } \right]dx}  \nonumber \\
&=& \int\limits_0^\pi  {\overline a_s^j \left( x \right)\exp \left[ {2i\sum\limits_{l = 0}^3 {\gamma _l \int\limits_0^x {\overline a_s^{l + 1} \left( u \right)} du} } \right]dx}  + \int\limits_0^\pi  {a_s^j \left( x \right)\exp \left[ { - 2i\sum\limits_{l = 0}^3 {\gamma _l \int\limits_0^x {a_s^{l + 1} \left( y \right)} dy} } \right]dx}  \label{use3a} \\
   \nonumber \\
&=& \overline {\int\limits_0^\pi  {a_s^j \left( x \right)\exp \left[ { - 2i\sum\limits_{l = 0}^3 {\gamma _l \int\limits_0^x {a_s^{l + 1} \left( u \right)} du} } \right]dx} }  + \int\limits_0^\pi  {a_s^j \left( x \right)\exp \left[ { - 2i\sum\limits_{l = 0}^3 {\gamma _l \int\limits_0^x {a_s^{l + 1} \left( y \right)} dy} } \right]dx}  \label{compcon} \\
&=& 2\operatorname{Re} \left\{ {\int\limits_0^\pi  {a_s^j \left( x \right)\exp \left[ { - 2i\sum\limits_{l = 0}^3 {\gamma _l \int\limits_0^x {a_s^{l + 1} \left( y \right)} dy} } \right]dx} } \right\} \label{qcdin3t3}
\end{eqnarray}

In (\ref{use3a}) we have used the property of (\ref{invol3}) and in (\ref{compcon}) we taken out complex conjugate from the individual terms and conjugated the whole integral.
\begin{eqnarray} 
{I_j^{\left( 3 \right)}  = 2\operatorname{Re} \left\{ {\int\limits_0^\pi  {a_s^j \left( x \right)\exp \left[ { - 2i\sum\limits_{l = 0}^3 {\gamma _l \int\limits_0^x {a_s^{l + 1} \left( y \right)} dy} } \right]dx} } \right\}}
\label{qcdin3t3b}
\end{eqnarray}

So the integral (\ref{qcdin3t3b}) is a purely real number.

We now consider the second integral 
\begin{eqnarray}     
I_{j;n}^{\left( 2 \right)} &=& \int\limits_{ - \pi }^\pi  {a_s^j \left( x \right)\exp \left[ {inx - 2i\sum\limits_{l = 0}^3 {\gamma _l \int\limits_0^x {a_s^{l + 1} \left( y \right)} dy} } \right]dx}  \nonumber \\
&=& \underline {\int\limits_{ - \pi }^0 {a_s^j \left( x \right)\exp \left[ {inx - 2i\sum\limits_{l = 0}^3 {\gamma _l \int\limits_0^x {a_s^{l + 1} \left( y \right)} dy} } \right]dx} }  + \int\limits_0^\pi  {a_s^j \left( x \right)\exp \left[ {inx - 2i\sum\limits_{l = 0}^3 {\gamma _l \int\limits_0^x {a_s^{l + 1} \left( y \right)} dy} } \right]dx}  
\label{qcdin2t1}
\end{eqnarray}

we make a change of variables $x \mapsto x =  - v$ $\Rightarrow$ $dx =  - dv $ and we obtain
\begin{eqnarray}  
I_{j;n}^{\left( 2 \right)}  = \int\limits_0^\pi  {a_s^j \left( { - v} \right)\exp \left[ { - inv - 2i\sum\limits_{l = 0}^3 {\gamma _l \underline {\int\limits_0^{ - v} {a_s^{l + 1} \left( y \right)} dy} } } \right]dv}  + \int\limits_0^\pi  {a_s^j \left( x \right)\exp \left[ {inx - 2i\sum\limits_{l = 0}^3 {\gamma _l \int\limits_0^x {a_s^{l + 1} \left( y \right)} dy} } \right]dx} 
\label{qcdin2t2}
\end{eqnarray}

we make another change of variables with $y \mapsto y =  - u $ $\Rightarrow$ $dy =  - du $  and renaming $v$ as $x$ which leads to
\begin{align}
I_{j;n}^{\left( 2 \right)}  &= \int\limits_0^\pi  {a_s^j \left( { - x} \right)\exp \left[ { - inx + 2i\sum\limits_{l = 0}^3 {\gamma _l \int\limits_0^x {a_s^{l + 1} \left( { - u} \right)} du} } \right]dx}  + \int\limits_0^\pi  {a_s^j \left( x \right)\exp \left[ {inx - 2i\sum\limits_{l = 0}^3 {\gamma _l \int\limits_0^x {a_s^{l + 1} \left( y \right)} dy} } \right]dx}  \nonumber \\
&= \int\limits_0^\pi  {\overline a _s^j \left( x \right)\exp \left[ { - inx + 2i\sum\limits_{l = 0}^3 {\gamma _l \int\limits_0^x {\overline a _s^{l + 1} \left( u \right)} du} } \right]dx}  + \int\limits_0^\pi  {a_s^j \left( x \right)\exp \left[ {inx - 2i\sum\limits_{l = 0}^3 {\gamma _l \int\limits_0^x {a_s^{l + 1} \left( y \right)} dy} } \right]dx}  \label{use2a} \\
&= \overline {\int\limits_0^\pi  {a_s^j \left( x \right)\exp \left[ {inx - 2i\sum\limits_{l = 0}^3 {\gamma _l \int\limits_0^x {a_s^{l + 1} \left( u \right)} du} } \right]dx} }  + \int\limits_0^\pi  {a_s^j \left( x \right)\exp \left[ {inx - 2i\sum\limits_{l = 0}^3 {\gamma _l \int\limits_0^x {a_s^{l + 1} \left( y \right)} dy} } \right]dx}  \label{compcon2} \\
&= 2\operatorname{Re} \left\{ {\int\limits_0^\pi  {a_s^j \left( x \right)\exp \left[ {inx - 2i\sum\limits_{l = 0}^3 {\gamma _l \int\limits_0^x {a_s^{l + 1} \left( y \right)} dy} } \right]dx} } \right\} 
\label{qcdin2t3}	
\end{align}

In (\ref{use2a}) we have used the property of (\ref{invol3}) and in (\ref{compcon2}) we have taken out the complex conjugate from the individual terms and conjugated the whole integral.

\begin{eqnarray}
{I_{j;n}^{\left( 2 \right)}  = 2\operatorname{Re} \left\{ {\int\limits_0^\pi  {a_s^j \left( x \right)\exp \left[ {inx - 2i\sum\limits_{l = 0}^3 {\gamma _l \int\limits_0^x {a_s^{l + 1} \left( y \right)} dy} } \right]dx} } \right\}}
\label{qcdin2t3b}
\end{eqnarray}

The integral (\ref{qcdin2t3b}) is again also a purely real number.

Now we finally consider the first integral. This is a problematic integral as it leads to a counter-intuitive conclusion.
\begin{align}
I_j^{\left( 1 \right)}  &= \int\limits_{ - \pi }^\pi  {\left( {\pi  + x} \right)a_s^j \left( x \right)\exp \left[ {ix - 2i\sum\limits_{l = 0}^3 {\gamma _l \int\limits_0^x {a_s^{l + 1} \left( y \right)} dy} } \right]dx}  \nonumber \\
&= \underline {\int\limits_{ - \pi }^0 {\left( {\pi  + x} \right)a_s^j \left( x \right)\exp \left[ {ix - 2i\sum\limits_{l = 0}^3 {\gamma _l \int\limits_0^x {a_s^{l + 1} \left( y \right)} dy} } \right]dx} } +\nonumber\\
&+ \int\limits_0^\pi  {\left( {\pi  + x} \right)a_s^j \left( x \right)\exp \left[ {ix - 2i\sum\limits_{l = 0}^3 {\gamma _l \int\limits_0^x {a_s^{l + 1} \left( y \right)} dy} } \right]dx} \label{qcdin1t1}	
\end{align}

we make a change of variables $x \mapsto x =  - v$ $\Rightarrow$ $dx =  - dv $ and we obtain

\begin{align}
I_j^{\left( 1 \right)}  =& \int\limits_0^\pi  {\left( {\pi  - v} \right)a_s^j \left( { - v} \right)\exp \left[ { - iv - 2i\sum\limits_{l = 0}^3 {\gamma _l \underline {\int\limits_0^{ - v} {a_s^{l + 1} \left( y \right)} dy} } } \right]dv}  +\nonumber\\
&+ \int\limits_0^\pi  {\left( {\pi  + x} \right)a_s^j \left( x \right)\exp \left[ {ix - 2i\sum\limits_{l = 0}^3 {\gamma _l \int\limits_0^x {a_s^{l + 1} \left( y \right)} dy} } \right]dx} 
\label{qcdin1t2}	
\end{align}

we make another change of variables with $y \mapsto y =  - u $ $\Rightarrow$ $dy =  - du $  and renaming $v$ as $x$ which leads to

\begin{align}
	I_j^{\left( 1 \right)} =& \int\limits_0^\pi  {\left( {\pi  - x} \right)a_s^j \left( { - x} \right)\exp \left[ { - ix + 2i\sum\limits_{l = 0}^3 {\gamma _l \int\limits_0^x {a_s^{l + 1} \left( { - u} \right)} du} } \right]dx}  +\nonumber \\
	&+ \int\limits_0^\pi  {\left( {\pi  + x} \right)a_s^j \left( x \right)\exp \left[ {ix - 2i\sum\limits_{l = 0}^3 {\gamma _l \int\limits_0^x {a_s^{l + 1} \left( y \right)} dy} } \right]dx}  \nonumber \\
	=& \int\limits_0^\pi  {\left( {\pi  - x} \right)\overline a _s^j \left( x \right)\exp \left[ { - ix + 2i\sum\limits_{l = 0}^3 {\gamma _l \int\limits_0^x {\overline a _s^{l + 1} \left( u \right)} du} } \right]dx}  +\nonumber\\
	&+ \int\limits_0^\pi  {\left( {\pi  + x} \right)a_s^j \left( x \right)\exp \left[ {ix - 2i\sum\limits_{l = 0}^3 {\gamma _l \int\limits_0^x {a_s^{l + 1} \left( y \right)} dy} } \right]dx}  \label{use1a}\\
	=& \int\limits_0^\pi  {\left( {\pi  - x} \right)\overline {a_s^j \left( x \right)\exp \left[ {ix - 2i\sum\limits_{l = 0}^3 {\gamma _l \int\limits_0^x {a_s^{l + 1} \left( u \right)} du} } \right]} dx}  +\nonumber\\
	&+ \int\limits_0^\pi  {\left( {\pi  + x} \right)a_s^j \left( x \right)\exp \left[ {ix - 2i\sum\limits_{l = 0}^3 {\gamma _l \int\limits_0^x {a_s^{l + 1} \left( y \right)} dy} } \right]dx} \label{compconj1} \\
	=& \pi \left[ {\overline {\int\limits_0^\pi  {a_s^j \left( x \right)\exp \left[ {ix - 2i\sum\limits_{l = 0}^3 {\gamma _l \int\limits_0^x {a_s^{l + 1} \left( u \right)} du} } \right]dx} }  + \int\limits_0^\pi  {a_s^j \left( x \right)\exp \left[ {ix - 2i\sum\limits_{l = 0}^3 {\gamma _l \int\limits_0^x {a_s^{l + 1} \left( u \right)} du} } \right]dx} } \right] \nonumber \\
	&+ \left[ {\int\limits_0^\pi  {xa_s^j \left( x \right)\exp \left[ {ix - 2i\sum\limits_{l = 0}^3 {\gamma _l \int\limits_0^x {a_s^{l + 1} \left( y \right)} dy} } \right]dx}  - \overline {\int\limits_0^\pi  {xa_s^j \left( x \right)\exp \left[ {ix - 2i\sum\limits_{l = 0}^3 {\gamma _l \int\limits_0^x {a_s^{l + 1} \left( y \right)} dy} } \right]dx} } } \right] \label{gather} \\
	=& 2\pi \operatorname{Re} \left\{ {\int\limits_0^\pi  {a_s^j \left( x \right)\exp \left[ {ix - 2i\sum\limits_{l = 0}^3 {\gamma _l \int\limits_0^x {a_s^{l + 1} \left( u \right)} du} } \right]dx} } \right\} +\nonumber\\
	&+ 2i\operatorname{Im} \left\{ {\int\limits_0^\pi  {xa_s^j \left( x \right)\exp \left[ {ix - 2i\sum\limits_{l = 0}^3 {\gamma _l \int\limits_0^x {a_s^{l + 1} \left( y \right)} dy} } \right]dx} } \right\} \nonumber \\ \label{link0} \\
	=& \pi I_{j;1}^{\left( 2 \right)}  + 2i\operatorname{Im} \left\{ {\int\limits_0^\pi  {xa_s^j \left( x \right)\exp \left[ {ix - 2i\sum\limits_{l = 0}^3 {\gamma _l \int\limits_0^x {a_s^{l + 1} \left( y \right)} dy} } \right]dx} } \right\}\label{link}	
\end{align}

In (\ref{use1a}) we have used the property of (\ref{invol3}) and in (\ref{compconj1}) we have taken out the complex conjugate from the individual terms and conjugated the whole integral. In (\ref{gather}) we have gathered the integrals and their complex conjugates and in (\ref{link}), we noted that the first term of (\ref{link0}) is just the case of (\ref{qcdin2t3b}) with $n=1$.
\begin{eqnarray}  
{I_j^{\left( 1 \right)}  = \pi I_{j;1}^{\left( 2 \right)}  + 2i\operatorname{Im} \left\{ {\int\limits_0^\pi  {xa_s^j \left( x \right)\exp \left[ {ix - 2i\sum\limits_{l = 0}^3 {\gamma _l \int\limits_0^x {a_s^{l + 1} \left( y \right)} dy} } \right]dx} } \right\}}
\label{qcdin1t3b}
\end{eqnarray}

The integral (\ref{qcdin1t3b}) is a part real and part imaginary number! This is problematic as the integral (\ref{i1}) has an additional factor of $1/i$ on the outside of the integral. From the result of (\ref{qcdin1t3b}) we can conclude that the result of the calculation for $\delta_5$ will have an imaginary component to it. This is bad as the quantity we are trying to calculate is a purely real number!


\section{Coefficient fixing of $\Delta(z)$}
\numberwithin{equation}{section}
The purpose of introducing a kernel (\ref{kernel}) in the calculation of the $\delta_{5}(R)$ was to minimise the effect of the resonance terms. This is done by insisting that $\Delta(M^{2}_1)=0$ and $\Delta(M^{2}_2)=0$ where $M^{2}_1$ and $M^{2}_2$ are the squared masses of the resonances of the Kaon particle
\begin{eqnarray}
 \Delta \left( {{\rm M}_1^2 } \right) = 1 - a_0 {\rm M}_1^2  - a_1 {\rm M}_1^4  = 0 \\ 
 \Delta \left( {{\rm M}_2^2 } \right) = 1 - a_0 {\rm M}_2^2  - a_1 {\rm M}_2^4  = 0 
\end{eqnarray}

solving the above equations simultaneously leads to
\begin{eqnarray} 
a_0  &=& \frac{{{\rm M}_{\rm 1}^{\rm 2}  + {\rm M}_{\rm 2}^{\rm 2} }}{{{\rm M}_{\rm 1}^{\rm 2} {\rm M}_{\rm 2}^{\rm 2} }} \\ 
a_1  &=&  - \frac{1}{{{\rm M}_{\rm 1}^{\rm 2} {\rm M}_{\rm 2}^{\rm 2} }} 
\end{eqnarray}

The masses are excited states of the Kaon \cite{pdgdata} which are $M_1 =1.40$ GeV and $M_2 =1.94$ GeV. This leads to 
\begin{eqnarray} 
a_0  &=& 0.7759/ \rm{  GeV^2} \nonumber \\ 
a_1  &=&  -0.1355/  \rm{  GeV^4}  \nonumber
\end{eqnarray}

For the Pion the excited states have masses which are $M_1 = 1.30$ GeV and $M_2 =1.81$ GeV. This leads to 
\begin{eqnarray} 
a_0  &=& 0.8949/ \rm{  GeV^2} \nonumber \\ 
a_1  &=&  -0.1794/  \rm{  GeV^4}  \nonumber
\end{eqnarray}
\section{Perturbative QCD Integrals}
\label{sec:PerturbativeQCDIntegrals}
\numberwithin{equation}{section}

General integrals of the form ${\rm I}_{{\rm nk}}$ are required to do the calculations with
\[
{\rm I}_{{\rm nk}}  = \frac{1}{{2\pi i}}\oint\limits_{C'(r)} {z^n \left[ {\ln \left( { - z} \right)} \right]^k dz} 
\]

where $C'(r)$ is the contour chosen to be a circle of radius $r$ parameterized from $-\pi$ to $\pi$ and $n,k$ $\in \mathbb{N}$.

For $n\geq0$ and $k\geq0$
\[
\sigma _{jn}  = \left\{ {\begin{array}{*{20}lcr}
   {\left( { - 1} \right)^{j + 1} {\rm ; n} \in {\rm 2\mathbb{N}}}  \\
   {\left( { - 1} \right)^j {\hspace{0.35cm} \rm ;n} \in {\rm 2\mathbb{N} + 1}}  \\
\end{array}} \right.
\]
\[
\beta _{knj}  = \frac{{k!\left( {n + 1} \right)^{k - j} }}{{\left( {k - j} \right)!}}
\]
\[a = \rm ln          \emph{r} \]
\begin{eqnarray}
{\rm I}_{{\rm nk}}  = \frac{{\left( { - r} \right)^{n + 1} }}{{2\pi i\left( {n + 1} \right)^{k + 1} }}\sum\limits_{j = 0}^{k} {\beta _{knj} \sigma _{jn} \left[ {\left( {a + i\pi } \right)^{k - j}  - \left( {a - i\pi } \right)^{k - j} } \right]} 
\label{I}
\end{eqnarray}

Using (\ref{I}) we computed the following integrals
\begin{eqnarray} 
{\rm I}_{01}  &=& r \\ 
 {\rm I}_{11}  &=& \frac{1}{2}r^2  \\
 {\rm I}_{21}  &=& \frac{1}{3}r^3   \\
 {\rm I}_{02}  &=& 2r\left[ {\ln r - 1} \right]  \\
 {\rm I}_{12}  &=& r^2 \left[ {\ln r - \frac{1}{2}} \right]  \\
 {\rm I}_{22}  &=& \frac{2}{9}r^3 \left[ {3\ln r - 1} \right] \\
 {\rm I}_{03}  &=& r\left[ {3\left( {\ln ^2 r - 2\ln r} \right) + 6 - \pi ^2 } \right]  \\
 {\rm I}_{13}  &=& r^2 \left[ {\frac{3}{2}\left( {\ln ^2 r - \ln r} \right) + \frac{3}{4} - \frac{1}{2}\pi ^2 } \right] \\ 
 {\rm I}_{23}  &=& r^3 \left[ {\left( {\ln ^3 r - 2\ln r} \right) + \frac{2}{9} - \frac{1}{3}\pi ^2 } \right] \\ 
  {\rm I}_{{\rm 04}}  &=& r\left[ {4\ln ^3 r - 12\ln ^2 r + 4\left( {6 - \pi ^2 } \right)\ln r + 4\left( {\pi ^2  - 6} \right)} \right] \\ 
 {\rm I}_{{\rm 14}}   &=& r^2 \left[ {2\ln ^3 r - 3\ln ^2 r + \left( {3 - 2\pi ^2 } \right)\ln r + \pi ^2  - \frac{3}{2}} \right] \\ 
 {\rm I}_{{\rm 24}}   &=& r^3 \left[ {\frac{4}{3}\left( {\ln ^3 r - \ln ^2 r + \left( {\frac{2}{3} - \pi ^2 } \right)\ln r} \right) + \frac{4}{3}\left( {\pi ^2  - \frac{2}{3}} \right)} \right] \\
 {\rm I}_{{\rm 05}}  &=& r\left[ {5\ln ^4 r - 20\ln ^3 r + 10\left( {\ln r - 2} \right)\left( {6 - \pi ^2 } \right)\ln r + \pi ^4  - 20\pi ^2  + 120} \right] \\ 
 {\rm I}_{{\rm 15}}  &=& r^2 \left[ {\frac{5}{2}\ln ^4 r - 5\ln ^3 r + 5\left( {\ln r - 1} \right)\left( {\frac{3}{2} - \pi ^2 } \right)\ln r + \frac{{\pi ^4 }}{2} - \frac{5}{2}\pi ^2  + \frac{{15}}{4}} \right] \\ 
 {\rm I}_{{\rm 25}}  &=& r^3 \left[ {\frac{5}{3}\ln ^4 r - \frac{{20}}{9}\ln ^3 r + \frac{{10}}{3}\left( {\ln r - \frac{2}{3}} \right)\left( {\frac{6}{9} - \pi ^2 } \right)\ln r + \frac{{\pi ^4 }}{3} - \frac{{20}}{{27}}\pi ^2  + \frac{{40}}{{81}}} \right] 
\end{eqnarray}


\section{The Resonance Integrals}
\label{TheResonanceIntegrals}

To compute the resonance contribution, $\left. {\delta _5 \left(R\right)} \right|_{RES}$, to the FESR we need to compute integrals of the form
\begin{eqnarray} 
\int\limits_{9M_\pi ^2 }^R {\frac{1}
{z}\Delta \left( z \right)z\Theta \left( z \right)BW_k \left( z \right)dz} 
\label{resform}
\end{eqnarray}

with the Breit-Wigner profile
\begin{eqnarray} 
BW_k \left( z \right) = \frac{{M_k^2 \left( {M_k^2  + \Gamma _k^2 } \right)}}
{{\left( {z - M_k^2 } \right)^2  + M_k^2 \Gamma _k^2 }}
\label{bwform}
\end{eqnarray}

defining for convenience the following 
\[
\begin{gathered}
  \alpha ^2  \equiv M_k^2 \left( {M_k^2  + \Gamma _k^2 } \right) \hfill \\
  \beta ^2  \equiv M_k^2  \hfill \\
  \gamma ^2  \equiv M_k^2 \Gamma _k^2  \hfill \\ 
\end{gathered} 
\]

Now substituting (\ref{bwform}) into (\ref{resform})
\begin{eqnarray}   
\int {\Delta \left( z \right)\frac{{\alpha ^2 }}{{\left( {z - \beta ^2 } \right)^2  + \gamma ^2 }}dz} = \alpha ^2 \left\{ {\int {\frac{1}{{\left( {z - \beta ^2 } \right)^2  + \gamma ^2 }}dz}  - a_0 		\int {\frac{z}{{\left( {z - \beta ^2 } \right)^2  + \gamma ^2 }}dz}  - a_1 \int {\frac{{z^2 }}{{\left( {z - \beta ^2 } \right)^2  + \gamma ^2 }}dz} } \right\}
\label{resform2}
\end{eqnarray}

the first integral can be easily integrated as
\begin{eqnarray}
\int {\frac{1}
{{\left( {z - \beta ^2 } \right)^2  + \gamma ^2 }}dz}  &=& \frac{1}{{\gamma ^2 }}\int {\frac{1}{{\left( {\frac{{z - \beta ^2 }}{\gamma }} \right)^2  + 1}}dz}  \nonumber \\
		&=& \frac{1}{\gamma }\arctan \left( {\frac{{z - \beta ^2 }}{\gamma }} \right)
		\label{resint1}
\end{eqnarray}

the second and third integrals require some work
\begin{eqnarray}   
\int {\frac{z}{{\left( {z - \beta ^2 } \right)^2  + \gamma ^2 }}dz}  &=& \int {\frac{{z - \beta ^2  + \beta ^2 }}{{\left( {z - \beta ^2 } \right)^2  + \gamma ^2 }}dz}  \nonumber \\
   						&=& \int {\frac{{z - \beta ^2 }}{{\left( {z - \beta ^2 } \right)^2  + \gamma ^2 }}dz}  + \int {\frac{{\beta ^2 }}{{\left( {z - \beta ^2 } \right)^2  + \gamma ^2 }}dz}  \nonumber \\
   						&=& \frac{1}{2}\ln \left[ {\left( {z - \beta ^2 } \right)^2  + \gamma ^2 } \right] + \frac{{\beta ^2 }}{\gamma }\arctan \left( {\frac{{z - \beta ^2 }}{\gamma }} \right) 
		\label{resint2}
\end{eqnarray}

For the third integral we make use of a substitution $u = z - \beta ^2 $ $\Rightarrow$ $dz=du$ so
\begin{eqnarray}   
\int {\frac{{z^2 }}{{\left( {z - \beta ^2 } \right)^2  + \gamma ^2 }}dz}  &=& \int {\frac{{u^2  + 2u\beta ^2  + \beta ^4 }}{{u^2  + \gamma ^2 }}du}  \nonumber \\
   &=& \int {\frac{{u^2 }}{{u^2  + \gamma ^2 }}du}  + 2\beta ^2 \int {\frac{u}{{u^2  + \gamma ^2 }}du}  + \beta ^4 \int {\frac{1}{{u^2  + \gamma ^2 }}du}  \nonumber \\
   &=& \int {\frac{{u^2  + \gamma ^2  - \gamma ^2 }}{{u^2  + \gamma ^2 }}du}  + 2\beta ^2 \int {\frac{u}{{u^2  + \gamma ^2 }}du}  + \beta ^4 \int {\frac{1}{{u^2  + \gamma ^2 }}du}  \nonumber \\
   &=& u + \left( {\beta ^4  - \gamma ^2 } \right)\int {\frac{1}{{u^2  + \gamma ^2 }}du}  + 2\beta ^2 \int {\frac{u}{{u^2  + \gamma ^2 }}du}  \nonumber \\
   &=& u + \left( {\frac{{\beta ^4  - \gamma ^2 }}{{\gamma ^2 }}} \right)\int {\frac{1}{{\left( {\frac{u}{\gamma }} \right)^2  + 1}}du}  + 2\beta ^2 \int {\frac{u}{{u^2  + \gamma ^2 }}du}  \nonumber \\
   &=& u + \left( {\frac{{\beta ^4  - \gamma ^2 }}{\gamma }} \right)\arctan \left( {\frac{u}{\gamma }} \right) + \beta ^2 \ln \left( {u^2  + \gamma ^2 } \right) \nonumber \\
   &=& z - \beta ^2  + \left( {\frac{{\beta ^4  - \gamma ^2 }}{\gamma }} \right)\arctan \left( {\frac{{z - \beta ^2 }}{\gamma }} \right) + \beta ^2 \ln \left[ {\left( {z - \beta ^2 } \right)^2  + \gamma ^2 } \right] 
		\label{resint3}
\end{eqnarray}

now substituting the integrals (\ref{resint1} - \ref{resint3}) into the right-hand side of (\ref{resform2})
\begin{align}
&\alpha ^2 \left\{ {\int {\frac{1}{{\left( {z - \beta ^2 } \right)^2  + \gamma ^2 }}dz}  - a_0 \int {\frac{z}{{\left( {z - \beta ^2 } \right)^2  + \gamma ^2 }}dz}  - a_1 \int {\frac{{z^2 }}
		{{\left( {z - \beta ^2 } \right)^2  + \gamma ^2 }}dz} } \right\} \nonumber \\
=& \alpha ^2 \left\{ {\frac{1}{\gamma }\left[ {1 - a_0 \beta ^2  - a_1 \left( {\beta ^4  - \gamma ^2 } \right)} \right]\arctan \left( {\frac{{z - \beta ^2 }}{\gamma }} \right) - \left( {\frac{1}
		{2}a_0  + a_1 \beta ^2 } \right)\ln \left[ {\left( {z - \beta ^2 } \right)^2  + \gamma ^2 } \right] - a_1 \left( {z - \beta ^2 } \right)} \right\}
\label{resformRHS}	
\end{align}
getting rid of the $\alpha, \beta$ and $\gamma$ by substituting in their definitions into (\ref{resformRHS}) and now defining a new function $G_k \left(z\right)$ as the right-hand side of (\ref{resformRHS}) we get 

\begin{align}
\!\!\!\!G_k \left( z \right) = M_k^2 \left( {M_k^2  + \Gamma _k^2 } \right) &\Biggl\{ \frac{1}{{M_k \Gamma _k }}\left[ {1 - a_0 M_k^2  - a_1 \left( {M_k^4  - M_k^2 \Gamma _k^2 } \right)} \right]\arctan \left( {\frac{{z - M_k^2 }}
		{{M_k \Gamma _k }}} \right) +\nonumber\\
		&- \left( {\frac{1}
		{2}a_0  + a_1 M_k^2 } \right)\ln \left[ {\left( {z - M_k^2 } \right)^2  + M_k^2 \Gamma _k^2 } \right] - a_1 \left( {z - M_k^2 } \right) \Biggr\}\nonumber\\  
\label{resintfinal}	
\end{align}

we have ignored the constant of integration in (\ref{resintfinal}) since (\ref{resform}) is a definite integral and when we evaluate using (\ref{resintfinal}), what ever constant that was there would subtract to zero.

Finally leading to 
\begin{eqnarray}
\int\limits_{9M_\pi ^2 }^R {\frac{1}
{z}\Delta \left( z \right)z\Theta \left( z \right)BW_k \left( z \right)dz}  = G_k \left( R \right) - G_k \left( {9M_\pi ^2 } \right)
		\label{resintfinalform}
\end{eqnarray}


\section{Exact solution of the RGE for $\alpha_s$  up to fifth order}
\label{sec:ExactSolutionOfTheRGEForAlphaSUpToFifthOrder}

A problem that has been on the authors mind is that of an exact solution to the Renormalization Group Equation(RGE) . Finding exact solutions has turned into a slight obsession! So on a lighter note we attempt to produce an exact solution to the RGE for the strong coupling $\alpha_s$

The evolution of $\alpha_s$ is governed by the differential equation
\begin{eqnarray}
\frac{{d{\rm a}_{\rm s} (x)}}{{dx}} =  - i\sum\limits_{n = 0}^3 {\beta _n {\rm a}_{\rm s}^{{\rm n + 2}} (x)} 
\label{asode}
\end{eqnarray}

which is expanded into
\begin{eqnarray}
\frac{{d{a_ s}(x)}}{dx} =  - i\left[ \beta _0 a_s ^2  + \beta _1 a_s ^3  + \beta _2 a_s ^4  + \beta _3 a_s ^ 5 \right]
\end{eqnarray}

factoring out $\beta _3 a_s ^2$ we obtain
\begin{eqnarray}
\frac{{d{a_ s}(x)}}{dx}  =  - i\beta _3 a_s ^2 \left[ a_s ^3 + \frac{\beta _2}{\beta _3 }a_s ^2  + \frac{\beta _1 }{\beta _3 } a_ s  + \frac{\beta _0 }{\beta _3  } \right]
\end{eqnarray}

for convenience we rename $a_s \rightarrow z$ and let
\begin{eqnarray}
\begin{array}{l}
 a = \frac{\beta _2 } {\beta _3 } \\ \\
 b = \frac{\beta _1 } {\beta _3 } \\ \\
 c = \frac{\beta _0 } {\beta _3 } \\ 
 \end{array}
\end{eqnarray}

we then obtain
\begin{eqnarray}
\frac{{dz(x)}}{{dx}} =  - i\beta _3 z^2 \left[ {z^3  + az^2  + bz + c} \right]
\label{NODE}
\end{eqnarray}

So this is the differential equation that must be solved to find the evolution of the strong coupling $z$
we note at this point that there is a cubic polynomial on the right-hand side of the above equation
\begin{eqnarray}
p\left( z \right) = z^3  + az^2  + bz + c
\end{eqnarray}

for
\begin{eqnarray} 
\mathop {\lim }\limits_{z \to  + \infty } p\left( z \right) \to  + \infty  \\ 
 \mathop {\lim }\limits_{z \to  - \infty } p\left( z \right) \to  - \infty  
\end{eqnarray}

now by the Intermediate Value Theorem/Bolzano's Theorem \cite{IVT} we are guaranteed of the existence of at least one root of $p(z)$. Let $z=\alpha$ be this root so $p(\alpha)=0$, then we can proceed with long division on
$p(z)$ and write $p(z)$ as a product of lower order polynomials i.e.
\[
p\left( z \right) = \left( {z - \alpha } \right)\left[ {z^2  + ez + f} \right]
\]

with
\[
\begin{array}{l}
 e = a + \alpha  \\ 
 f = b + \alpha a + \alpha ^2  \\ 
 \end{array}
\]

So getting back to the problem of the solution to the differential equation we can now write (\ref{NODE}) as
\[
\frac{{dz(x)}}{{dx}} =  -i \beta _3 z^2 \left( {z - \alpha } \right)\left[ {z^2  + ez + f} \right]
\]

This is a separable differential equation so
\begin{eqnarray}
\int { - i\beta _3  \textit{ } dx}  = \int {\frac{{dz}}{{z^2 \left( {z - \alpha } \right)\left[ {z^2  + ez + f} \right]}}} 
\label{SODE}
\end{eqnarray}

The left hand side is trivially integrated but the right-hand side requires some work, we note that the integrand can be expressed as
\[
\frac{1}{{z^2 \left( {z - \alpha } \right)\left[ {z^2  + ez + f} \right]}} = \frac{{Az + D}}{{z^2 }} + \frac{B}{{\left( {z - \alpha } \right)}} + \frac{{Cz + E}}{{\left[ {z^2  + ez + f} \right]}}
\]

with a suitable choice of the coefficients $A,B,C,D,E$ we shall expand more about how they are chosen, by simplifying the above expression and using a common denominator we have:
\begin{align}
&\frac{1}{{z^2 \left( {z - \alpha } \right)\left[ {z^2  + ez + f} \right]}} \nonumber\\
=& \frac{{(A + B + C)z^4  + \left[ {(e - \alpha )A + eB - \alpha C + D + E} \right]z^3  + \left[ {(f - e\alpha )A + fB + (e - \alpha )D - \alpha E} \right]z^2 }}{{z^2 \left( {z - \alpha } \right)\left[ {z^2  + ez + f} \right]}}\nonumber\\
+& \frac{{\left[ { - \alpha fA + (f - e\alpha )D} \right]z - \alpha fD}}{{z^2 \left( {z - \alpha } \right)\left[ {z^2  + ez + f} \right]}}
\label{integrand}	
\end{align}
since the numerator of left-hand side of (\ref{integrand}) is a polynomial of order zero the numerator of the right-hand side also has to be a polynomial of order zero so we have to have
\begin{eqnarray}
 A + B + C &=& 0 
 \label{cond1}\\ 
 (e - \alpha )A + eB - \alpha C + D + E &=& 0 \nonumber \\ 
 (f - e\alpha )A + fB + (e - \alpha )D - \alpha E &=& 0 \nonumber \\ 
  - \alpha fA + (f - e\alpha )D &=& 0 \nonumber \\ 
  - \alpha fD &=& 1 \nonumber 
\end{eqnarray}

we now have a system of five equations with five unknowns which can be solved uniquely to give
\begin{eqnarray} 
 A &=&  - \frac{{f - e\alpha }}{{\alpha ^2 f^2 }} \\ 
 B &=& \frac{1}{{\alpha ^2 \left( {\alpha ^2  + e\alpha  + f} \right)}} \\ 
 C &=& \frac{{f - e^2  - e\alpha }}{{f^2 \left( {\alpha ^2  + e\alpha  + f} \right)}} \\ 
 D &=&  - \frac{1}{{\alpha f}} \\ 
 E &=& \frac{{\left( {f - e^2 } \right)\alpha  + \left( {2f - e^2 } \right)e}}{{f^2 \left( {\alpha ^2  + e\alpha  + f} \right)}} 
\end{eqnarray}

so the integral
\[
\int {\frac{{dz}}{{z^2 \left( {z - \alpha } \right)\left[ {z^2  + ez + f} \right]}}}
\label{rhs}
\]

can be expressed as 
\[
\int {\left[ {\frac{{Az + D}}{{z^2 }} + \frac{B}{{\left( {z - \alpha } \right)}} + \frac{{Cz + E}}{{\left[ {z^2  + ez + f} \right]}}} \right]dz} 
\]

and then simplifying a bit
\[
\int {\left[ {\frac{A}{z} + \frac{D}{{z^2 }} + \frac{B}{{\left( {z - \alpha } \right)}} + \frac{C}{2}\frac{{2z + e - e}}{{\left[ {z^2  + ez + f} \right]}} + \frac{E}{{\left[ {z^2  + ez + f} \right]}}} \right]dz} 
\]

then separating the fourth term and combining with the fifth
\[
\int {\left[ {\frac{A}{z} + \frac{D}{{z^2 }} + \frac{B}{{\left( {z - \alpha } \right)}} + \frac{C}{2}\frac{{2z + e}}{{\left[ {z^2  + ez + f} \right]}} + \left[ {E - \frac{1}{2}eC} \right]\frac{1}{{\left[ {z^2  + ez + f} \right]}}} \right]dz} 
\]

now completing the square on the denominator of the last term we obtain
\[
\int {\left[ {\frac{A}{z} + \frac{D}{{z^2 }} + \frac{B}{{\left( {z - \alpha } \right)}} + \frac{C}{2}\frac{{2z + e}}{{\left[ {z^2  + ez + f} \right]}} + \left[ {E - \frac{1}{2}eC} \right]\frac{1}{{\left[ {\left( {z + \frac{1}{2}e} \right)^2  + \frac{{4f - e^2 }}{4}} \right]}}} \right]dz} 
\]

if \[
{\frac{{4f - e^2 }}{4}} > 0
\]

then
\begin{eqnarray}
\int {\left[ {\frac{A}{z} + \frac{D}{{z^2 }} + \frac{B}{{\left( {z - \alpha } \right)}} + \frac{C}{2}\frac{{2z + e}}{{\left[ {z^2  + ez + f} \right]}} + \left[ {E - \frac{1}{2}eC} \right]\frac{4}{{4f - e^2 }}\frac{1}{{\left[ {\frac{4}{{4f - e^2 }}\left( {z + \frac{1}{2}e} \right)^2  + 1} \right]}}} \right]dz} 
\label{suitform}
\end{eqnarray}

we can now integrate (\ref{suitform}) exactly to give
\[
A\ln z - \frac{D}{z} + B\ln \left( {z - \alpha } \right) + \frac{1}{2}C\ln \left( {z^2  + ez + f} \right) + \frac{{2\left( {E - \frac{1}{2}eC} \right)}}{{\sqrt {4f - e^2 } }}\arctan \left[ {\frac{{2\left( {z + \frac{1}{2}e} \right)}}{{\sqrt {4f - e^2 } }}} \right] + G
\]

where G is the constant of integration, we can tidy the above equation by combining the logarithmic terms together to give
\begin{eqnarray}
\ln \left[ {z^A \left( {z - \alpha } \right)^B \left( {z^2  + ez + f} \right)^{C/2} } \right] - \frac{D}{z} + \frac{{2\left( {E - \frac{1}{2}eC} \right)}}{{\sqrt {4f - e^2 } }}\arctan \left[ {\frac{{2\left( {z + \frac{1}{2}e} \right)}}{{\sqrt {4f - e^2 } }}} \right] + G
\label{rhsfin}
\end{eqnarray}

now (\ref{rhsfin}) is the right-hand side of (\ref{SODE})
which gives
\begin{eqnarray}
 -i \beta _3  x = \ln \left[ {z^A \left( {z - \alpha } \right)^B \left( {z^2  + ez + f} \right)^{{C/2}} } \right] - \frac{D}{z} + \frac{{2\left( {E - \frac{1}{2}eC} \right)}}{{\sqrt {4f - e^2 } }}\arctan \left[ {\frac{{2\left( {z + \frac{1}{2}e} \right)}}{{\sqrt {4f - e^2 } }}} \right] + G
\label{sol}
\end{eqnarray}

and now changing back to the original notation
\begin{eqnarray}
 -i \beta _3 x = \ln \left[ a_s ^A \left( a_s  - \alpha  \right)^B \left( {a_s ^2  + e a_s  + f} \right)^{C/2 } \right] - \frac{D}{{a_s }} + \frac{{2\left( {E - \frac{1}{2}eC} \right)}}{{\sqrt {4f - e^2 } }}\arctan \left[ {\frac{{2\left( {a_s  + \frac{1}{2}e} \right)}}{{\sqrt {4f - e^2 } }}} \right] + G
\label{finsol}
\end{eqnarray}

Now equation (\ref{asode}) was obtained from the following change of variables $y=y_0 e^{ix}$ now reverting back to the original notation for $a_s (y)$ we obtain
\begin{eqnarray}
 - \beta _3 \ln\left( \frac{y}{y_0} \right) = \ln \left[ a_s ^A \left( a_s  - \alpha  \right)^B \left( {a_s ^2  + e a_s  + f} \right)^{C/2 } \right] - \frac{D}{{a_s }} + \frac{{2\left( {E - \frac{1}{2}eC} \right)}}{{\sqrt {4f - e^2 } }}\arctan \left[ {\frac{{2\left( {a_s  + \frac{1}{2}e} \right)}}{{\sqrt {4f - e^2 } }}} \right] + G
\label{finsol1}
\end{eqnarray}

This is an exact solution to the RGE for $a_s(y)$ up to fifth order $a_s(y)$! 

We have tried with very little success to invert (\ref{finsol1}) to get $a_s(y)$ but we have had no success. To check intuitively if this gives the right behavior we expect that for large $y$ which corresponds to large energies, the coupling will tend to zero
so

\begin{align*}
 \mathop {\lim }\limits_{y \to \infty } {\rm LHS} &= \mathop {\lim }\limits_{y \to \infty } \left[ { - \beta _3 \ln\left( \frac{y}{y_0} \right)} \right] \to  - \infty \nonumber \\ 
\mathop {\lim }\limits_{a_s  \to 0^ +  } {\rm RHS} &= \mathop {\lim }\limits_{a_s  \to 0^ +  } \Biggl[ A\ln a_s  - \frac{D}{{a_s }} + B\ln \left( {a_s  - \alpha } \right) + \frac{1}{2}C\ln \left( {a_s ^2  + e a_s  + f} \right) +\nonumber\\
&+ \frac{{2\left( {E - \frac{1}{2}eC} \right)}}{{\sqrt {4f - e^2 } }}\arctan \left[ {\frac{{2\left( {a_s  + \frac{1}{2}e} \right)}}{{\sqrt {4f - e^2 } }}} \right] + G \Biggr]	 \nonumber \\ 
 &= \mathop {\lim }\limits_{a_s  \to 0^ +  } \left[ {A\ln a_s  - \frac{D}{{a_s }}} \right] \to  - \infty \nonumber
\end{align*}

so the limits of left and right-hand side agrees and behaves according to our intuition for the high energies/small coupling. On the other hand we expect that for small energies the coupling tends to large values,
\begin{align}
 \mathop {\lim }\limits_{y \to y_0 ^+ } {\rm LHS} &= \mathop {\lim }\limits_{y \to y_0 ^+} \left[ { - \beta _3 \ln\left( \frac{y}{y_0} \right)} \right] \to 0 \nonumber \\ 
\mathop {\lim }\limits_{a_s  \to \infty } {\rm RHS} &= \mathop {\lim }\limits_{a_s  \to \infty } \left[ {\ln \left[ {a_s} ^A \left( {a_s  - \alpha } \right)^B \left( {a_s ^2  + e a_s  + f} \right)^{C/2} \right] - \frac{D}{{a_s }} + \frac{{2\left( {E - \frac{1}{2}eC} \right)}}{{\sqrt {4f - e^2 } }}\arctan \left[ {\frac{{2\left( {a_s   + \frac{1}{2}e} \right)}}{{\sqrt {4f - e^2 } }}} \right] + G} \right] \nonumber \\ 
&= \mathop {\lim }\limits_{a_s  \to \infty } \ln \left[ a_s ^A \left( {a_s  - \alpha } \right)^B \left( {a_s ^2  + e a_s  + f} \right)^{C/2} \right] \nonumber \\ 
&= \mathop {\lim }\limits_{a_s  \to \infty } \ln \left[ a_s ^A a_s ^B \left( {1 - \frac{\alpha }{{a_s }}} \right)^B a_s ^C \left( {1 + \frac{e}{{a_s }} + \frac{f}{{a_s ^2 }}} \right)^{C/2} \right] \nonumber \\ 
 &= \mathop {\lim }\limits_{a_s  \to \infty } \ln \left[ a_s ^{A + B + C} \left( {1 - \frac{\alpha }{{a_s }}} \right)^B \left( {1 + \frac{e}{{a_s }} + \frac{f}{{a_s ^2 }}} \right)^{{C/2} } \right] \nonumber
\end{align}
but from (\ref{cond1}): $A+B+C=0$ we see that
\begin{eqnarray}
 \mathop {\lim }\limits_{a_s  \to \infty } {\rm RHS} &=& \mathop {\lim }\limits_{a_s  \to \infty } \ln \left[ \left( {1 - \frac{\alpha }{{a_s }}} \right)^B \left( {1 + \frac{e}{a_s } + \frac{f}{a_s ^2 }} \right)^{C/2} \right] \nonumber \\ 
 \mathop {\lim }\limits_{a_s  \to \infty } {\rm RHS} &=& \mathop {\lim }\limits_{a_s  \to \infty } \ln \left[ \left( {1 - 0} \right)^B \left( {1 + 0 + 0} \right)^{C/2} \right] \to 0 \nonumber 
\end{eqnarray}

so the limits of left and right-hand side agrees and behaves according to our intuition for the small energies/high coupling! We have to be careful when taking the limit of the LHS as $y \rightarrow 0^+$ as the LHS tends to $+\infty$. This is just the result of the truncation of the series expansion of the QCD $\beta(a_s)$ function and if we had managed to include all the terms, this divergence would not show up!

This (\ref{finsol}) is the general class of solutions, to fix $G$ we need $y, a_s(y)$ at a particular point.

This exact solution rests upon the fact that we can find the root $\alpha$ such that $p(\alpha)=0$, but how is this done? The method to find the root of the cubic was developed by the mathematician
Cardano to provide an exact solution to the roots of a cubic function.

\clearpage


\bibliographystyle{unsrt}
\bibliography{MScBib}

\begin{thebibliography}{10}

\bibitem{Gross1}
David~H Politzer.
\newblock Reliable perturbative results for strong interactions?
\newblock {\em Physical Review Letters}, 30(26):1346, 1973.

\bibitem{Gross2}
David~J Gross and Frank Wilczek.
\newblock Ultraviolet behavior of non-abelian gauge theories.
\newblock {\em Physical Review Letters}, 30(26):1343, 1973.

\bibitem{shifman1}
Mikhail~A Shifman, Arkady~I Vainshtein, and Valentin~I Zakharov.
\newblock {QCD} and resonance physics. theoretical foundations.
\newblock {\em Nuclear Physics B}, 147(5):385--447, 1979.

\bibitem{shifman2}
Mikhail~A Shifman, AI~Vainshtein, and Valentin~I Zakharov.
\newblock {QCD} and resonance physics. applications.
\newblock {\em Nuclear Physics B}, 147(5):448--518, 1979.

\bibitem{Jamin0}
Matthias Jamin.
\newblock Flavour-symmetry breaking of the quark condensate and chiral
  corrections to the {G}ell-{M}ann--{O}akes--{R}enner relation.
\newblock {\em Physics Letters B}, 538(1-2):71--76, 2002.

\bibitem{cad1}
CA~Dominguez, A~Ramlakan, and K~Schilcher.
\newblock Ratio of strange to non-strange quark condensates in {QCD}.
\newblock {\em Physics Letters B}, 511(1):59--65, 2001.

\bibitem{PCAC1}
William~I Weisberger.
\newblock Renormalization of the weak axial-vector coupling constant.
\newblock {\em Physical Review Letters}, 14(25):1047, 1965.

\bibitem{PCAC2}
Stephen~L Adler.
\newblock Consistency conditions on the strong interactions implied by a
  partially conserved axial-vector current.
\newblock {\em Physical Review}, 137(4B):B1022, 1965.

\bibitem{PCAC3}
Stephen~L Adler.
\newblock Sum rules for the axial-vector coupling-constant renormalization in
  $\beta$ decay.
\newblock {\em Physical Review}, 140(3B):B736, 1965.

\bibitem{Lawrie}
Ian~D Lawrie.
\newblock {\em A unified grand tour of theoretical physics}.
\newblock Taylor \& Francis, 2012.

\bibitem{grassman}
Lewis~H Ryder.
\newblock {\em Quantum field theory}.
\newblock Cambridge university press, 1996.

\bibitem{Colangelo}
Pietro Colangelo and Alexander Khodjamirian.
\newblock {QCD} sum rules, a modern perspective.
\newblock In {\em At The Frontier of Particle Physics: Handbook of QCD (in 3
  Volumes)}, pages 1495--1576. World Scientific, 2001.

\bibitem{JordansLemma1}
Elias~M Stein and Rami Shakarchi.
\newblock Real analysis, {P}rinceton {L}ectures in {A}nalysis {III}, 2005.

\bibitem{JordansLemma2}
George~B Arfken, Hans~J Weber, and Frank~E Harris.
\newblock {\em Mathematical methods for physicists: a comprehensive guide}.
\newblock Academic press, 2011.

\bibitem{JordansLemma3}
Louis~Legendre Pennisi, Louis~I Gordon, and Sim Lasher.
\newblock Elements of complex variables.
\newblock 1963.

\bibitem{tech4}
Fernando Bod{\'\i}-Esteban, J~Bordes, and J~Penarrocha.
\newblock B and b s decay constants from moments of finite energy sum rules in
  {QCD}.
\newblock {\em The European Physical Journal C-Particles and Fields},
  38:277--281, 2004.

\bibitem{tech5}
J~Penarrocha and K~Schilcher.
\newblock {QCD} duality and the mass of the charm quark.
\newblock {\em Physics Letters B}, 515(3-4):291--296, 2001.

\bibitem{Jamin}
Martin Beneke and Matthias Jamin.
\newblock $\alpha_{s}$ and the $\tau$ hadronic width: fixed-order,
  contour-improved and higher-order perturbation theory.
\newblock {\em Journal of High Energy Physics}, 2008(09):044, 2008.

\bibitem{paper}
Cesareo~A Dominguez, Nasrallah~F Nasrallah, and Karl Schilcher.
\newblock Strange quark condensate from {QCD} sum rules to five loops.
\newblock {\em Journal of High Energy Physics}, 2008(02):072, 2008.

\bibitem{tech1}
Cesareo~A Dominguez, Nasrallah~F Nasrallah, Raoul R{\"o}ntsch, and Karl
  Schilcher.
\newblock Strange quark mass from finite energy {QCD} sum rules to five loops.
\newblock {\em Journal of High Energy Physics}, 2008(05):020, 2008.

\bibitem{tech2}
CA~Dominguez, NF~Nasrallah, RH~R{\"o}ntsch, and K~Schilcher.
\newblock Up-and down-quark masses from finite-energy {QCD} sum rules to five
  loops.
\newblock {\em Physical Review D}, 79(1):014009, 2009.

\bibitem{paper2}
KG~Chetyrkin, AL~Kataev, and FV~Tkachov.
\newblock Higher-order corrections to $\sigma_{tot}$($e^{+}+e^{-}\to$ hadrons)
  in quantum chromodynamics.
\newblock {\em Physics Letters B}, 85(2-3):277--279, 1979.

\bibitem{paper3}
Michael Dine and Jonathan Sapirstein.
\newblock Higher-order quantum chromodynamic corrections in $e^+ e^-$
  annihilation.
\newblock {\em Physical Review Letters}, 43(10):668, 1979.

\bibitem{paper4}
William Celmaster and Richard~J Gonsalves.
\newblock Analytic calculation of higher-order quantum-chromodynamic
  corrections in $e^+ e^-$ annihilation.
\newblock {\em Physical Review Letters}, 44(9):560, 1980.

\bibitem{paper5}
Timo van Ritbergen, Jozef Antoon~M Vermaseren, and Sergey~A Larin.
\newblock The four-loop beta-function in quantum chromodynamics.
\newblock {\em arXiv preprint hep-ph/9701390}, 1997.

\bibitem{paper6}
KG~Chetyrkin, CA~Dominguez, D~Pirjol, and K~Schilcher.
\newblock Mass singularities in light quark correlators: The strange quark
  case.
\newblock {\em Physical Review D}, 51(9):5090, 1995.

\bibitem{tech3}
Jos{\'e} Bordes, Jos{\'e} Pe{\~n}arrocha, and Karl Schilcher.
\newblock B and bs decay constants from {QCD} duality at three loops.
\newblock {\em Journal of High Energy Physics}, 2004(12):064, 2005.

\bibitem{tech6}
J~Bordes, J~Penarrocha, and K~Schilcher.
\newblock Bottom quark mass and {QCD} duality.
\newblock {\em Physics Letters B}, 562(1-2):81--86, 2003.

\bibitem{menke}
Sven Menke.
\newblock On the determination of alpha\_s from hadronic tau decays with
  contour-improved, fixed order and renormalon-chain perturbation theory.
\newblock {\em arXiv preprint arXiv:0904.1796}, 2009.

\bibitem{dominguez2}
Jos{\'e} Bordes, CA~Dominguez, P~Moodley, J~Penarrocha, and K~Schilcher.
\newblock Chiral corrections to the {SU}(2)$\times${SU}(2)
  gell-mann-oakes-renner relation.
\newblock {\em Journal of High Energy Physics}, 2010(5):1--16, 2010.

\bibitem{gluonc}
Thomas~G Steele, JC~Breckenridge, M~Benmerrouche, V~Elias, and AH~Fariborz.
\newblock {QCD} laplace sum rules and the $\pi(1300)$ resonance.
\newblock {\em arXiv preprint hep-ph/9706473}, 1997.

\bibitem{pagels}
Heinz Pagels and A~Zepeda.
\newblock Where are the corrections to the {G}oldberger-{T}reiman relation?
\newblock {\em Physical Review D}, 5(12):3262, 1972.

\bibitem{pdgdata}
K~Nakamura, C~Amsler, Particle~Data Group, et~al.
\newblock Particle physics booklet.
\newblock {\em Journal of Physics G: Nuclear and Particle Physics},
  37(7A):075021, 2010.

\bibitem{CADMass}
Cesareo~A Dominguez, Nasrallah~F Nasrallah, Raoul R{\"o}ntsch, and Karl
  Schilcher.
\newblock Strange quark mass from finite energy {QCD} sum rules to five loops.
\newblock {\em Journal of High Energy Physics}, 2008(05):020, 2008.

\bibitem{dominguez3}
CA~Dominguez and E~De~Rafael.
\newblock Light quark masses in {QCD} from local duality.
\newblock {\em Annals of Physics}, 174(2):372--400, 1987.

\bibitem{lqcdpre}
A.~Bazavov et~al.
\newblock {MILC results for light pseudoscalars}.
\newblock {\em PoS}, CD09:007, 2009.

\bibitem{Amoros}
G~Amorosa.
\newblock {QCD} isospin breaking in meson masses, decay constants and quark
  mass ratios.
\newblock {\em Nucl. Phys. B}, 174(2):372--400, 602,87 (2001).

\bibitem{pascual}
Pedro Pascual and Rolf Tarrach.
\newblock {\em {QCD}: Renormalization for the Practitioner}.
\newblock Springer, 1984.

\bibitem{hey}
Ian Johnston~Rhind Aitchison and Anthony~JG Hey.
\newblock {\em Gauge theories in particle physics, Volume II: {QCD} and the
  Electroweak Theory}.
\newblock CRC Press, 2003.

\bibitem{Daviera}
Michel Davier, Andreas H{\"o}cker, and Zhiqing Zhang.
\newblock The physics of hadronic tau decays.
\newblock {\em Reviews of modern physics}, 78(4):1043, 2006.

\bibitem{Daviera2}
M~Davier, S~Descotes-Genon, Andreas H{\"o}cker, B~Malaescu, and Z~Zhang.
\newblock The determination of $\alpha_s$ from $\tau$ decays revisited.
\newblock {\em The European Physical Journal C}, 56:305--322, 2008.

\bibitem{IVT}
Howard Anton, Irl~C Bivens, and Stephen Davis.
\newblock {\em Calculus: early transcendentals}.
\newblock John Wiley \& Sons, 2010.

\end{thebibliography}

\end{document}